\DocumentMetadata{} 
\documentclass[acmtiis,natbib=false]{acmart}

\AtBeginDocument{%
    }

\setcopyright{acmlicensed}
\copyrightyear{2018}
\acmYear{2018}
\acmDOI{XXXXXXX.XXXXXXX}

\acmJournal{JACM}
\acmVolume{37}
\acmNumber{4}
\acmArticle{111}
\acmMonth{8}

\RequirePackage[
    datamodel=acmdatamodel,
    style=acmauthoryear,
]{biblatex}

\usepackage{xurl}
\usepackage{multirow}
\usepackage{colortbl}
\usepackage{tabularx}
\usepackage{tabulary}
\usepackage{enumitem}
\usepackage{amsfonts}
\usepackage{amsmath}
\usepackage[nameinlink]{cleveref}
\usepackage{calc}
\usepackage{changepage}
\usepackage{wrapfig}
\usepackage{subcaption}
\usepackage{ragged2e}
\usepackage{lipsum}
\usepackage[normalem]{ulem}
\usepackage{overpic}
\usepackage[many]{tcolorbox}
\usepackage{listings}
\usepackage{siunitx}
\usepackage{fontawesome5}

\makeatletter
\def\dbvisparagraphi{\@startsection{paragraph}{4}{\z@}{1.5ex plus 0.5ex minus .2ex}{-0ex}{\normalsize\bf}}
\newcommand{\dbvisparagraph}[1]{\phantomsection\dbvisparagraphi{#1}\,\,---\,}
\makeatother

\makeatletter
\newcommand{\cleantitle}[1]{\def\@cleantitle{#1}}
\AtBeginDocument{\hypersetup{pdftitle=\@cleantitle}}
\makeatother
\hypersetup{%
    pdftitle={TEMP}%
}

\crefname{table}{table}{tables}
\crefname{figure}{figure}{figures}
\crefname{equation}{equation}{equations}
\crefname{part}{part}{parts}
\crefname{section}{section}{sections}
\crefname{subsection}{section}{section}
\crefname{subsubsection}{section}{section}
\crefname{lstlisting}{listing}{listings}

\Crefname{table}{Table}{Tables}
\Crefname{figure}{Figure}{Figures}
\Crefname{equation}{Equation}{Equations}
\Crefname{part}{Part}{Parts}
\Crefname{section}{Section}{Sections}
\Crefname{subsection}{Section}{Sections}
\Crefname{subsubsection}{Section}{Sections}
\Crefname{lstlisting}{Listing}{Listings}

\usepackage[scale=0.9]{opensans}
\definecolor{MiroBlack}{HTML}{1a1a1a}

\def\arraystretch{1.25}       
\definecolor{ColorSummary}{HTML}{A42C2C}
\definecolor{ColorDefinition}{HTML}{408E2F}
\definecolor{ColorExample}{HTML}{4F85C3}

\newcommand{\rawboxdonotuse}[4]{
    \vspace{0.5em}%
    \noindent%
    \begin{minipage}{\linewidth}%
    \begin{tabularx}{\linewidth}{ >{\columncolor{#2!30!white}\centering\arraybackslash}m{#1} >{\columncolor{#2!10!white}}X }%
    #3 & #4%
    \end{tabularx}%
    \end{minipage}%
}

\newcommand{\sumbox}[2]{
    \rawboxdonotuse{1.5cm}{ColorSummary}{\includegraphics[width=1em]{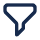}\linebreak\footnotesize Summary\linebreak\textbf{Sec.\,#1}}{\small#2}
}

\newcommand{\exbox}[2]{
    \rawboxdonotuse{2.5cm}{ColorExample}{\includegraphics[width=1em]{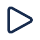}\linebreak\footnotesize Example\linebreak\textbf{#1}}{\small#2}
}

\newcommand{\defbox}[2]{
    \rawboxdonotuse{2.7cm}{ColorDefinition}{\includegraphics[width=1em]{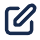}\linebreak\footnotesize Definition\linebreak\textbf{#1}}{\small#2}
}

\newcounter{requirementcategoryi}
\renewcommand*{\therequirementcategoryi}{R\arabic{requirementcategoryi}}

\newlist{requirementcategory}{description}{2}
\setlist[requirementcategory,1]{before={\setcounter{requirementcategoryi}{0}},%
    labelindent=0em, leftmargin=0em, rightmargin=0em,%
    nolistsep,%
    font={\phantomsection\normalfont\normalsize\bfseries\refstepcounter{requirementcategoryi}(\therequirementcategoryi)~}}

\newcounter{datacategoryi}
\newcounter{datacategoryii}[datacategoryi]
\renewcommand*{\thedatacategoryi}{D\arabic{datacategoryi}}
\renewcommand*{\thedatacategoryii}{\thedatacategoryi.\arabic{datacategoryii}}

\newlist{datacategory}{description}{2}
\setlist[datacategory,1]{before={\setcounter{datacategoryi}{0}},%
    labelindent=0em, leftmargin=0em, rightmargin=0em,%
    nolistsep,%
    font={\phantomsection\normalfont\normalsize\bfseries\refstepcounter{datacategoryi}(\thedatacategoryi)~}}
\setlist[datacategory,2]{%
    labelindent=0em, leftmargin=0em, rightmargin=0em,%
    nolistsep,%
    font={\phantomsection\normalfont\normalsize\bfseries\refstepcounter{datacategoryii}(\thedatacategoryii)~}}

\newcounter{dimensioncategoryi}
\renewcommand*{\thedimensioncategoryi}{I\arabic{dimensioncategoryi}}

\newlist{dimensioncategory}{description}{2}
\setlist[dimensioncategory,1]{before={\setcounter{dimensioncategoryi}{0}},%
    labelindent=0em, leftmargin=0em, rightmargin=0em,%
    nolistsep,%
    font={\phantomsection\normalfont\normalsize\bfseries\refstepcounter{dimensioncategoryi}(\thedimensioncategoryi)~}}

\newcounter{mechanismcategoryi}
\renewcommand*{\themechanismcategoryi}{M\arabic{mechanismcategoryi}}

\newlist{mechanismcategory}{description}{2}
\setlist[mechanismcategory,1]{before={\setcounter{mechanismcategoryi}{0}},%
    labelindent=0em, leftmargin=0em, rightmargin=0em,%
    nolistsep,%
    font={\phantomsection\normalfont\normalsize\bfseries\refstepcounter{mechanismcategoryi}(\themechanismcategoryi)~}}

\setdescription{labelindent=1em,leftmargin=2em}

\DeclareMathOperator{\Var}{Var}

\newlength\myheight%
\newlength\mydepth%
\settototalheight\myheight{Xygp}%
\newcommand*\inlinegraphics[1]{%
    \settototalheight\myheight{Xygp}%
    \settodepth\mydepth{Xygp}%
    \raisebox{-\mydepth}{\includegraphics[height=\myheight]{#1}}%
}%

\definecolor{GetFG}{HTML}{2f8132}
\definecolor{GetBG}{HTML}{dff2e0}
\definecolor{PostFG}{HTML}{186faf}
\definecolor{PostBG}{HTML}{dfeaf2}
\makeatletter
\newtcbox{\gettcbox}{on line,fontupper=\footnotesize\sffamily,boxrule=0.5pt,arc=2pt,coltext=GetFG,colback=GetBG,colframe=GetBG,boxsep=0pt,left=4pt,right=4pt,top=2pt,bottom=2pt}
\makeatother
\makeatletter
\newtcbox{\posttcbox}{on line,fontupper=\footnotesize\sffamily,boxrule=0.5pt,arc=2pt,coltext=PostFG,colback=PostBG,colframe=PostBG,boxsep=0pt,left=4pt,right=4pt,top=2pt,bottom=2pt}
\makeatother
\newcommand{\get}{\gettcbox{GET}}
\newcommand{\post}{\posttcbox{POST}}

\definecolor{CustomLstBackground}{HTML}{F7F7F7}
\definecolor{CustomLstForeground}{HTML}{560088}
\definecolor{CustomLstKeyword}{HTML}{134EB2}
\definecolor{CustomLstComment}{HTML}{F6981E}
\definecolor{CustomLstIdentifier}{HTML}{212121}
\definecolor{CustomLstString}{HTML}{B7141F}
\lstdefinestyle{PythonCustomLst}{
    belowcaptionskip=1\baselineskip,
    breaklines=false,
    language=Python,
    showstringspaces=false,
    backgroundcolor=\color{CustomLstBackground},
    basicstyle=\footnotesize\ttfamily,
    keywordstyle=\bfseries\color{CustomLstKeyword},
    commentstyle=\itshape\color{CustomLstComment},
    identifierstyle=\color{CustomLstIdentifier},
    stringstyle=\color{CustomLstString},
}

\begin{document}

\title{\textbf{iNNspector:} Visual, Interactive Deep Model Debugging}
\cleantitle{iNNspector: Visual, Interactive Deep Model Debugging}
\author{Thilo Spinner}
\affiliation{%
    \institution{ETH Zurich}
    \city{Zurich}
    \country{Switzerland}}
\email{thilo.spinner@inf.ethz.ch}

\author{Daniel Fürst}
\affiliation{%
    \institution{University of Konstanz}
    \city{Konstanz}
    \country{Germany}}
\email{daniel.fuerst@uni-konstanz.de}

\author{Mennatallah El-Assady}
\affiliation{%
    \institution{ETH Zurich}
    \city{Zurich}
    \country{Switzerland}}
\email{menna.elassady@ai.ethz.ch}


\begin{abstract}
    Deep learning model design, development, and debugging is a process driven by best practices, guidelines, trial-and-error, and the personal experiences of model developers.
At multiple stages of this process, performance and internal model data can be logged and made available.
However, due to the sheer complexity and scale of this data and process, model developers often resort to evaluating their model performance based on abstract metrics like accuracy and loss.
We argue that a structured analysis of data along the model’s architecture and at multiple abstraction levels can considerably streamline the debugging process.
Such a systematic analysis can further connect the developer’s design choices to their impacts on the model behavior, facilitating the understanding, diagnosis, and refinement of deep learning models.
Hence, in this paper, we (1) contribute a conceptual framework structuring the data space of deep learning experiments.
Our framework, grounded in literature analysis and requirements interviews, captures design dimensions and proposes mechanisms to make this data explorable and tractable.
To operationalize our framework in a ready-to-use application, we (2) present the iNNspector system.
iNNspector enables tracking of deep learning experiments and provides interactive visualizations of the data on all levels of abstraction from multiple models to individual neurons.
Finally, we (3) evaluate our approach with three real-world use-cases and a user study with deep learning developers and data analysts, proving its effectiveness and usability.

\end{abstract}



\received{20 February 2007}
\received[revised]{12 March 2009}
\received[accepted]{5 June 2009}

\maketitle

\section{Introduction}
\label{inn:sec:introduction}
Despite the recent prevalence of deep learning algorithms, the model-building process is still a trial-and-error one, primarily guided by the developer's experience or a few known best practices~\citep{Liu2017TowardsBetterAnalysisMachineLearningModels}.
Typically, different architecture-- and hyperparameter configurations are tested manually or by automated architecture search \citep{Elsken2019NeuralArchitectureSearch}, from which the best-performing model is then selected and further refined.
Assessment and selection are often based on performance metrics, such as loss or accuracy, which highly aggregate information and leave the model's inner workings opaque~\citep{Schneider2021CockpitPracticalDebugging}.
However, conflicting with deep-learning model's black-box character, a deep understanding of the model's inner workings is essential for well-informed model diagnosis, verification, and refinement.
Furthermore, such understanding is fundamental for trust-building and model verification, e.g., for its deployment in safety-critical environments such as in the automotive or medical sector.

Techniques for explainable artificial intelligence~\citep{Guidotti2018SurveyMethodsExplaining} try to tackle this challenge;
however, many approaches are model-agnostic, leaving the model's inner workings opaque, and operate on a local level, explaining decisions only for single input samples.
To achieve a deep understanding of the model and establish a connection between its behavior and the underlying architecture, a more holistic view on the model and its development process is needed.
Therefore, we argue that the structured \emph{debugging of neural networks} should become a routine in the daily workflow of model development and --design.
Complementary to the debugging in traditional software development \citet{Layman2013DebuggingRevisitedUnderstanding}, we define the debugging of deep learning models as follows:

\defbox{Debugging of DL Models}{
    We define the debugging of deep learning models as the process of identifying and fixing errors in the \textbf{model's architecture}, \textbf{behavior}, or \textbf{training data} either during \textbf{training} or \textbf{post-training} by iteratively calibrating the understanding of the model behavior through the structured \textbf{analysis} of its \textbf{architecture, hyperparameters, performance metrics, learned features, activations,} and \textbf{outputs}.
}

In contrast to traditional software debugging, there often is no definite connection between an error and its source.
Therefore, error identification in deep learning models might involve several techniques used separately or in combination.
For instance, \textbf{local output analysis} can help identify instances where the model performs poorly or provides incorrect predictions and then determine why these errors occur.
In contrast, \textbf{global output analysis} can reveal patterns in the model's behavior over the entire input data space, e.g., to identify biases or overfitting.
Methods for \textbf{interpretability and explainability} provide additional tools to interpret the decision-making process of a model.
For instance, attention maps in convolutional neural networks can help understand which parts of the input are most influential in the model's decisions.
While the previous techniques focus on in-- and output data and its model-internal representations, \textbf{architecture analysis} focuses on the model's computational graph.
For example, the graph can be analyzed to identify unusual or erroneous combinations of layers or operations.
Particularly relevant to assess the success of a model are \textbf{performance statistics}.
Overfitting, for instance, can be detected by checking the learning curve, which plots the model's performance on the training and validation datasets over training epochs.
Finally, \textbf{model statistics} use statistical methods to understand the model's internal state, such as weights and activations or output characteristics, for example, to identify model biases.
\textbf{Visualizations} support the abovementioned techniques, helping the analyst understand the results.
E.g., a visual representation of the architecture graph can reveal suspicious patterns~\citep{Wongsuphasawat2018VisualizingDataflowGraphs}, or a plot of activations can reveal a neuron's focus~\citep{Yosinski2015UnderstandingNeuralNetworks}.

This way of debugging allows the developer a holistic view of the model and, thus, is equally relevant in cases where there are no apparent errors in the model.
If the model is not behaving according to the developer's intention, systematic debugging also allows for the identification of error sources, indicating possible areas for model refinement.

Existing commercial and open source systems supporting model explainability and debugging \citep{TensorBoard2020TensorboardTensorflowsVisualization,WeightsBiases2021WeightsBiases} do \emph{generalize} well to different data, models, and tasks.
However, they come with narrow limitations regarding the richness of available tools, the data that is accessible after finishing the training stage, and their provided visualizations and interactions, leaving the debugging process shallow.
In contrast, there has been a multitude of visual analytics systems for explainable deep learning in the last years \citep{Endert2017StateArtIntegrating,Hohman2018VisualAnalyticsDeep}, allowing for \emph{thorough} investigation of models.
They provide highly specialized tools, visualizations, and interactions, with the downside of substantial limitations regarding their transferability to different data, architectures, or tasks.
With this work, we argue for the need to combine \emph{generalizability} and \emph{thoroughness} to bridge the prevalent research gap in state-of-the-art techniques for model debugging.
Hence, we aim to answer the following research questions arising from the identified research gap:
\emph{which steps and data in the model development workflow are of interest to the developer for debugging?
Following that, what are the mechanisms, interfaces, and visualizations a system should implement to make these units of analysis accessible?}

To enable this kind of systematic debugging, the machine learning community has to get away from tedious single-item inspection, often requiring the manual setup of logging mechanisms, time-consuming re-training of the model, and single-use data visualization pipelines.
Instead, all data arising in a machine learning experiment must be readily available during the debugging stage.
Additionally, human-interpretable representations of this data should be effortlessly generated by the machine learning developer, minimizing the entry barrier for systematic model debugging.

\noindent
To tackle these challenges, in this paper, we contribute:

\dbvisparagraph{1) Debugging Framework}
We capture and structure the design space of systematic model debugging, focusing on the entirety of data generated in machine learning experiments and the necessary mechanisms to explore it.
We provide an integrated conceptual framework that combines the structured components into an overarching approach.
The framework captures the different aspects to be considered and establishes fundamental guidelines for designing real-world model debugging applications.
To date, such comprehensive characterization and systematization of the debugging process is missing in the field.

\dbvisparagraph{2) System Implementation}
With iNNspector, we present a comprehensive system for the systematic debugging of deep learning experiments.
The system instantiates the mechanisms and guidelines formalized in our conceptual debugging framework.
The system is released as open-source software and is designed as a platform for real-world model debugging scenarios, going beyond the sole demonstration of our conceptual framework's actionability.
iNNspector is a new kind of debugging system, providing generalizing access to \emph{all} data arising in the deep learning workflow through implementing appropriate visualizations, navigation patterns, and an extensible tool palette.

\dbvisparagraph{3) Requirements Analysis and Evaluation Study}
Our conceptual debugging framework and system implementation are grounded in current research gaps, and a qualitative requirements study with real-world model developers.
We capture the developers' everyday debugging workflows and --needs, revealing significant challenges in the state-of-the-art.
Finally, we evaluate our system with model developers, assessing the applicability and usefulness of our approach and how it closes the identified gaps.

\begin{figure}[!t]
    \begin{overpic}[width=\textwidth]{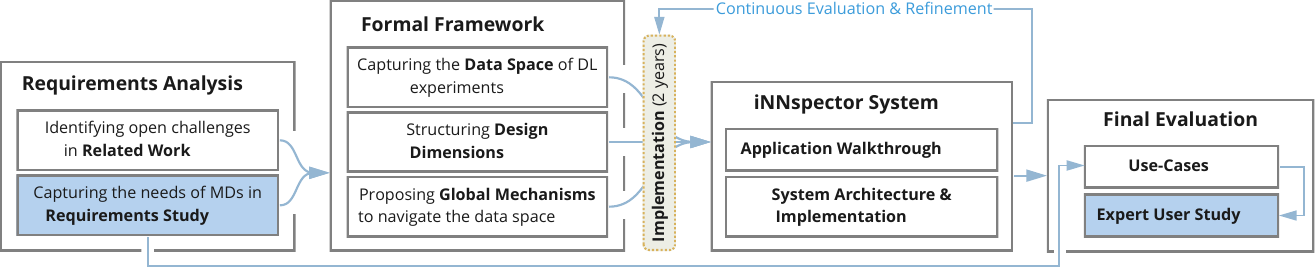}
        \opensans\scriptsize 
        \put(18.8,13.7){\scalebox{1.15}{\color{MiroBlack}{\textbf{(\ref{inn:sec:requirements})}}}}
        \put(41.75,18.1){\scalebox{1.15}{\color{MiroBlack}{\textbf{(\ref{inn:sec:framework})}}}}
        \put(71.7,12.25){\scalebox{1.15}{\color{MiroBlack}{\textbf{(\ref{inn:sec:system})}}}}
        \put(96.0,10.9){\scalebox{1.15}{\color{MiroBlack}{\textbf{(\ref{inn:sec:evaluation})}}}}

        \put(14.6,8.6){\color{MiroBlack}{\textbf{(\ref{inn:subsec:research-gap})}}}
        \put(16.0,3.6){\color{MiroBlack}{\textbf{(\ref{inn:subsec:requirements-study})}}}

        \put(38.5,13.4){\color{MiroBlack}{\textbf{(\ref{inn:subsec:data})}}}
        \put(38.5,8.45){\color{MiroBlack}{\textbf{(\ref{inn:subsec:design-dimensions})}}}
        \put(42.4,3.6){\color{MiroBlack}{\textbf{(\ref{inn:subsec:mechanisms})}}}

        \put(71.8,8.75){\color{MiroBlack}{\textbf{(\ref{inn:subsec:application-walkthrough})}}}
        \put(69.15,3.6){\color{MiroBlack}{\textbf{(\ref{inn:subsec:implementation})}}}

        \put(92.15,7.4){\color{MiroBlack}{\textbf{(\ref{inn:sec:use-cases})}}}
        \put(94.6,3.7){\color{MiroBlack}{\textbf{(\ref{inn:sec:evaluation})}}}
    \end{overpic}
    \caption[Workflow to substantiate, design, implement, and evaluate iNNspector]{
        The workflow we follow to substantiate, design, implement, and evaluate the iNNspector system for systematic model debugging, which also reflects in the structure of this paper.
    }
    \label{inn:fig:paper-structure-and-workflow}
\end{figure}
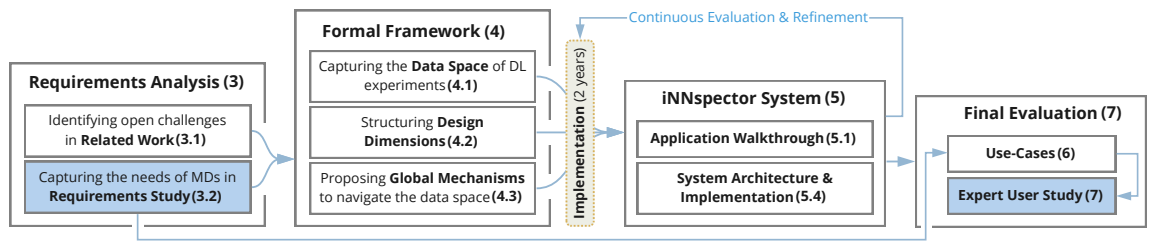

\vskip 1em
To establish our approach, we follow the workflow of substantiating, designing, implementing, and evaluating the iNNspector system for systematic model debugging, diagramed in~\cref{inn:fig:paper-structure-and-workflow}.
Starting with an extensive requirements analysis (\cref{inn:sec:requirements}), we identify open research challenges from related work (\cref{inn:subsec:research-gap}) and capture the needs of model developers in a requirements study (\cref{inn:subsec:requirements-study}).
Following, we structure these requirements into a concecptual framework (\cref{inn:sec:framework}), covering the different aspects that have to be considered for the design of a system supporting systematic model debugging.
Namely, this includes the data that is relevant during the debugging stage (\cref{inn:subsec:data}), the design dimensions that have to be considered in such a system (\cref{inn:subsec:design-dimensions}), and the mechanisms the system needs to implement to make the data explorable by the analyst (\cref{inn:subsec:mechanisms}).
Based on the conceptual framework, we present the iNNspector system for systematic model debugging (\ref{inn:sec:system}).
In a detailed application walkthrough (\cref{inn:subsec:application-walkthrough}), we explain the interface, visualizations, and interactions the system implements, linking them back to the requirements analysis and the framework.
A short description of the system architecture and --implementation (\cref{inn:subsec:implementation}) explains how iNNspector can be integrated into the existing model development workflow.
We evaluate our approach based on use cases (\cref{inn:sec:use-cases}), showcasing all components of the system.
The use cases are the foundation for an expert user study (\cref{inn:sec:evaluation}), which evaluates the usability and usefulness of system and framework.

\section{Related Work}
\label{inn:sec:related-work}
\sumbox{\ref{inn:sec:related-work}}{
    We identify several subtopics touching the field of deep model debugging, providing a broad spectrum of tools and methodologies, each with its strengths and limitations.
    In contrast to our approach, the state-of-the-art techniques only tackle sub-parts of deep learning experiments, constraining them to specific domains, models, tasks, or stages of the model development workflow.
}

The work related to the debugging of neural networks can be divided into several subtopics, including techniques for automated machine learning, explainable artificial intelligence, visual analytics for machine learning, and systems for tracking and assessing machine learning experiments.
Note that we chose to structure our literature review of explainable artificial intelligence into three subsections, namely algorithm-centered XAI (covering the technical ML perspective), human-centered XAI (covering the human-computer interaction perspective), and application-centered XAI (covering the visual analytics perspective).

\subsection{Automated Machine Learning}
\label{inn:subsec:automated-machine-learning}

\sumbox{\ref{inn:subsec:automated-machine-learning}}{
    AutoML techniques aim to automate model building and tuning.
    The subfields include hyperparameter optimization, meta-learning, and neural architecture search.
    While these techniques offer automation of model refinement, they still require systematic debugging for human control and trust-building.
}

In contrast to the manual model building, --diagnosis, and --refinement process, techniques for automated machine learning (\emph{AutoML}) strive to automate the time-consuming hyperparameter search by systematically probing the search space, as summarized by~\cite{Hutter2019AutomatedMachineLearning}.
They divide the field into three subtopics: \emph{hyperparameter optimization} is the most basic task in AutoML, addressing the search for parameter configurations (e.g., layer capacities, optimization algorithms, regularization) that maximize the model's performance.
Here, recently, Bayesian optimization-based approaches~\citep{Bergstra2011AlgorithmsHyperParameter}, multi-fidelity methods~\citep{Li2017HyperbandNovelBandit}, and a combination of both~\citep{Falkner2018BohbRobustEfficient} proved successful.
In contrast, \emph{meta learning}, or \emph{learning to learn}, strives to speed up learning a new task by learning from existing models.
In deep learning, prominent examples are transfer learning~\citep{Thrun1998LearningLearnIntroduction}, where models trained on a source task are used as a starting point to learn a target task, and few-shot learning~\citep{Wang2021GeneralizingFewExamples}, where prior experiences with similar tasks are exploited to train a model on a new task for which only few training samples are available.
Finally, \emph{neural architecture search} focuses on the automated search for model architectures~\citep{Elsken2019NeuralArchitectureSearch}, making it the most complex task of AutoML.
Several search strategies exist; evolutionary algorithms~\citep{Floreano2008NeuroevolutionArchitecturesLearning} have a long history in the field~\citep{Miller1989DesigningNeuralNetworks} and still prove successful~\citep{Real2019RegularizedEvolutionImage}.
More recently, \citeauthor{Zoph2017NeuralArchitectureSearch}'s advancements using reinforcement learning~\citep{Zoph2017NeuralArchitectureSearch} brought new momentum to the field.
Other approaches rely on Bayesian optimization~\citep{Bergstra2013MakingScienceModel} or random search~\citep{Li2020RandomSearchReproducibility}.

While AutoML techniques can automate large parts of the model refinement and parameter tuning process, systematic debugging is nevertheless relevant.
Tracking search space coverage, trust-building, and model verification are essential when the model-building process is left to an opaque optimization algorithm.
For instance, to rule out overfitting of the AutoML algorithm~\citep{Cawley2010OverFittingModel} or when model explainability is mandatory~\citep{Gunning2016ExplainableArtificialIntelligence}.

\subsection{Algorithm-Centered XAI}
\label{inn:subsec:rw-algorithm-centered-xai}

\sumbox{\ref{inn:subsec:rw-algorithm-centered-xai}}{
    Algorithm-centered XAI techniques tackle the black-box nature of neural networks by providing (external) explanations of their predictions.
    The approaches can be categorized based on attributes like being local or global, intrinsic or post-hoc, and model-agnostic or model-specific.
    While useful, Algorithm-centered XAI techniques are criticized for their fragility and limited applicability for architectural adjustments, which our approach addresses by linking model behavior to its architecture.
}

Explainable artificial intelligence (XAI) provides techniques to break up the black-box nature of deep learning models, enabling insights into their working~\citep{Gunning2016ExplainableArtificialIntelligence}.
It, therefore, spans an essential subtopic of neural network debugging.
\citet{Adadi2018PeekingBlackBox} survey popular XAI techniques and classify them according to mutual properties, such as \emph{local} (explaining single predictions) vs.~\emph{global} (explaining the model as a whole).
Furthermore, they distinguish between \emph{intrinsic} (the model is inherently explainable, e.g., \emph{decision trees}) vs.~\emph{post-hoc} (the model is a black-box and needs external explanation, e.g., \emph{DNNs}) and \emph{model agnostic} (the explanation works for different model types) vs.~\emph{model specific} (the technique only works for specific model types).
This classification is widely accepted in related literature~\citep{Spinner2020ExplainerVisualAnalytics,Molnar2022InterpretableMachineLearning}.
Simultaneously, \citet{Guidotti2018SurveyMethodsExplaining} present an extensive state-of-the-art report on XAI, formalizing the problem of explaining black-box models.
They classify explanations based on the problem faced, the type of explanator, the black box model that the explanator can open, and the type of data used as input by the black-box model.
With explAIner, \citet{Spinner2020ExplainerVisualAnalytics} structure the process of explanation, integrating the various XAI techniques into an overarching framework and system.

Guided by \citeauthor{Molnar2022InterpretableMachineLearning}'s neatly structured overview on interpretable machine learning~\citep{Molnar2022InterpretableMachineLearning}, in the following, we summarize the state-of-the-art XAI techniques most prominent in deep learning.
All presented techniques are post-hoc since deep learning models have a black-box character by default.
Here, we focus on local, attribution-based techniques.
In contrast, global techniques focus on the explanation of learned features;
due to their strong relation to the network architecture, we cover them as a subtopic of~\cref{inn:subsec:neural-network-visualization}.
Local attribution methods can be further divided into model-agnostic and model-specific techniques.
Model-agnostic methods, such as LIME~\citep{Ribeiro2016WhyShouldI}, Anchors~\citep{Ribeiro2018AnchorsHighPrecision}, or SHAP~\citep{Lundberg2017UnifiedApproachInterpreting}, perturbate the input to generate explanations, allowing to assess the importance of individual input features.
In contrast, counterfactuals~\citep{Wachter2017CounterfactualExplanationsOpening} probe the decision boundary of a model by finding the smallest possible change in the input to generate a desirable outcome, e.g., to flip the predicted class.
Saliency Maps~\citep{Simonyan2014DeepConvolutionalNetworks} and GradCAM~\citep{Selvaraju2017GradCamVisual} are examples of gradient-based techniques and are, therefore, model-specific.
They generate pixel attributions as a special form of feature attribution.

Notably, the described techniques are widely criticized for being fragile~\citep{Ghorbani2019InterpretationNeuralNetworks}, prone to manipulations~\citep{Dombrowski2019ExplanationsCanBe}, missleading~\citep{Lipton2018MythosModelInterpretability}, and barely actionable towards architectural changes~\citep{Spinner2020ExplainerVisualAnalytics}.
Our present work tackles these issues by targeting a holistic view of neural networks, inherently connecting the model's behavior to its architecture.
The discussed algorithm-centered XAI techniques can be considered potential tools in our overarching debugging framework, spanning a sub-task of neural network debugging.

\subsection{Neural Network Visualization}
\label{inn:subsec:neural-network-visualization}

\sumbox{\ref{inn:subsec:neural-network-visualization}}{
    Various tools exist for visualizing neural network architectures and learned representations.
    Our work extends the architecture view by incorporating a higher-order graph to show relationships between different model generations, providing a comprehensive view for debugging.
    Also, we simplify access to learned representations through single-click tools.
}

With the abundance of network architectures and --types, various approaches to visualize these architectures arose.
\citet{Patel2021ToolsDesignVisualize} presents a comprehensive overview of tools to visualize neural networks.
With NN-SVG, \citet{LeNail2019NnSvgPublication} provide a tool to generate SVG visualizations in the fully-connected--, LeNet--~\citep{Lecun1998GradientBasedLearning}, and AlexNet-styles~\citep{Krizhevsky2017ImagenetClassificationDeep}.
TensorSpace.js\footnote{https://tensorspace.org/} combines interactive 3D visualizations of the networks with explanations of the model's prediction process.
TensorBoard~\citep{TensorBoard2020TensorboardTensorflowsVisualization} visualizes the computational graph of TensorFlow models down to single operations.
In contrast, Netron~\citep{Roeder2022NetronVisualizerNeural} supports various DL frameworks and generates more abstract, layer-based graph representations.
\citet{Wang2019VisualGenealogyDeep} present an interactive geneology of prevalent network types and --architectures.

Besides or in conjunction with architecture visualization, many approaches focus on the visualization of learned representations, particularly in computer vision~\citep{Seifert2017VisualizationsDeepNeural}.
Common are pixel displays of activations~\citep{Gruen2016TaxonomyLibraryVisualizing} or kernels~\citep{Brachmann2016UsingConvolutionalNeural}, and heat maps~\citep{Samek2017EvaluatingVisualizationWhat}.
Approaches for feature visualization~\citep{Olah2017FeatureVisualization} strive to explain the functionality of individual neurons, channels, or layers through activation maximization, putting it at the junction of feature-based network visualization and global, algorithm-centered XAI.
\citet{Hohman2020SummitScalingDeep} combine feature visualizations with displays of the network architecture in the form of attribution graphs, associating individual dataset classes with strongly activated subnetworks.
While these approaches can provide valuable insights, they are often hard to access and require expert knowledge to implement.

The present work combines architecture visualizations of layers and neurons with visualizations of learned representations.
For the first time, we extend the visualization of model architectures with a higher-order graph, expressing the relationship between models and differences between model generations.
Furthermore, our toolbox-based approach allows single-click visualizations of learned representations, making them easily accessible.

\subsection{Human-Centered XAI}
\label{inn:subsec:human-centered-xai}

\sumbox{\ref{inn:subsec:human-centered-xai}}{
    Human-centered XAI focuses on integrating human understanding into the explanation process.
    Our work aligns with this by making model debugging an interactive process reconciling human intentions with algorithmic behavior.
}

Human-centered explainable AI (HCXAI)~\citep{Ehsan2020HumanCenteredExplainable} strives to fix the shortcoming of algorithm centered techniques discregarding the human's role in the explanation process.
Recently, there has been a surge in calls for explainability that does not only open the black-box but also considers the individual and social factors of the human interacting with the black-box~\citep{Ehsan2022HumanCenteredExplainable}.
The field of HCXAI is currently evolving;
recent positions~\citep{Singh2021LexFrameworkOperationalising}, design guidelines~\citep{Liao2020QuestioningAiInforming}, frameworks~\citep{Gobbo2022XaiPrimercomVisual}, and studies~\citep{SmithRenner2020NoExplainabilityAccountability} outline possible future directions.
Our work takes up the concept of tightly integrating the human into the machine learning loop;
model debugging inherently is about matching human intentions against the behavior of the algorithm~\citep{Reiss2014ChallengeHelpingProgrammer}.

\subsection{Application-Centered XAI}
\label{inn:subsec:application-centered-xai}

\sumbox{\ref{inn:subsec:application-centered-xai}}{
    Application-centered XAI involves visual analytics systems specialized to data, tasks, or models.
    The strong specialization of existing systems often limits reusability across different tasks and data, which our approach overcomes by structuring the debugging process in a general-purpose framework for model debugging.
}

Closely related to HCXAI are visual analytics techniques for ML and model explanation.
In recent years, various visual analytics systems for machine learning have been presented, spanning a wide range from particular application areas to tools covering one or multiple stages of the machine learning workflow~\citep{Yuan2020SurveyVisualAnalytics}.
In contrast to the algorithm-centered techniques discussed in~\cref{inn:subsec:rw-algorithm-centered-xai}, these systems not only provide visualizations of models and data but integrate the human into the interactive ML loop~\citep{Sacha2019Vis4mlOntologyVisual}.
\citet{Hohman2018VisualAnalyticsDeep} review and structure recent approaches based on the 5 \emph{W}'s and \emph{How}.
They identify different entities that can be explained (i.e., \emph{``What?''}), such as computational graph \& network architecture~\citep{Wongsuphasawat2018VisualizingDataflowGraphs}, learned model parameters~\citep{Liu2018AnalyzingTrainingProcesses}, individual computational units~\citep{Kahng2018ActivisVisualExploration}, or neurons in high-dimensional space~\citep{Strobelt2018LstmvisToolVisual}.

Most of these systems are highly specialized to data, tasks, or models, preventing re-usability and hindering application in practice.
Our present work tackles these limitations by proposing mechanisms to make the entirety of data explorable and presenting an extensible system implementing those mechanisms.
\citet{Baeuerle2022SymphonyComposingInteractive} identify a similar research gap and propose \emph{Symphony}, a framework for composing interactive interfaces
for machine learning.
While iNNspector is designed as a standalone tool, Symphony integrates into external platforms like Jupyter notebooks.
Also, the systems differ in their way of accessing, navigating, and visualizing data, e.g., Symphony provides no mechanisms to access or navigate structural data (c.f.,~\cref{inn:subsec:mechanisms}).

In their DL debugging framework \emph{Cockpit}, \citet{Schneider2021CockpitPracticalDebugging} gather higher-order information on the gradients during training and visualize them to identify common bugs in data processing, parameter selection, and model architecture.
To this end, they define different quantities tracked and computed during training and visualize them in plots called ``instruments.''
While both Cockpit and iNNspector provide plots of internal model dynamics, they differ in their focus.
Cockpit provides specialized instruments visualizing advanced internal dynamics during training.
In contrast, iNNspector provides a broader range of instruments, covering the entire deep learning workflow, complementing the plots of model metrics with visualizations of model architectures, in- and output data (e.g., images and classification results), and model-internal representations (e.g., kernels or activations).

\subsection{Systems \& Toolkits}
\label{inn:subsec:systems-and-toolkits}

\sumbox{\ref{inn:subsec:systems-and-toolkits}}{
    Various systems and toolkits exist for tracking and assessing machine learning experiments.
    While these systems provide different functionalities for model management, visualization, and assessment, a holistic view of the deep learning workflow must be included.
    Our conceptual framework structures the debugging process and proposes components and mechanisms to guide the development of such systems.
}

Keeping track of experiments is an essential task in the daily machine learning workflow.
Systems like TensorBoard~\citep{TensorBoard2020TensorboardTensorflowsVisualization} or Weights~\& Biases \citep{WeightsBiases2021WeightsBiases} provide various tools for model management, --assessment, and --comparison.
Furthermore, TensorBoard offers tools for visualizing architecture and data flow of a model~\citep{Wongsuphasawat2018VisualizingDataflowGraphs}, as well as explainability techniques~\citep{Wexler2019WhatIfTool}.
TensorWatch~\citep{Shah2019SystemRealTime} is a debugging and visualization tool that integrates into Jupyter\footnote{https://jupyter.org/} Notebooks.
The Error Analysis toolkit~\citep{Microsoft2021ErrorAnalysisToolkit} provides tools to identify and diagnose data cohorts with high error rates.
Specialized in the field of natural language processing, the Language Interpretability Tool~\citep{Tenney2020LanguageInterpretabilityTool} supports model understanding based on what-if-probing.

\vskip 1em
The discussed techniques and systems cover model debugging concerning specific domains, models, or tasks.
In contrast, this paper targets the systematic debugging of deep learning experiments (1) across all stages and (2) over any data arising in the deep learning workflow.
The conceptual framework we present structures the data and design space and provides tools and techniques to cover the spaces accordingly, without introducing constraints to domain, model, or task.

\section{The Need For Systematic Model Debugging}
\label{inn:sec:requirements}
To effectively tackle our goal of facilitating and establishing the systematic debugging of deep learning experiments in everyday machine learning workflows, we start by capturing research gaps and open challenges in existing academic and industry works.
We ground the design decisions of our approach in requirement interviews with real-world deep learning developers.
We gather shortcomings in their existing tool chains, collect real-world use cases where systematic debugging would have been crucial, and capture requirements for a system enabling such debugging.

\subsection{Research Gap and Open Challenges}
\label{inn:subsec:research-gap}

\sumbox{\ref{inn:subsec:research-gap}}{
    There exist multiple research gaps and challenges in the domain of explaining and debugging neural networks.
    Current \textbf{visual analytics} approaches are \textbf{highly specialized} either in terms of network types, domains, or goals.
    While this specialization is necessary for deep insights, \textbf{it limits} the techniques' \textbf{generalizability and transferability} to broader deep learning workflows.
    Existing \textbf{XAI techniques} mainly offer local explanations and are often model-agnostic.
    They do \textbf{not effectively explain the entire space of model inputs and outputs} and \textbf{cannot trace errors back to model architecture}.
    While \textbf{neuron- or layer-wise explanations} can provide such global views, they are limited by their \textbf{lack of integrated access to} both \textbf{the technique itself} and \textbf{lower levels of the network}.
    \textbf{Model tracking} is underrepresented in existing work, lacking visual representations of model iterations and changes between these iterations.
    \textbf{Implementation complexity} affects all of the above: elaborate data provisioning and intricate implementation act as a barrier, hindering the widespread adoption of debugging and explainability techniques.
    The identified challenges emphasize the need for a general-purpose framework to address these gaps.
}

\noindent
Systematic model debugging has recently gained momentum in the machine learning community as the foundation for trust-building, informed decision making, and effective model refinement~\citep{Lakkaraju2019DebuggingMachineLearning}.
In the following, we identify research gaps in the state-of-the-art, substantiating why this kind of debugging currently is not common practice and what open challenges have to be resolved to close these gaps.

\dbvisparagraph{Overspecialization of Current Visual Analytics Techniques}
Current research addressing the debugging of machine learning models predominantly focuses on specialized network types, domains, and analysis goals~\citep{Endert2017StateArtIntegrating,Hohman2018VisualAnalyticsDeep,Yuan2020SurveyVisualAnalytics}.

\exbox{Specialization of VA in ML}{
    Most of the recently presented approaches target a highly specific problem; thus, the presented VA techniques are also highly specialized.
    For example, \citet{Kahng2019GanLabUnderstanding} propose an approach for investigating and debugging GANs \textit{(specialization on network type)},
    \citet{Strobelt2019Seq2seqVisVisual} present a tool for the debugging of sequence-to-sequence models for text translation \textit{(specialization on network-- and data type)},
    and \citet{Cabrera2019FairvisVisualAnalytics} focus on detecting intersectional bias \textit{(specialization on analysis goal)}.
}

\noindent

It should be noted that the specialization of these tools is essential to facilitate the deep insights they can provide and to solve the specified tasks, and, therefore, is not a shortcoming.
However, our approach aims at a general technique enabling model debugging in the everyday deep learning workflow.
In this setting, the specialization of the presented tools limits transferability and generalizability, rendering them inappropriate for our goal.
Particularly, with iNNspector, we strive to structure the debugging process in a way that allows for the integration of specialized techniques while still providing a general-purpose framework for systematic model debugging.

\dbvisparagraph{Limitations of State-of-the-Art XAI Methods}
Prevalent XAI techniques primarily provide local explanations~\citep{Adadi2018PeekingBlackBox}, rendering them incapable of providing explanations over the entire space of possible model in- and outputs.
Furthermore, the explanations generated by these approaches are often model agnostic, leaving the inner working of the model and its influence on the model behavior opaque.

\exbox{Limitations of Current XAI Methods}{
    LIME~\citep{Ribeiro2016WhyShouldI} and LRP~\citep{Bach2015PixelWiseExplanations} are two popular XAI techniques that provide local explanations for a model's predictions, i.e., they explain the model's behavior for a single input sample.
    The results, often used to annotate the input with a heatmap, can reveal the most relevant features for the prediction.
    This conveys two significant limitations:
    First, only known inputs can be inspected.
    A prominent example is Tesla's autopilot misinterpreting the moon as a traffic light~\citep{JordanTeslaTech2021IncidentNumber145}, which could only be identified as problematic beforehand by LIME and LRP if images of the moon were part of the test data.
    Second, the explanations are post-hoc and, therefore, do not provide any information on \emph{why} the model behaves the way it does.
}

\noindent
This leaves such XAI techniques limited regarding the systematic model debugging since, typically, the identified errors cannot be projected back onto architectural issues and possible refinements.
The narrow constraints of existing XAI implementations in their applicability to specific data types and model architectures further complicate their usage as a general debugging tool.

\dbvisparagraph{Missing Access to Neurons}
Neuron- or layer-wise explanations, such as feature visualization~\citep{Olah2017FeatureVisualization} or attribution graphs~\citep{Hohman2020SummitScalingDeep} are a powerfull global alternative to the data-specific, local explanation of single samples.
Here, integrated access to both the techniques themselves and to lower levels of the network seems currently the most limiting factor.

\dbvisparagraph{Insufficient Model Tracking}
The problem of model tracking is mostly ignored by related work and only covered on a technical level by commercial or open-source tools, relying on tabular representations~\citep{WeightsBiases2021WeightsBiases} or Git logs~\citep{Kuprieiev2022DvcDataVersion}.
However, visual representations of model iterations and changes between iterations are currently missing.

\dbvisparagraph{Lack of Accessibility}
Finally, all of the previously mentioned aspects share the fundamental challenge of requiring elaborate data provisioning and intricate implementation, preventing the widespread use of debugging and explainability techniques in the daily workflow of deep learning developers.
Thus, ease of access and adaptability to individual tasks and domains is crucial to establishing the systematic debugging of deep learning models as a common practice.

\subsection{Capturing Developer Requirements}
\label{inn:subsec:requirements-study}

\sumbox{\ref{inn:subsec:requirements-study}}{
    We interviewed three model developers to identify requirements for a system aiding in deep learning debugging.
    The results show that:
    (1) Model management is crucial, especially change detection and model comparison.
    (2) Developers need access to high-level performance metrics and runtime visualizations.
    (3) Different levels of architectural detail are desired for comparing or closely analyzing models.
    (4) User guidance and automated mistake detection are considered useful.
    (5) Varied interest exists in visualizing neuron activations and network weights.
}

To capture the needs and expectations towards a system facilitating systematic debugging of deep learning experiments, we conduct structured requirements interviews with three real-world model developers.
In each interview, we capture the field of expertise of the developers, including tasks, data types and model architectures they typically deal with, as well as their toolchain and workflow.
Subsequently, the interview focuses on the developers' current debugging workflow, covering habits, used tools, and, particulary, concrete challenges and use cases they experienced in their work where an in-depth debugging was necessary to finally resolve issues with the model.
The use cases used for the final evaluation of our approach (see~\cref{inn:sec:use-cases}) are inspired by the experiences and workflows our developers reported in the requirement interviews.
As foundation for the design of our system, the interview continues with open-ended questions on expectations towards such a system.
Particularly, we ask about the parts of the model the developers are interested in during debugging (e.g., different parts of the architecture or metrics), and how common tasks (e.g., model comparison) could be solved in such a system.
Finally, we present early ideas of how such a system could work, particularly, how we plan to use the structural backbone (see~\cref{inn:subsec:mechanisms}, \ref{inn:mechanism:structural-backbone}) and navigation over different levels of model abstraction (see~\cref{inn:subsec:mechanisms}, \ref{inn:mechanism:levels-of-abstraction}) to provide access to all entities of an experiment.

Summarizing the results of our study, we identify the following requirements from the perspective of the model developers.
For additional results of the requirements study, including descriptions of the developer's current workflows, the data relevant for debugging, and concrete debugging use cases, please refer to~\cref{inn:apx:sec:requirements-study}.

\begin{requirementcategory}
    \item[Model Management ---\label{inn:requirement:model-management}]
    At the top of their iterative model development and refinement workflow, the model developers name model management a crucial task.
    While this also involves the organization and storage of source code and models, the most highlighted functionalities in this task are change detection and model comparison.
    All interviewed developers use, besides TensorBoard, customized workflows for model management, often relying on simple tools like filenames, folder structures, and command-line scripts.

    \item[Performance Metrics ---\label{inn:requirement:performance-metrics}]
    For high-level model overview and comparison, the model developers want to assess the performance of models based on the visualization of summarizing metrics, like loss or accuracy.
    Also, visualizing the model's runtime performance is requested for time-critical environments.

    \item[Architecture Representation ---\label{inn:requirement:architecture-representation}]
    The developers wish for an abstract architecture representation when comparing multiple models.
    For the closer analysis of one specific model, more architectural details and information about the model's parameterization are requested.
    When viewing a single model, the flow of data in the model is relevant to the developers, as well as the individual layer capacities and operations.

    \item[Inspection of Weights and Activations ---\label{inn:requirement:weights-activations}]
    When the analysis gets more detailed, e.g., when searching for specific issues in a model, the developers ask for visualizations of underlying data.
    Thereby, they seem more interested in the activations of neurons on single samples or whole dataset classes than in the trainable weights of the network.
    For both, they identify use cases where they need to inspect the distribution of values to detect issues with bottlenecks, gradients, or network capacity.
    However, the tasks and requirements involving data inspection and visualization varies strongly between developers.
    While some prefer a summarizing representation of data with no interest in weights or activations, others describe use cases in their development workflow where the detailed depiction of such data plays a crucial role.
    Notably, the developers mention that all network parts are relevant for model explainability.
    For the analysis of weights and activations, an even closer focus on specific layers is wished for.
    Particular interest is shown in visualizing the development of such variables over time, for example, for convolutional kernels.

    \item[Guidance and Abnormality Detection ---\label{inn:requirement:guidance-and-abnormality}]
    User guidance is mentioned as a useful extension by the developers to cover the vast search space that a systematic model analysis opens.
    Especially the automated detection of careless mistakes, such as accidentally applying an activation function twice after a layer, is a frequent requirement.
\end{requirementcategory}

\section{Formal Framework for the Systematic Debugging of DL Experiments}
\label{inn:sec:framework}
Several fundamental requirements and mechanisms have to be considered to facilitate the systematic debugging of deep learning experiments, which we strive to formalize with the conceptual framework presented in this section.
The framework provides guidelines and serves as a checklist for designing systems for the debugging of deep learning experiments.

We structure this section by the two substantial aspects that have to be considered for the design of such a system: the various types of data arising in deep learning experiments (\cref{inn:subsec:data}) and the necessary mechanisms and visual representations required to make this data effortlessly accessible by the deep learning developer during debugging (\cref{inn:subsec:mechanisms}).

\subsection{Capturing the Data Space of Deep Learning Experiments}
\label{inn:subsec:data}

\sumbox{\ref{inn:subsec:data}}{
    Logging quality metrics and model checkpoints is insufficient for effective model debugging.
    Thus, we define the \textbf{data space} of a machine learning experiment as all data arising during the machine learning workflow and classify this data according to its \textbf{origin}, \textbf{dimensionality}, and \textbf{variability over time} into four main categories: \textbf{structural data}, \textbf{scalar data}, \textbf{high-dimensional data}, and \textbf{functions}.
    This classification builds the foundation of our conceptual framework by formalizing the data that should be made available to the deep learning developer during debugging.
}

\noindent
The assessment of model quality and training progression is usually based on quality metrics logged during the training process.
Together with model checkpoints, which encode the model instance at a certain point for later use, they form the entirety of data being typically logged during machine learning experiments.
However, we argue that for systematic model debugging, various other data arising in the DL workflow is relevant and should be considered during the analysis phase.

Therefore, extending the \emph{data} term used in the machine learning community, usually referring to the sole in-- and output data (which we will refer to as \emph{dataset}), we define the \emph{data space} of a deep learning experiment as follows:

\defbox{Data Space of\linebreak DL Experiments}{
    We define the \textbf{data space} of a deep learning experiment as the entirety of data arising in a deep learning experiment.
    A \textbf{deep learning experiment} denotes the entire model building, diagnosis, and refinement workflow over multiple model iterations to solve a particular task.
}

\noindent
This data can be categorized according to its origin, dimensionality, and variability over time.
Additionally, we classify the data into the categories \textit{structural}, \textit{scalar}, \textit{high-di\-men\-sion\-al}, and \textit{functions}.
In the following, we give a detailed characterization of those categories, building the foundation for our proposed framework.
\Cref{inn:tab:data-types} provides an overview.

\begin{table}[t]
    \centering
    \renewcommand{\arraystretch}{1.0}
\setlength{\tabcolsep}{5pt}

\begin{tabularx}{\linewidth}{llXlX}
    \toprule
    \textbf{Category}                                                     & \textbf{Domain}                     & \textbf{Sub-category}                           & \textbf{Variability} & \textbf{Examples}                           \\ \midrule
    \multirow{2}{*}{\textbf{Structural (\ref{inn:data:structural})}}      & \multirow{2}{*}{$G = (V, E)$}                                & Tree of Models                                                           & constant                  & \multicolumn{1}{c}{---}                       \\ \cmidrule(l) {3-5}
    &                                     & Architecture Graph                              & constant             & \multicolumn{1}{c}{---}                     \\ \midrule
    \multirow{2}{*}{\textbf{Scalar (\ref{inn:data:scalar})}}              & \multirow{2}{*}{$\mathbb{R}$}                                & Hyperparameters                                                          & constant                  & Batch Size                                    \\ \cmidrule(l) {3-5}
    &                                     & Metrics                                         & variable             & Loss, Accuracy                              \\ \midrule
    \multirow{4.6}{*}{\textbf{$n$-dimensional (\ref{inn:data:high-dim})}} & \multirow{5.0}{*}{$\mathbb{R}^{n}$}                          & In-- and Output\newline ($\sim$ \emph{Dataset})                                                                & constant                               & Time-Series ($1$D)\newline Images~($2/3 $D)      \\ \cmidrule(l) {3-5}
    &                                     & Weights                                         & variable             & Bias ($1$D)\newline Dense Kernel~($2$D)     \\ \cmidrule(l) {3-5}
    &                                     & Activations                                     & variable             & \multicolumn{1}{c}{---}                     \\ \midrule
    \multirow{3.8}{*}{\shortstack[l]{\textbf{Function (\ref{inn:data:function})} \\\footnotesize c.f., Architecture Graph}}                                                                      & \multirow{4}{*}{$\mathbb{R}^{n} \rightarrow \mathbb{R}^{n}$}                                     & Structural                                    & constant             & Reshape\newline Concatenate     \\ \cmidrule(l) {3-5}
    &                                     & Mathematical                                    & constant             & Convolution\newline Activation Function     \\
    \bottomrule
\end{tabularx}
    \caption[Overview of the types of data arising in the DL workflow]{
        Overview of the different types of data arising in the deep learning workflow, each having different domains, origins, and instantiations.
        If the data is changing during training it is categorized as \emph{variable}, otherwise as \emph{constant}.
    }
    \label{inn:tab:data-types}
\end{table}

\begin{datacategory}
    \item[Structural ($G = (V, E)$) ---\label{inn:data:structural}]
    Data entities that logically form a graph.
    The graph structure can be either inherent to the data or originate from the temporal evolution of its entities (cf. a family tree).
    Following this definition, there are primarily two sources of structural data in the ML workflow, one being the evolution of models over time and the other being the architectural graph of each model.

    \begin{datacategory}
        \item[Tree of Models ---\label{inn:data:structural-tom}]
        Starting from an initial architecture, the iterative diagnosis and refinement process will lead to ever-new model generations with slight modifications compared to the ancestor generation.
        Addressing~\textbf{(\ref{inn:requirement:model-management})}, we propose to model this process as a directed graph, with nodes representing model architectures and links indicating a parent-child relationship.
        It should be noted that this graph does not always stricly form a tree since desirable characteristics of multiple models in the ancestor generation might be merged into a new architecture, embodying multiple inheritance.
        \item[Architecture Graph ---\label{inn:data:structural-ag}]
        The architecture graph represents the order of operations applied to data while flowing through the model as a directed graph.
        This graph representation, targeting~\textbf{(\ref{inn:requirement:architecture-representation})}, can be of varying granularity; typically, nodes encode the building blocks modern ML frameworks offer for model building, such as layers or functions, while links indicate the data flow between nodes.
    \end{datacategory}

    Structural data is considered constant over time.
    While this assumption seems evident for the architectural graph of a model, it might at first be contra-intuitive for the temporal evolution of models.
    However, the different architectures could have been determined since the beginning of the experiment.
    They do not depend on training, nor do individual nodes change over time.
    Therefore, we consider the tree of models final under the most recent model iteration.

    \item[Scalar ($\mathbb{R}$) ---\label{inn:data:scalar}]
    Real number describing the model configuration (hyperparameters) or the model state at a single point in time (metrics).
    \begin{datacategory}
        \item[Hyperparameters]
        \emph{Learning rate} or the \emph{batch size} are a form of scalar data describing properties which significantly influence the performance of the model,
        despite neither being part of the architecture nor the trained instance.
        They are (usually) only defined once and, therefore, considered constant over time.
        \item[Metrics]
        \emph{Accuracy} and \emph{loss}~\textbf{(\ref{inn:requirement:performance-metrics})} are typical representatives of this category.
        However, other scalar descriptors might be relevant for debugging, such as a model's execution time or its memory consumption.
        Metrics are considered variable over time, i.e., changing throughout the training process.
    \end{datacategory}

    \item[High-Dimensional ($\mathbb{R}^{n}$) ---\label{inn:data:high-dim}]
    Vectors or matrices representing human-interpretable, but also abstract data entities.
    For example, visual representations of image data can be readily understood by humans, whereas abstract data distributions might require transformations or targeted visualizations to make them approachable by a human.
    High-dimensional data covers~\textbf{(\ref{inn:requirement:weights-activations})}.
    It often consists of several thousand individual values and therefore places high demands on storage, memory, and calculations.
    In the context of the DL workflow, we distinguish different sub-categories of high-dimensional data, depending on their origin:

    \begin{datacategory}
        \item[In- / Output ---]
        Data that is fed into the model or matched against the model output.
        In-- and output data describe either single samples or aggregations
        used for model training and --inference.
        It is considered constant over time, since the individual samples do not change during training.
        In-- and output data is $1$ to $n$-dimensional.
        Examples include time series ($1$D) or images ($2/3$D).
        \item[Weights ---]
        Data that is inherent to the current network instance.
        We refer to weights as all the network variables that are learned during the training process.
        Therefore, they are considered variable over time.
        The dimensionality of weights can range from $1$D (e.g., biases), over $2$D (e.g., dense kernels) to $n$D (e.g., convolution kernels).
        Their values are often initialized according to a certain strategy, e.g., by sampling from a Gaussian distribution.
        \item[Activations ---]
        Data that is dependent both on the network instance and the network input.
        Specifically, activations are the response of each network entity under the data fed through the network.
        Since activations combine input data and weights, they change over the training process of the network and, therefore, are considered variable over time.
        Activations can be human-interpretable by default (e.g., in convolutional networks) or abstract $1$ to $n$-dimensional distributions (e.g., dense networks).
    \end{datacategory}

    \item[Function ($\mathbb{R}^{n} \rightarrow \mathbb{R}^{n}$) ---\label{inn:data:function}]
    Mappings between number ranges or matrix shapes.
    Functions can be seen as the elements forming the \emph{architecture graph}.
    A function's output is only dependent on its input; thus, functions are considered constant over the training time.
    In terms of neural networks, we distinguish two sub-categories:

    \begin{datacategory}
        \item[Structural Functions ---]
        Functions that modify the shape or structure of data entities with the primary goal to render them compatible with mathematical functions.
        Examples include concatenation or reshaping.
        \item[Mathematical Functions ---]
        Mathematical operations, mapping an input to a specific output.
        Regarding neural networks, many mathematical functions take learned weights as additional parameters.
        The weights are iteratively adjusted during training to make the function results resemble the training data distribution as close as possible.
        Examples include matrix multiplications, convolution operations, or activation functions.
    \end{datacategory}

\end{datacategory}

\noindent
While the variability of each of the introduced data categories might at first seem secondary, it is fundamental for the conceptual framework we derive in the following.
Particularly, (1) constant data entities can be used as a backbone for variable data and (2) we do not have to resolve the progression over the course of training when visualizing them.

\subsection{Structuring Design Dimensions for Representation and Interaction}
\label{inn:subsec:design-dimensions}

\sumbox{\ref{inn:subsec:design-dimensions}}{
    To provide the necessary flexibility for a general-purpose debugging framework, we propose a modular approach for creating \emph{debugging components}.
    A debugging component is a user-interface element visualizing one of the discussed data entities.
    Guiding the design of debugging components, we derive different aspects and constraints that have to be considered to make data accessible to the user during the debugging process, called \emph{design dimensions}.
    Overall, we identify $6$ design dimensions that influence the instantiation and appearance of a debugging component.
}

A \emph{debugging component} is a modular user-interface element visualizing one of the discussed data entities.
Following, we describe how a debugging component is instantiated and which dimensions influence its appearance.

Besides the previously covered data space, the systematic debugging of DL experiments involves various other dimensions.
\Cref{inn:tab:vis-dimensions} gives an overview of those dimensions, including their characteristics and providing examples for each.
Depending on the use case and the individual machine learning workflow, other important characteristics may arise, complementing the pre-identified aspects.
In the following, a short description of each dimension is given to substantiate the dimensions and their characteristics.

\subsubsection{Design Dimensions for Data Representation and Interaction}
\label{inn:subsubsec:design-dimensions}

\begin{dimensioncategory}
    \item[Task ---\label{inn:dim:task}]
    The goal of the analysis and, thereby, the debugging process.
    The analysis task is strongly related to the underlying machine learning task, i.e., the task for which the model is designed.
    For example, in a setting where high classification accuracy is demanded, \textbf{model assessment} might be the primary analysis task.
    In contrast, if the model is designed to be used in a security-critical environment, formal \textbf{verification} will likely be of absolute priority.
    The analysis task is also influenced by the stage of the machine learning workflow.
    Usually, multiple architectures are compared against each other during the initial model selection, focusing on model \textbf{assessment} and \textbf{--comparison}.
    In a later stage, when the rough architecture is finished and should be fine-tuned on the machine learning task, \textbf{correctness checking} is needed to verify if the model behaves according to human intention.
    For example, local XAI methods might be used in this stage, or distributions in the latent space of the model could be investigated.

    \item[Level of Abstraction ---\label{inn:dim:lofa}]
    The granularity in which the model is inspected.
    The level of abstraction strongly depends on the stage of the debugging process.
    In the beginning, an overview over \textbf{multiple models} might be required, e.g., to assess and compare their performance.
    Then, exciting details like abnormalities in their architectures or a high loss can be tracked down into \textbf{single models}.
    For an even closer debugging, architectural details or kernel distributions of single \textbf{layers} might be observed to understand where a particular model behavior originates.
    Finally, for highly specialized use cases, even an inspection of single \textbf{neurons} or \textbf{weights} might be needed, e.g., to assess the performance of pruning strategies or by using advanced XAI techniques, such as activation maximization with a deep dream approach.

    \begin{table}[t]
        \centering
        \renewcommand{\arraystretch}{1}
\setlength{\tabcolsep}{6pt}

\begin{tabularx}{\linewidth}{>{\hsize=0.75\hsize}X>{\hsize=0.75\hsize}X>{\hsize=1.5\hsize}X}
    \toprule
    \textbf{Dimension}                                                      & \textbf{Characteristics}                  & \textbf{Examples}                                   \\
    \midrule
    \textbf{Data} & \multicolumn{2}{c}{\emph{See \hyperref[inn:subsec:data]{\cref*{inn:subsec:data},~``\nameref*{inn:subsec:data}''.}}} \\
    \midrule
    \multirow{4}{*}{\textbf{Task (\ref{inn:dim:task})}}                     & Assessment                                & Assess model quality                                \\
    \cmidrule(l){2-3}
    & Verification / \newline correctness check & Verify that model behavior follows human intentions \\
    \cmidrule(l){2-3}
    & Comparison                                & Compare architctures                                \\
    \midrule
    \multirow{4}{*}{\shortstack[l]{\textbf{Level of} \\\textbf{abstraction (\ref{inn:dim:lofa})}}}                                                                        & Multi-model                              & Model change tracking                                    \\
    \cmidrule(l){2-3}
    & Single model                            & Quality measures                                       \\
    \cmidrule(l){2-3}
    & Layers / Units                         & Distributions                 \\
    \cmidrule(l){2-3}
    & Weights / Neurons                                & Activation maximization, deep dream                               \\
    \midrule
    \multirow{4}{*}{\textbf{Processing (\ref{inn:dim:processing})}}                                                                        & None (raw)                            & Images, text, scalars                              \\
    \cmidrule(l){2-3}
    & Transformation                               & Re-shaping, projection                                 \\
    \cmidrule(l){2-3}
    & Aggregation                   & Binning, clustering              \\
    \cmidrule(l){2-3}
    & Statistical descriptors                                   & Variance, min, max, density estimation               \\
    \midrule
    \multirow{2}{*}{\textbf{Representation (\ref{inn:dim:representation})}}                                                                        & Visualization                             & Charts, histograms, images                                        \\
    \cmidrule(l){2-3}
    & Verbalization                                   & Text, tables                            \\
    \midrule
    \multirow{3}{*}{\textbf{Dependencies (\ref{inn:dim:dependencies})}}                                                                        & Dataset                                     & Image, text, categorical                             \\
    \cmidrule(l){2-3}
    & Model                                     & Feed-forward, recurrent                             \\
    \cmidrule(l){2-3}
    & Layer                                     & Dense, covn., operation                             \\
    \bottomrule
\end{tabularx}

        \caption[iNNspector design dimensions]{
            The design dimensions influencing appropriate data representations.
            Despite them being mostly orthogonal to each other, dependencies might arise upon their combination.
            For example, while \textbf{binned} values could be represented through \textbf{visualization} or \textbf{verbalization} and could be used for \textbf{assessment} or \textbf{comparison}, they can never be applied to \textbf{structural} data.
        }
        \vspace{-0.5em}
        \label{inn:tab:vis-dimensions}
    \end{table}

    \item[Processing ---\label{inn:dim:processing}]
    Transformations to bring data into a human-readable format and reduce storage-- and compuational complexity.
    For example, a weight matrix constitutes an abstract data distribution, being opaque to a human in its \textbf{raw} form and, therefore, requiring appropriate abstraction.
    Usually, relevant features should be preserved and emphasized while reducing the total amount of data, which strongly relates to the analysis task: if the analyst is only interested in the range of values in the weight matrix, minimum and maximum as \textbf{statistical descriptors} already fulfill the analysis goal.
    Ideally, the appropriate processing transforms the data into a format that directly answers the analysis question (e.g., min or max to assess the value range), is inherently interpretable by a human (e.g., images or text), or conforms with the human's mental model of the data.
    For example, the distribution of latent activations could be \textbf{aggregated} using binning, telling the human if the data distribution follows his intuition on how the values should ideally be distributed in the latent space.

    \item[Representation ---\label{inn:dim:representation}]
    How the data is presented to the user.
    The ideal representation depends on various factors, such as the analysis task or the kind of data.
    Furthermore, human perception plays an essential role: data types that are inherently interpretable by a human, such as graphs or images, benefit from a \textbf{visual} representation, while text or tabular data are ideally represented as \textbf{verbalization}.

    \item[Dependencies ---\label{inn:dim:dependencies}]
    Some debugging techniques come with limitations regarding their applicability, referred to as dependencies.
    The characteristics of the other dimensions determine the dependencies; therefore, they form an inferred dimension, denying direct user control.
    For example, a \textbf{dataset dependency} arises upon the inspection of activations, which only exists under input data.
    This contrasts with an inspection of, e.g., the model architecture, which already exists at design time and, therefore, is \textbf{dataset agnostic}.
    Other inspections might only apply to a specific type of model or layer, forming \textbf{model--} or \textbf{layer} dependencies, respectively.
    Examples are investigating time steps in RNNs or the image-based inspection of convolution kernels.
\end{dimensioncategory}

\subsubsection{Combining Design Dimensions}

\begin{figure*}[t]
    \centering
    \includegraphics[width=\textwidth]{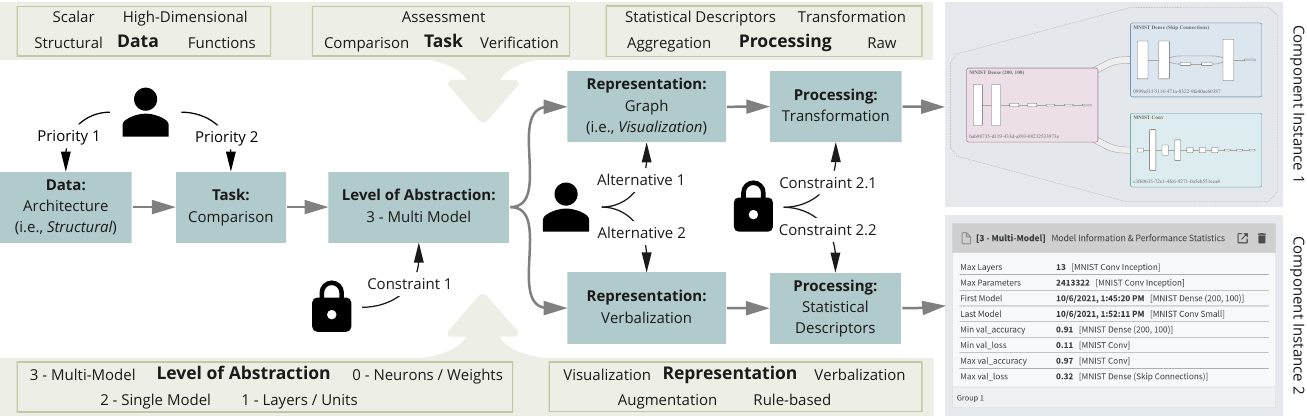}
    \caption[Exemplary instantiation of an iNNspector debugging component]{
        An exemplary instantiation of a debugging component.
        The user selects characteristics for preferred dimensions, possibly constraining other dimensions.
        Eventually, all dimensions are determined by iterating this process, and the component can be created.
    }
    \vspace{-0.5em}
    \label{inn:fig:combining-dimensions}
\end{figure*}

To combine the identified dimensions into an overarching guideline for the systematic debugging of machine learning experiments, we propose a systematization based on \emph{debugging components}.
A debugging component is a modular user-interface element, giving insights into one of the discussed data categories.
To instantiate a debugging component, one or multiple identified dimensions must be set to a particular characteristic, determining its final appearance.

\exbox{Debugging Component}{
    A classic line chart, showing the development of the loss over the training time, is a debugging component combining the following characteristics:
    \vskip 1em
    \begin{center}
        \def\arraystretch{1.0}
        \setlength{\tabcolsep}{5pt}
        \footnotesize
        \begin{tabular}{cccccc}
            \toprule
            \textbf{Data} & \textbf{Task} & \textbf{Level of Abstr.} & \textbf{Processing} & \textbf{Represent.} & \textbf{Dependencies}     \\
            \cmidrule(r){1-1}
            \cmidrule(lr){2-2}
            \cmidrule(lr){3-3}
            \cmidrule(lr){4-4}
            \cmidrule(lr){5-5}
            \cmidrule(l){6-6}
            Scalar        & Assessment    & Single Model             & Raw                 & Visual              & None                  \\
            \bottomrule
        \end{tabular}
    \end{center}
    \vskip 0.2em
}

\noindent
Our proposed framework describes the creation of debugging components as a modular approach, comparable to choosing elements from a toolbox and combining them.
Typically, the user has clear preferences on which aspects the debugging process should focus.
By choosing characteristics for preferred dimensions, the user can narrow down the appearance of the debugging component.
This process must be iterated until all dimensions are fixed by the user or by constraints arising from combinations of already-defined dimensions.
Therefore, with increasing specificity, more and more options will either be fixed or get ``greyed out'', until all dimensions are set to a particular value, determining the final appearance of the debugging component.

\Cref{inn:fig:combining-dimensions} shows an exemplary instantiation of a debugging component, following the described process.
Initially, no constraints are present, and the user can freely determine his preferred dimensions.
The \textbf{data} dimension is set to \textbf{structural} in the example since the user wants to assess the model architecture.
Next, \textbf{comparison} is set as the task, which infers the first constraint: for a comparison, multiple models must be considered simultaneously.
Therefore, a degree of freedom is removed in the \textbf{level of abstraction} dimension, automatically inferring the \textbf{multi-model} level.
There are still degrees of freedom in the dimensions \textbf{representation} and \textbf{processing}.
The user can now decide between a \textbf{visual} and a \textbf{verbal} data representation, which in both cases automatically constraints the \textbf{processing} dimension to \textbf{transformation} or \textbf{statistical descriptors}, respectively.
This leads to all dimensions being fixed, fully determining the debugging component's appearance and allowing its instantiation.

\subsection{Navigating the Data Space through Global Mechanisms}
\label{inn:subsec:mechanisms}

\sumbox{\ref{inn:subsec:mechanisms}}{
    Complementing the design dimensions that determine the appearance of a debugging component, we propose a set of global mechanisms to navigate the data space.
    The mechanisms are suggested functionalities a system for debugging DL experiments should provide.
    These mechanisms set our framework apart from existing debugging systems by proposing a comprehensive way to locate entities of interest and access their underlying data through them.
}

\exbox{Global Mechanisms vs.~Existing Systems}{
    Existing systems, such as TensorBoard, treat the logged data as independent entities that are not connected to each other.
    E.g., one tab shows the architecture graph, another tab shows scalars, and a third tab shows images.
    In contrast, our framework leverages the natural connection between the entities.
    E.g., models are composed of layers, which are composed of kernels and operations.
    Training metrics are related to a model, activations are related to layers, and weights are related to kernels.
    Therefore, our framework uses the architecture graph to navigate to the entity of interest, e.g., a convolutional layer.
    From there, the user can access the layer's weights and activations by applying tools to them, revealing their underlying data.
}

\noindent
In the following, we identify global mechanisms which the analysis toolchain can implement to make the previously structured data space explorable by the model developer.
Although these mechanisms should are considered independent of each other, they become particularly effective in combination.
\Cref{inn:fig:mechanisms} shows an overview of the mechanisms and how they work together to navigate the space.

\begin{mechanismcategory}
    \item[Units of Analysis \emph{(UoA)} ---\label{inn:mechanism:units-of-analysis}]
    The data space has to be recursively segmented into logically complete subcomponents, making them accessible by the analyst.
    For example, depending on the analysis task, the analyst might be interested in a single model, a specific layer, or even a single neuron of a model, representing units of analysis of different granularity.
    Analogously, such units of analysis can be determined for the majority of the identified data categories and sub-categories.

    \item[Structural Backbone ---\label{inn:mechanism:structural-backbone}]
    Linking abstract data to real-world entities facilitates integration into the analyst's mental model.
    For this, structural data can be used as a backbone, providing access to units of analysis in a navigable graph representation of the abstract data entities.
    Besides the visual representation of the model architecture, which is prevalent in existing tools \citep{TensorBoard2020TensorboardTensorflowsVisualization} and related works \citep{Wongsuphasawat2018VisualizingDataflowGraphs,LeCun1989BackpropagationAppliedHandwritten,Krizhevsky2017ImagenetClassificationDeep} and tracking of models in version control systems~\citep{Kuprieiev2022DvcDataVersion}, our framework proposes to additionally visualize the evolution of the models themselves in a tree-like structure, offering a powerful tool for model tracking and change analysis.

    \exbox{Applying Tools to UoAs}{
        For example, an analyst might want to debug a vision model by applying LRP~\citep{Bach2015PixelWiseExplanations} to its layers.
        He navigates to the model and layer of interest \textit{($\sim$ unit of analysis)} by descending into a visual representation of the model's architecture \textit{($\sim$ structural backbone)}.
        From there, he can apply the LRP tool to the layer, letting him select an input sample and automatically visualize the resulting saliency map.
    }

    \item[Levels of Abstraction ---\label{inn:mechanism:levels-of-abstraction}]
    The elements of the structural backbone inherently form a hierarchy, i.e., some units are sub-components of other units.
    This hierarchy is implemented in our framework as different levels of abstraction.
    The analyst can go from overview to details by navigating over these levels while keeping the global context.
    For example, if the analyst is interested in a specific filter kernel of a convolution operation, the graph can be navigated over namespaces, layers, and operations down to the kernel of interest.

    \item[Search (Highlighting) ---\label{inn:mechanism:search}]
    The vast data space spanned by machine learning experiments renders search mechanisms imperative.
    By searching for certain UoA types and, subsequently, highlighting them in the structural backbone, entities of interest can be effortlessly spotted and tracked down.
    For this, the highlights have to propagate through the structural backbone up to the highest levels of abstraction.
    E.g., by searching for convolutional and dense operations and propagating the results up to the tree of models, the analyst can effortlessly distinguish between dense and convolutional model architectures as an entry point to the analysis.

    \item[Interestingness ---\label{inn:mechanism:interestingness}]
    Complementing the search mechanism, which helps to locate data whose shape is known beforehand, interestingness detection can point the user to irregularities in the data~\citep{Geng2006InterestingnessMeasuresData}, covering~\textbf{(\ref{inn:requirement:guidance-and-abnormality})}.
    For example, the analyst might be interested in kernels whose distribution diverges significantly from a baseline.
    Interestingness measures could identify such entities as abnormal and point the user to the corresponding unit of analysis.
    Automated interestingness detection is essential since manual annotation is not feasible due to the large amount of data to cover.

    Analogous to the accepted classifications of interestingness measures in the context of data mining~\citep{Geng2006InterestingnessMeasuresData,Sharma2020ExpectedVsUnexpected}, we provide the following definition in the context of debugging neural networks.

    \defbox{Interestingness in DL Debugging}{
        In the context of debugging neural networks, we define \textbf{interestingness} as a quantifiable measure of \textbf{abnormality in the data}, described as the \textbf{deviation from a specified baseline}.
        Depending on the type of entity whose interestingness is measured, the baseline can be a fixed value, a distribution, or a rule set.
        The baseline can either be inferred from the data domain (objective), learned from other data or a ground truth (subjective), or defined by the users (user-defined).
    }

    In the following, we provide examples of interestingness measures in the context of debugging neural networks to concretize the provided definition.

    \exbox{Deviation from Learned Rule}{
        Neural networks are typically composed of pre-defined building blocks combined in a specific order, such as layers or operations.
        For example, a convolutional layer might be followed by a pooling layer in $85 \%$ of its occurrences in a dataset, while a reshape layer might be followed by another reshape layer in only $2 \%$ of its occurrences since it semantically does not make sense.
        From this observation, a rule can be learned that defines the expected order of operations, assigning a higher interestingness to the second example.
    }
    \vskip 0.5em
    \exbox{Deviation from Objective Distribution}{
        The weights of a neural network are typically initialized according to a specific strategy, e.g., by sampling from a Gaussian distribution.
        A weight that deviates significantly from the initial distribution during learning might indicate a problem in the training process, e.g., an exploding gradient.
    }

    \begin{figure*}[t]
        \centering
        \begin{overpic}[width=\textwidth]{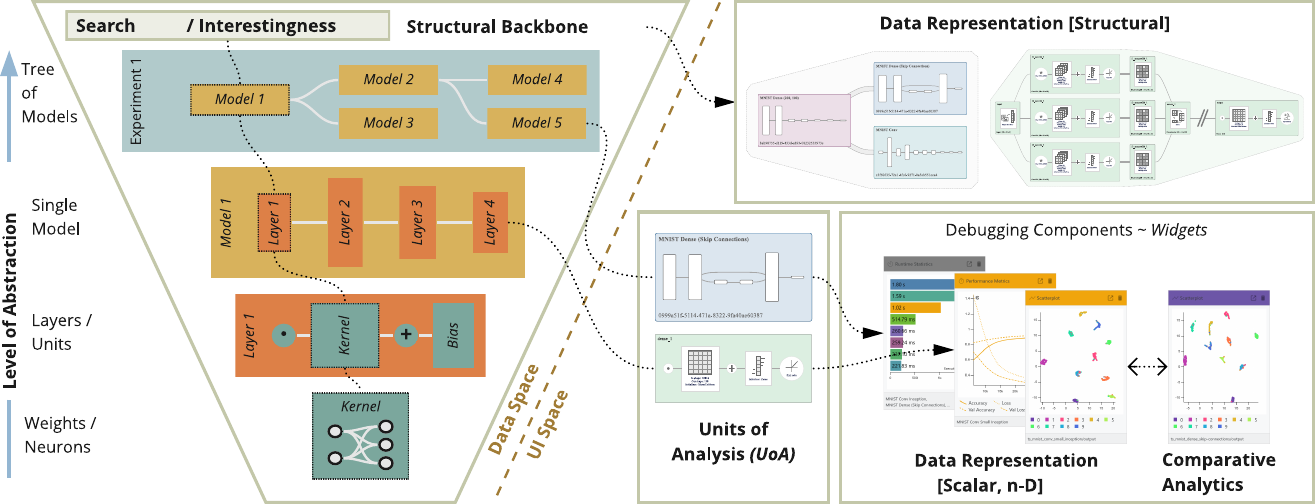}
            \opensans\scriptsize
            \put(45.2,35.9){\color{gray}{\textbf{(\ref{inn:mechanism:structural-backbone})}}}
            \put(10.6,35.95){\color{gray}{\textbf{(\ref{inn:mechanism:search})}}}
            \put(26.0,35.95){\color{gray}{\textbf{(\ref{inn:mechanism:interestingness})}}}
            \put(89.5,36.1){\color{gray}{\textbf{(\ref{inn:mechanism:data-representation})}}}
            \put(79.8,00.95){\color{gray}{\textbf{(\ref{inn:mechanism:data-representation})}}}
            \put(95.1,00.95){\color{gray}{\textbf{(\ref{inn:mechanism:comparative-analytics})}}}
            \put(54.6,01.4){\color{gray}{\textbf{(\ref{inn:mechanism:units-of-analysis})}}}
            \put(0.25,22.6){\rotatebox{90}{\color{gray}{\textbf{(\ref{inn:mechanism:levels-of-abstraction})}}}}
        \end{overpic}
        \caption[iNNspector's global mechanisms to navigate the data space]{
            Overview of the global mechanisms to navigate the data space.
            The structural data provides a backbone for navigation over multiple levels of abstraction.
            Entities of the structural data are referred to as units of analysis, which can have multiple debugging components with different characteristics attached.
            Interestingness measures and filters guide through the structural backbone to relevant units of analysis.
        }
        \label{inn:fig:mechanisms}
    \end{figure*}
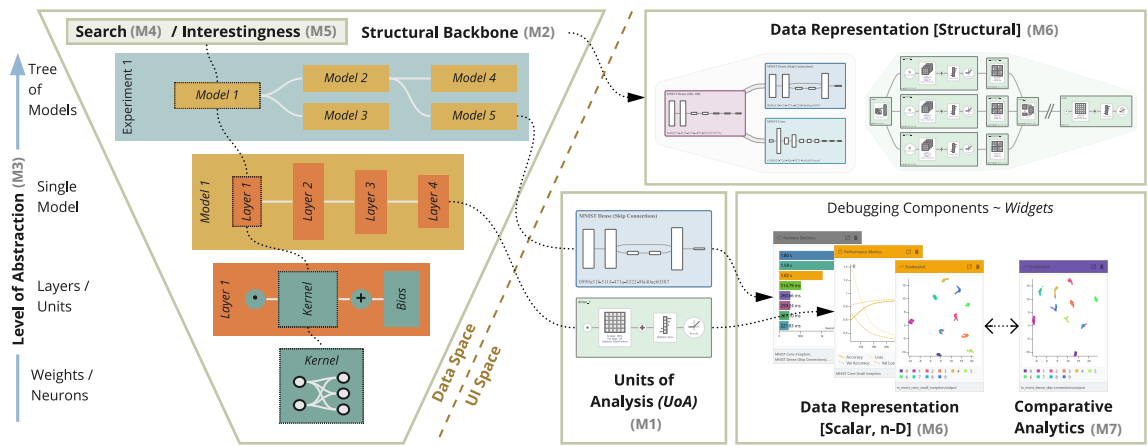

    \item[Appropriate Data Representation ---\label{inn:mechanism:data-representation}]
    Only a small share of the data arising in machine learning experiments is inherently interpretable by humans.
    Often, the data encodes complex relationships (e.g., graphs), is of high dimensionality (e.g., kernels), or of abstract type (e.g., time series).
    Therefore, processing and representation have a great impact on the accessibility of the data.
    Ideally, the data representation helps the user focus on the relevant details while insignificant or redundant information is discarded.
    Furthermore, the representation should be targeted towards human perception.
    For example, humans have an intuitive understanding of images and text; in contrast, making sense of a series of sensor values is inherently complicated.

    \item[Comparative Analytics ---\label{inn:mechanism:comparative-analytics}]
    Comparing data across different units of analysis is a fundamental use case in the debugging workflow, allowing the analyst to evaluate them against each other or reason over abnormality.
    Therefore, it is essential to enable the concurrent inspection of multiple UoAs, e.g., by implementing side-by-side views.
    In this scenario, multiple data representations from remote units of analysis might be simultaneously visible, e.g., when comparing two layers originating in different models.
    Therefore, navigational patterns for retrieving the underlying unit of analysis have to be provided, allowing the user to jump to a debugging component's respective unit of analysis.
\end{mechanismcategory}

\section{The iNNspector System for Systematic Model Debugging}
\label{inn:sec:system}
\begin{figure}[!t]
    \tcbset{on line,
        boxsep=0pt,
        boxrule=0pt,
        sharpish corners,
        left=0pt,right=0pt,top=0pt,bottom=0pt,
        colframe=black,colback=white,
        enhanced
    }
    \tcbox[fuzzy shadow={0mm}{0mm}{-0.3mm}{0.1mm}]{%
        \includegraphics[width=\linewidth]{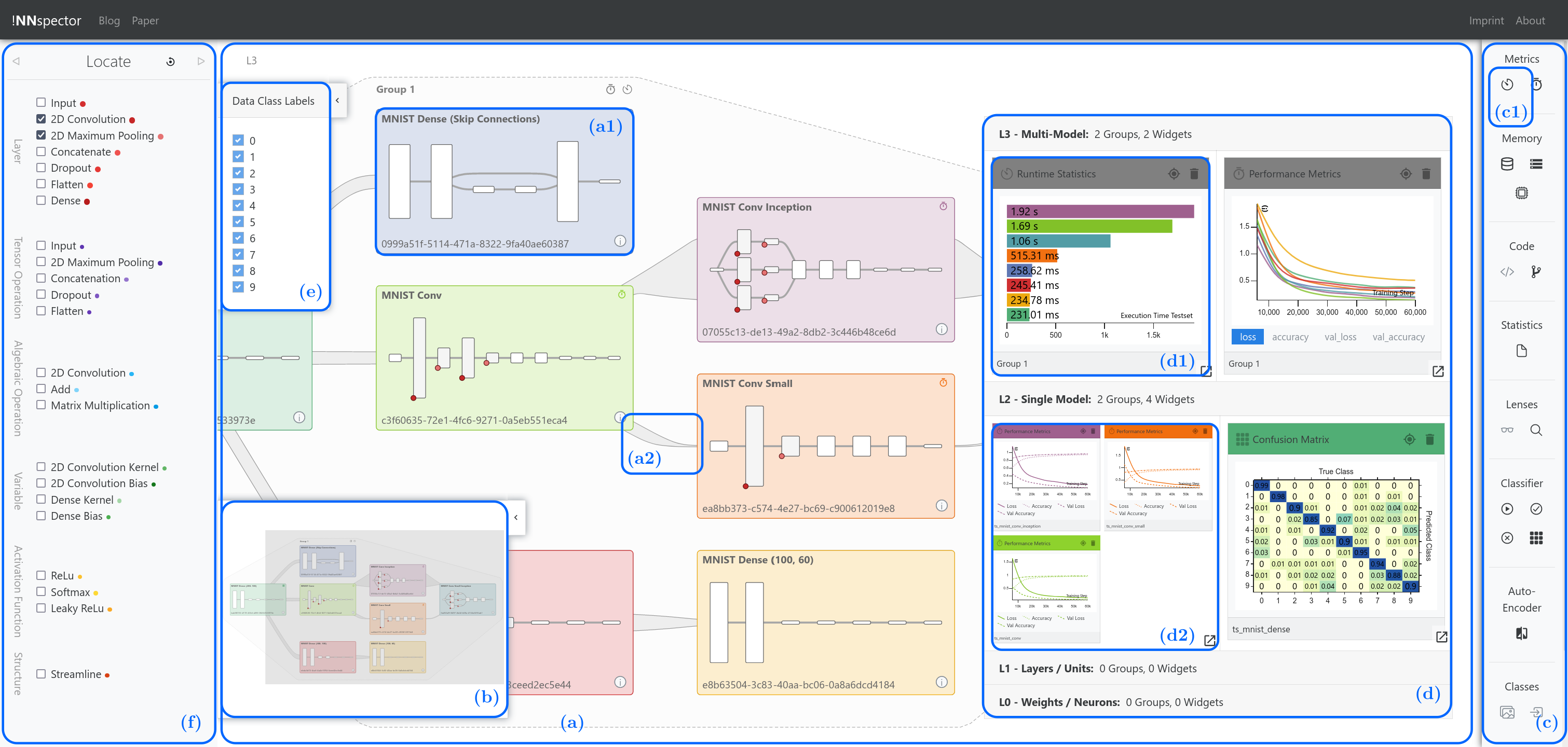}%
    }
    \caption[Screenshot of the iNNspector system frontend]{
        The iNNspector frontend.
        It is built around the inspection panel (a), showing nodes (a1) and links (a2) of the structural backbone on the current level of abstraction.
        The minimap (b) helps to navigate the viewport.
        Tools (c1) from the Toolbox (c) can be applied to units of analysis in the structural backbone to create widgets (d1), showing underlying data.
        Widgets are arranged in the widget panel (d), where they are organized according to their level of abstraction.
        Semantically related widgets can be combined into groups (d2).
        Widgets showing class-dependant data can be constrained to certain classes using the global class selector (e).
        The localization and interestingess panel (f) provides tools to identify units of interest.
    }
    \label{inn:fig:innspector-frontend}
\end{figure}

Materializing the before-proposed conceptual framework, we present the iNNspector system for systematic model debugging.
More than two years of conceptual ideas and implementation work have gone into the system.
The system is driven by three primary components: the custom keras checkpoint, collecting data for iNNspector during training, the backend, providing the data of the experiments via REST API, and the frontend, providing the actual iNNspector user interface.

\subsection{Application Walkthrough}
\label{inn:subsec:application-walkthrough}

In the following, we give a description of the iNNspector interface and its components in the form of an elaborate application walkthrough.
Thereby, we illustrate how iNNspector meets its claim of being a comprehensive system facilitating the systematic debugging of DL models.
By consequently linking system features to the respective data category (see~\cref{inn:subsec:data}), design dimension (see~\cref{inn:subsec:design-dimensions}), and global mechanism (see~\cref{inn:subsec:mechanisms}), we show how the system design is anchored in the requirements analysis (see~\cref{inn:sec:requirements}) and the foundations compiled into our conceptual framework (see \cref{inn:sec:framework}).

\dbvisparagraph{Inspection Panel}
iNNspector is built around a central graph view, referred to as \emph{inspection panel} (\cref{inn:fig:innspector-frontend}a).
It visually represents the structural data of the experiment, i.e., the tree of models, the model's architectures, and subsets of the weight-neuron network.
Thus, it implements the structural backbone~\textbf{(\ref{inn:mechanism:structural-backbone})} and builds the primary component to locate units of interest~\textbf{(\ref{inn:mechanism:units-of-analysis})} over different levels of abstraction~\textbf{(\ref{inn:mechanism:levels-of-abstraction})}.
Starting with the multimodel view (\hyperref[inn:paragraph:l3]{\ref*{inn:paragraph:l3}, L3}), showing the tree of models as it evolved during the experiment, the graph can be navigated to lower levels: by double-clicking a model and, subsequently, a layer, the user can descend over the single-model-view (\hyperref[inn:paragraph:l2]{\ref*{inn:paragraph:l2}, L2}) down to the neuron-weight-network view (\hyperref[inn:paragraph:l1]{\ref*{inn:paragraph:l1}, L1}).
The following paragraphs outline the specifics and functionalities of each layer.

\begin{figure}
    \centering
    \captionsetup[subfigure]{justification=centering}
    \begin{subfigure}{0.4\linewidth}
        \includegraphics[width=\linewidth]{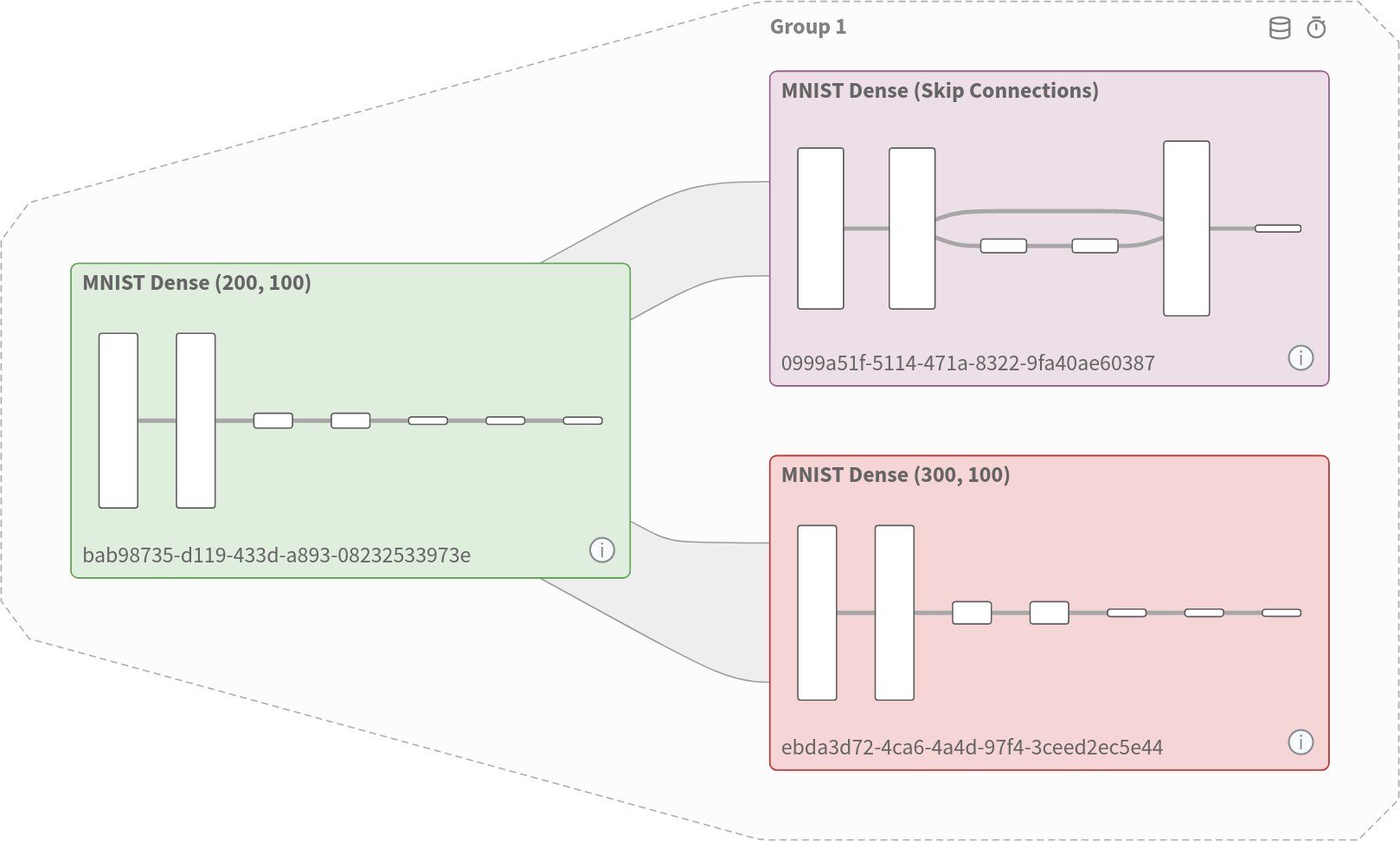}
        \caption{First experiment, with 3 models.}
        \label{inn:fig:innspector-l3-groups-a}
    \end{subfigure}
    \hfill
    \begin{subfigure}{0.55\linewidth}
        \includegraphics[width=\linewidth]{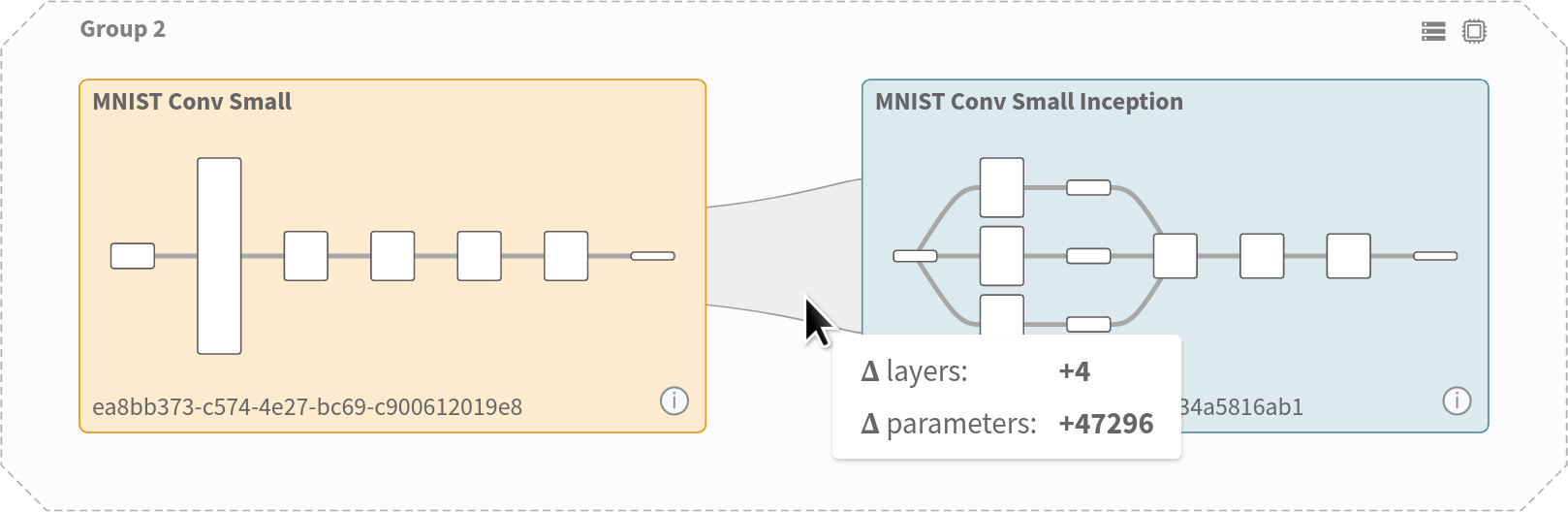}
        \caption{Second experiment, with 2 models.}
        \label{inn:fig:innspector-l3-groups-b}
    \end{subfigure}
    \caption[Inspection panel on L3, ``Multi-model'']{
        Inspection panel on L3, `Multi-model'.
        Experiment~(\subref{inn:fig:innspector-l3-groups-a}) has three models and experiment~(\subref{inn:fig:innspector-l3-groups-b}) two, each model represented by a unique color.
        Edges denote parent-child relationships, with tooltips and edge width indicating changes between model pairs.
    }
    \label{inn:fig:innspector-l3-groups}
\end{figure}

\dbvisparagraph{Multi-Model View [L3]\label{inn:paragraph:l3}}
The highest level of abstraction is the \emph{multimodel view}.
It shows all models in the iNNspector system, grouped into \emph{experiments}.
An experiment denotes a connected subset of models, i.e., a subset where each model shares a parent-child relationship with at least one other model.
\Cref{inn:fig:innspector-l3-groups} shows two experiments, the first consisting of three, the second of two models.
Each experiment is surrounded by a dashed outline, which forms a unit of analysis itself.
Each model is assigned a color based on a sequential color scale, which is re-used by other parts of the system to mark the part as belonging to a certain model, e.g., when descending into lower levels of abstraction.
Each model is depicted by a box, containing its friendly name, which is set by the programmer at development time (see~\ref{inn:paragraph:innspector-header}, ``\nameref{inn:paragraph:innspector-header}''), its unique ID, and an abstract representation of its architecture.
Each tree of models is arranged from left to right, with the leftmost model being the initial architecture and descendant models being ranked according to their depth in the tree.
The edges indicate a parent-child relationship, which is defined during development time, either manually by the programmer or by automatically branching models using the available tool from the toolbox.
The change in the edge width between models denotes the relative change in the number of trainable parameters, with the absolute values being shown in a tooltip when hovering the edge.

\begin{figure}[t]
    \centering
    \includegraphics[width=0.9\linewidth]{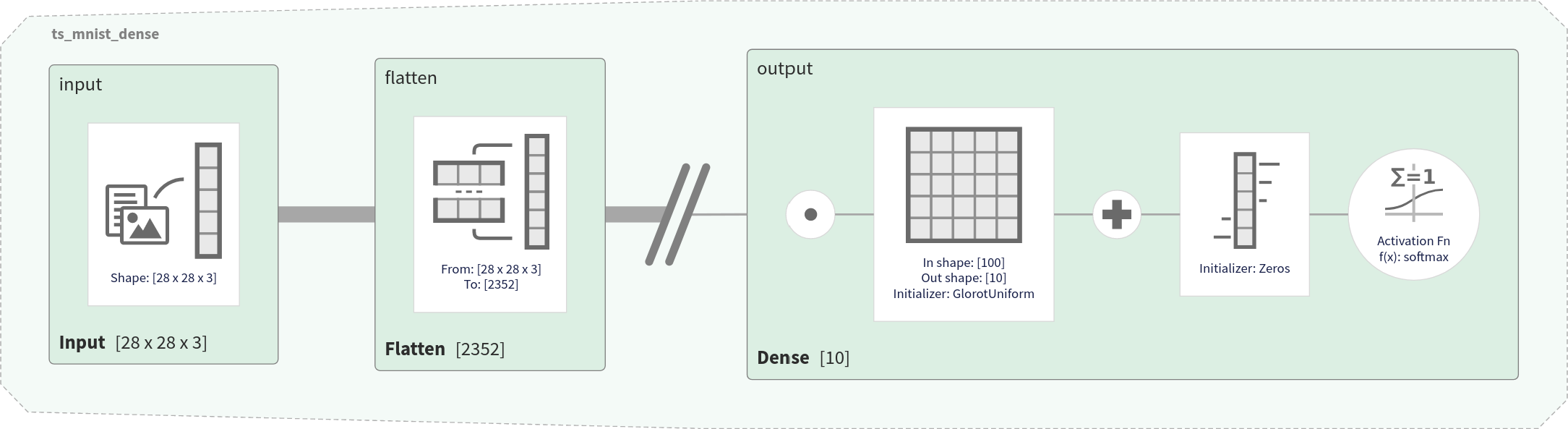}
    \caption[Inspection panel on L2, ``Single Model'']{
        Inspection panel on L2, ``Single Model''.
        The view shows the architecture of a model, with each box representing a layer and the edges denoting the data flow between layers.
        Elements inside the layer boxes represent operations applied to the layer input.
    }
    \label{inn:fig:innspector-l2-model}
\end{figure}

\dbvisparagraph{Single Model View [L2]\label{inn:paragraph:l2}}
By double-clicking a model, the user can descend by one level into the \emph{single model view.}
\Cref{inn:fig:innspector-l2-model} shows the (shortened) single model view of the upper left model visible in~\cref{inn:fig:innspector-l3-groups}.
It provides a graph visualization of the model's architecture, with each box representing a layer of the model and the edges denoting the data flow between layers.
The dashed outline denotes the model as a whole.
All layers universally show their name in the top left, and their type and output size in the bottom left.
Additionally, layers contain one or multiple inner elements, encoding information specific to their type and configuration.
These inner elements form a graph on themselves, which represents the way mathematical functions and variables are subsequently applied to the layer input.
E.g., the dense layer in \cref{inn:fig:innspector-l2-model} starts with a matrix multiplication \inlinegraphics{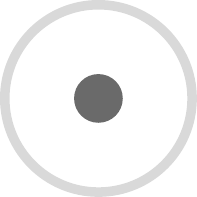} of the input with a kernel \inlinegraphics{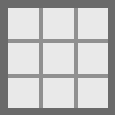}, continues with adding \inlinegraphics{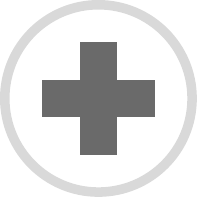} a bias \inlinegraphics{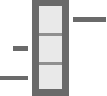}, and finally applies a softmax activation \inlinegraphics{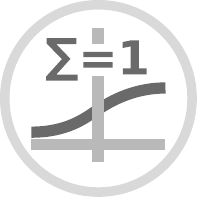} to the result.
The pictograms give a visual impression of the respective operation or variable, rendering the system useful for less experienced model developers or for educational purposes.
Additionally, operations and variables are augmented with information relevant for model development, such as input and output shapes, kernel initializers, or filter shapes.

\begin{figure}
    \centering
    \captionsetup[subfigure]{justification=centering}
    \begin{subfigure}{0.6\linewidth}
        \includegraphics[width=\linewidth]{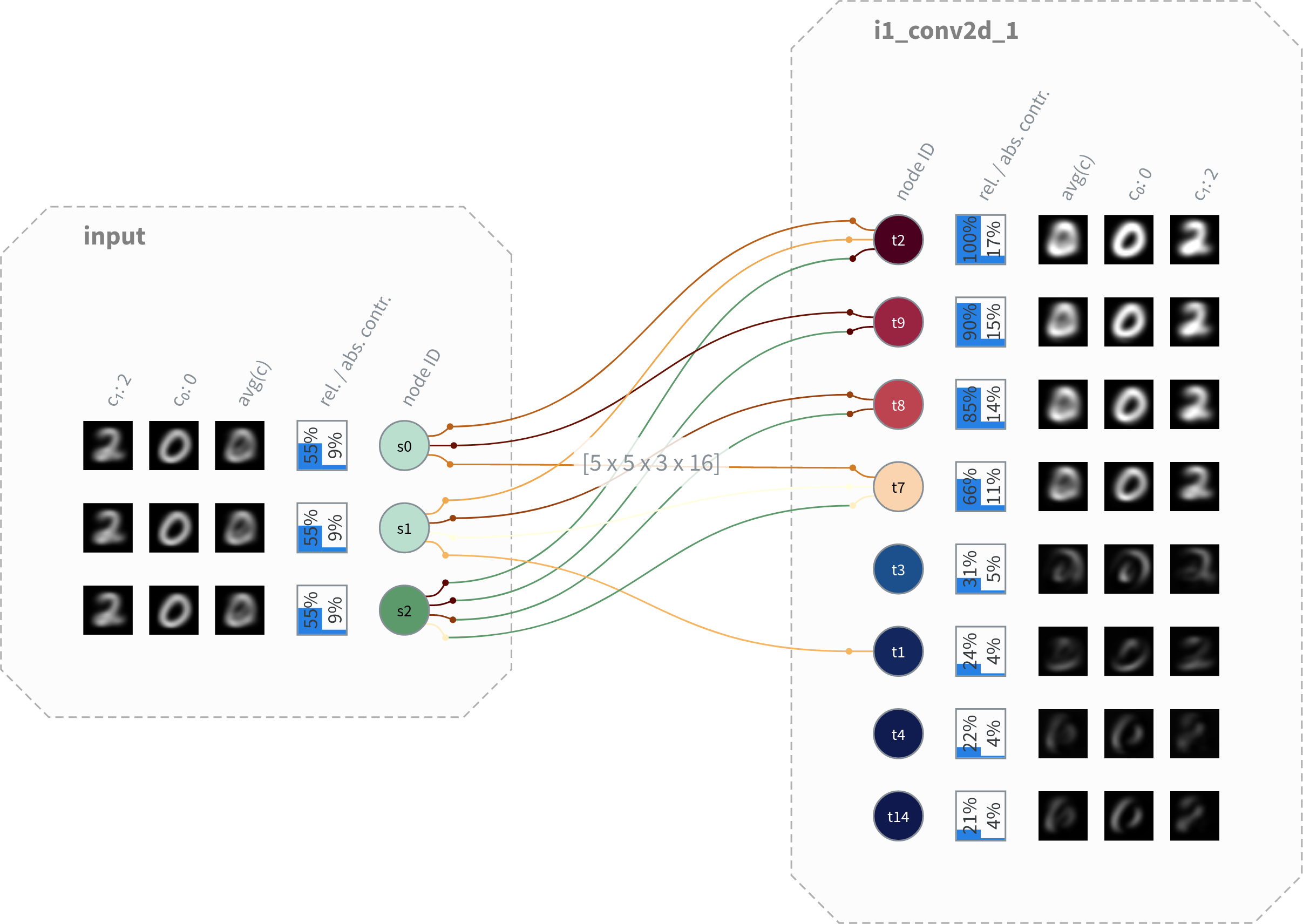}
        \caption{The neuron-weight-network view.}
        \label{inn:fig:innspector-l1-weight-network}
    \end{subfigure}
    \hfill
    \begin{subfigure}{0.35\linewidth}
        \includegraphics[width=\linewidth]{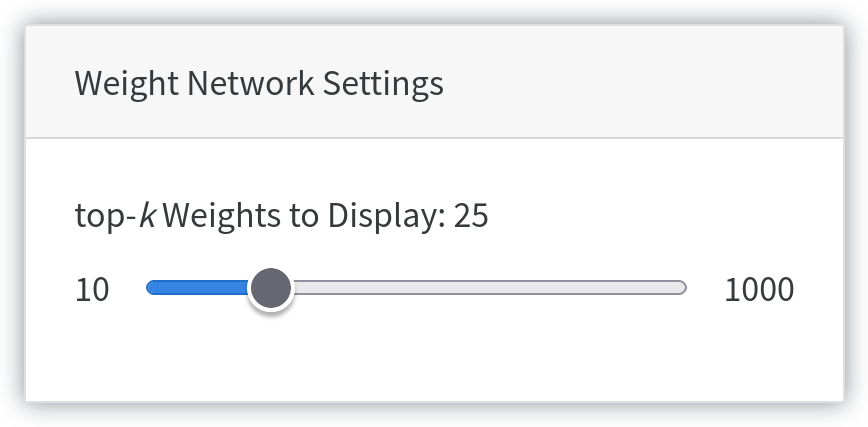}
        \caption{Slider to set the number~$k$\\of top weights to show.}
        \label{inn:fig:innspector-l1-weight-network-settings}
    \end{subfigure}
    \caption[Inspection panel on L1, ``Neuron-Weight-Network'']{
        Inspection panel on L1, ``Neuron-Weight-Network''.
        The view shows the top-$k$ most significant subset of neurons for both layers, according to their mean activation under the current class selection.
        The edges between the neurons represent the elements of the weight matrix connecting two neurons through a matrix multiplication or a convolution operation.
    }
    \label{fig:layer-one}
\end{figure}

\dbvisparagraph{Neuron-Weight-Network View [L1]\label{inn:paragraph:l1}}
By double-clicking a layer on L2, the user can descend to the final level L1, the \emph{neuron-weight network view}, of which an example is shown in~\cref{inn:fig:innspector-l1-weight-network}.
Like the name suggests, it focusses on the weights connecting the previous to the current layer.
For this, we visualize a subset of neurons for both layers, sorted by their mean activation under the current class selection (see~\ref{inn:paragraph:class-selector-panel}, ``\nameref{inn:paragraph:class-selector-panel}'').
The neurons are colored according to a linear color scale, mapping the lowest activation to dark blue and the highest activation to dark red.
Edges between the neurons represent the largest $k$ weights, i.e., the element of the weight matrix which connects the two neurons through a matrix multiplication or a convolution operation.
The value for $k$ can be set using a slider, which is shown in~\cref{inn:fig:innspector-l1-weight-network-settings}.
Since the number of neurons in a layer can quickly grow to the order of several thousands, we filter for the ones that (1) are connected through at least one of the top-$k$ weights, or (2) feature either one of the $10$ highest or $10$ lowest mean activations.
Neurons and edges can be highlighted by hovering or clicking them, which makes it easier to observe single connections in large networks.
Besides the activation color, we augment the neurons with additional information, such as their relative and absolute share of the summed activations currently viewed.
Furthermore, for activations having an image-interpretable format (i.e., two-dimensional grayscale or three-dimensional coloscale), we show their mean activation $c_{i}$ for $i = 0, \ldots, n - 1$ for the $n$ currently selected classes, as well as an average over the selected classes $\textnormal{avg}(c) = \frac{1}{n}\sum_{i = 0}^{n-1}{c_{i}}$.

\begin{figure}
    \centering
    \captionsetup[subfigure]{justification=centering}
    \begin{subfigure}{0.32\linewidth}
        \includegraphics[width=\linewidth]{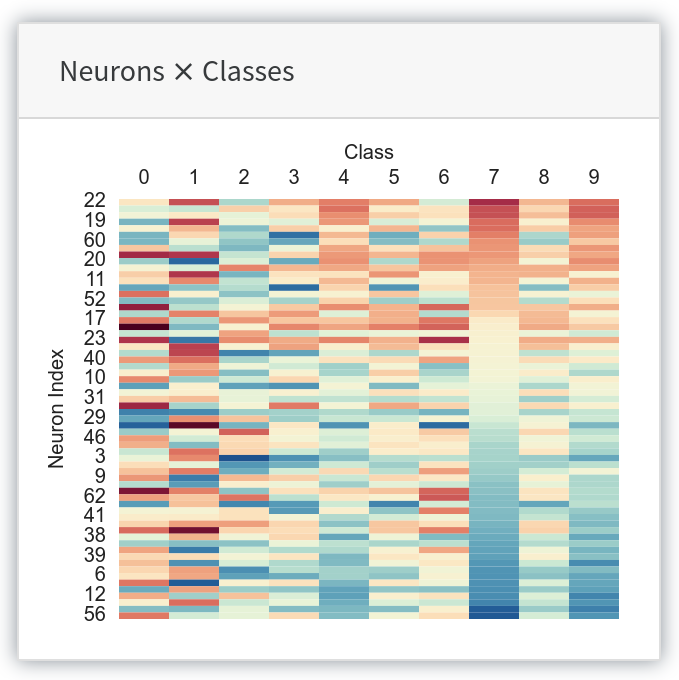}
        \caption{Neurons $\times$ Classes panel.}
        \label{inn:fig:innspector-l1-neuron-x-class}
    \end{subfigure}
    \hfill
    \begin{subfigure}{0.32\linewidth}
        \includegraphics[width=\linewidth]{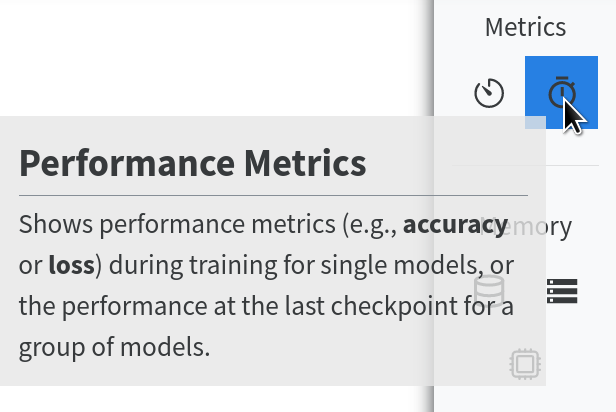}
        \caption{The iNNspector toolbox.}
        \label{inn:fig:innspector-toolbox}
    \end{subfigure}
    \hfill
    \begin{subfigure}{0.28\linewidth}
        \includegraphics[width=\linewidth]{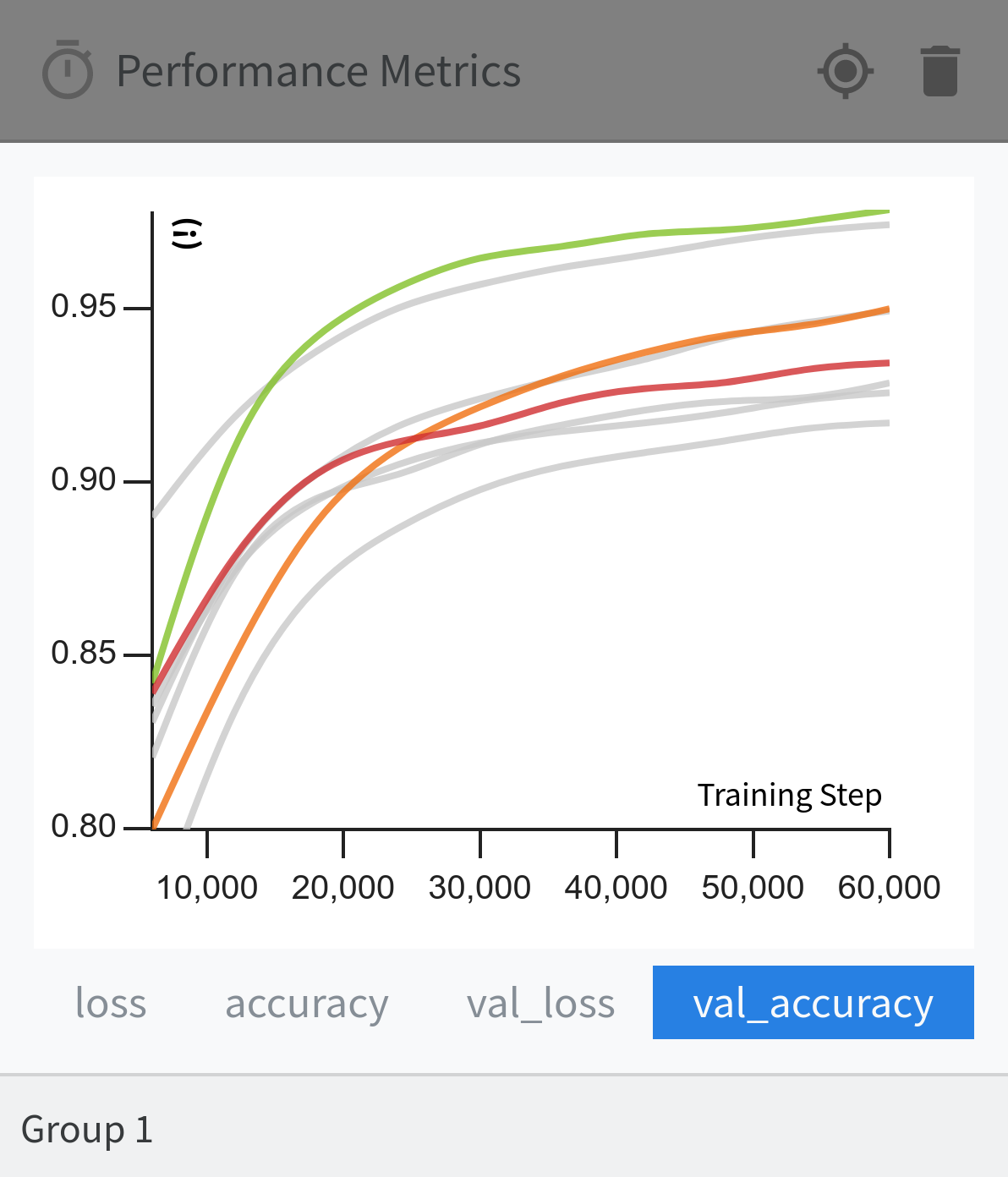}
        \caption{An exemplary widget.}
        \label{inn:fig:innspector-widget}
    \end{subfigure}
    \caption[Panels, tools, and widgets in the iNNspector UI]{
        Panels, tools, and widgets in the iNNspector UI.
        (\subref{inn:fig:innspector-l1-neuron-x-class}) A L1-specific panel showing the activation of each neuron with respect to each class.
        (\subref{inn:fig:innspector-toolbox}) The toolbox, providing tools to investigate units of interest. The tools can be applied to compatible elements in the inspection panel to create widgets.
        (\subref{inn:fig:innspector-widget}) An exemplary widget, showing the activation of a neuron with respect to each class.
    }
    \label{fig:innspector-tools-and-widgets}
\end{figure}

Since the neuron-weight-network can only visualize a subset of neurons and weights and always averages over the current class selection, we augment the L1 view with the \emph{Neurons $\times$ Classes panel}, shown in~\cref{inn:fig:innspector-l1-neuron-x-class}.
It provides a matrix view showing the activation of each individual neuron with respect to each class of the datset, rendering it especially useful for the debugging of failed classifications and disentanglement of latent spaces.
By clicking the header of a class column, all columns are re-sorted according to the clicked class, enabling correlation analysis across classes.
Notably, with decreasing level of abstraction, the data space we have to cover shrinks significantly.
Therefore, more data can be shown by specialized panels (e.g., \emph{Classes $\times$ Neurons}) and the structural view itself, reducing the number of tools that are needed to show underlying data.

\dbvisparagraph{Toolbox, Widgets}
\label{inn:paragraph:toolbox-widgets}
The \textbf{toolbox} (\cref{inn:fig:innspector-frontend}c) provides a variety of tools to investigate and annotate units of interest, extend the tree of models by new model variations, and -- most importantly -- visualize the underlying data of units of analysis.
Each tool is represented through a small icon, which, when hovered, offers a detailed description of the tool's functionality, as displayed in~\cref{inn:fig:innspector-toolbox}.
After selecting a tool, it stays active until applied to a unit of analysis, e.g., a model or a layer.
The tools in the toolbox resolve several dependencies~\textbf{(\ref{inn:dim:dependencies})} by narrowing their applicability to specific UoA types and levels of abstraction.
E.g., the \emph{Histogram} tool, which creates a histogram visualization of a data distribution, can only be applied to UoAs comprising high-dimensional data~\textbf{(\ref{inn:data:high-dim})}, such as kernels or activations of a layer.
In most cases, applying a tool to a UoA results in the creation of a \textbf{widget} (\cref{inn:fig:innspector-widget}).
A widget offers visual or verbal representations~\textbf{(\ref{inn:dim:representation})} of the underlying data of a unit of analysis.
This data can be scalar~\textbf{(\ref{inn:data:scalar})} or high-dimensional~\textbf{(\ref{inn:data:high-dim})}, rendering appropriate processing~\textbf{(\ref{inn:dim:processing})} essential.
Therefore, each tool defines which data to query and how to transform this data before forwarding it to the newly created widget.
The transformation pipeline is expressed as a list of transform operations that are consecutively applied by the backend to the queried data before returning them.
The data querying mechanisms follow a standardized format, making the system effortlessly expandable by custom tool.
For technical details on the data querying routes and the transformation grammar, see~\cref{inn:subsec:innspector-backend}.

\begin{figure}
    \centering
    \includegraphics[width=0.7\linewidth]{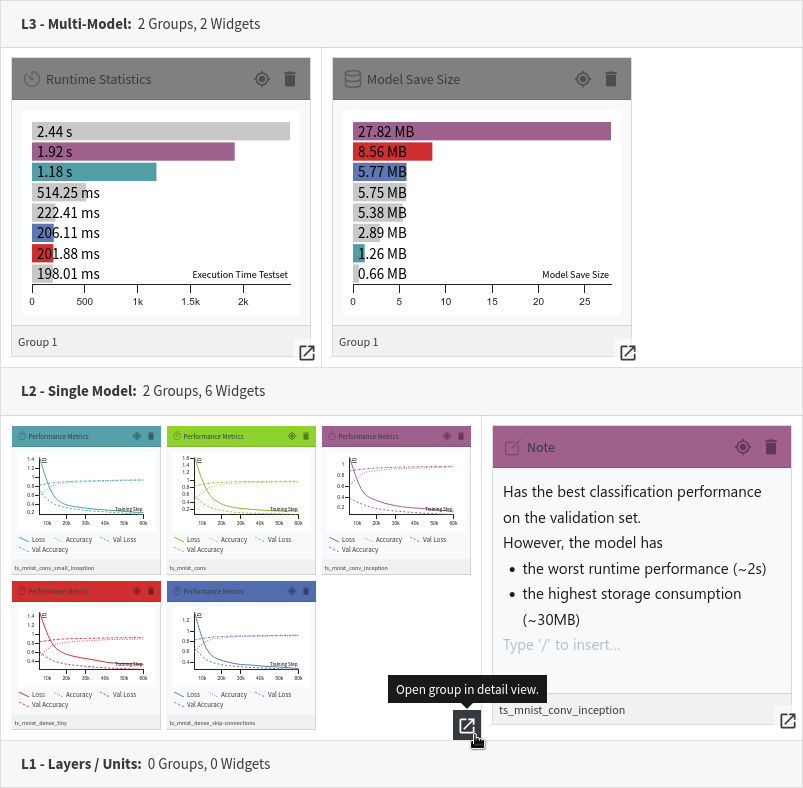}
    \caption[iNNspector's widget panel]{
        The iNNspector widget panel.
        Widgets are organized according to their level of abstraction.
        They can be grouped by dragging and dropping them onto other widgets or groups.
    }
    \label{inn:fig:innspector-widget-panel}
\end{figure}

We provide a comprehensive set of pre-defined widgets, supporting a variety of use cases and analysis scenarios.
The widgets can be classified into three groups: (1) generalizing representations, (2) data-specific representations, and (3) functional.
\textbf{Generalizing representations} provide typical charts for common types of data.
While they set strict conditions on the format of the data, they do not introduce any further dependencies~\textbf{(\ref{inn:dim:dependencies})}.
For example, we provide linecharts for time-dependent, scalar data or bar charts for single-time scalar data.
Complementary, we offer different types of histograms for high-dimensional data.
\textbf{Data-specific representations} are charts specialized to a certain type of model, layer, or dataset.
For example, for classifiers, we include a visualization of the class probability distribution, while for image-to-image models, we visualize the difference between input and reconstruction.
Both only work for the stated type of model.
Finally, functional widgets do not visualize underlying data of the experiment, but provide information for model management and documentation of the analysis process.
For instance, the \emph{Note} tool creates a widget featuring a markdown editor, while the \emph{Branch Model} tool generates a new iNNspector header, denoting a new parent-child relationship in the tree of models.
For details on the iNNspector header, refer to~\cref{inn:paragraph:innspector-header},~``\nameref{inn:paragraph:innspector-header}''.

Widgets can be extended through arbitrary interactions, e.g., to filter or exclude data series, or constrain the widget to a subset of UoAs.
Particularly, this includes linking and brushing of widgets and their contained data across all components of the iNNspector system.
See~``\nameref{inn:system:linking-and-brushing}'' for more details.

\begin{figure}[t]
    \centering
    \captionsetup[subfigure]{justification=centering}
    \begin{subfigure}[t]{0.70\linewidth}
        \includegraphics[width=\linewidth]{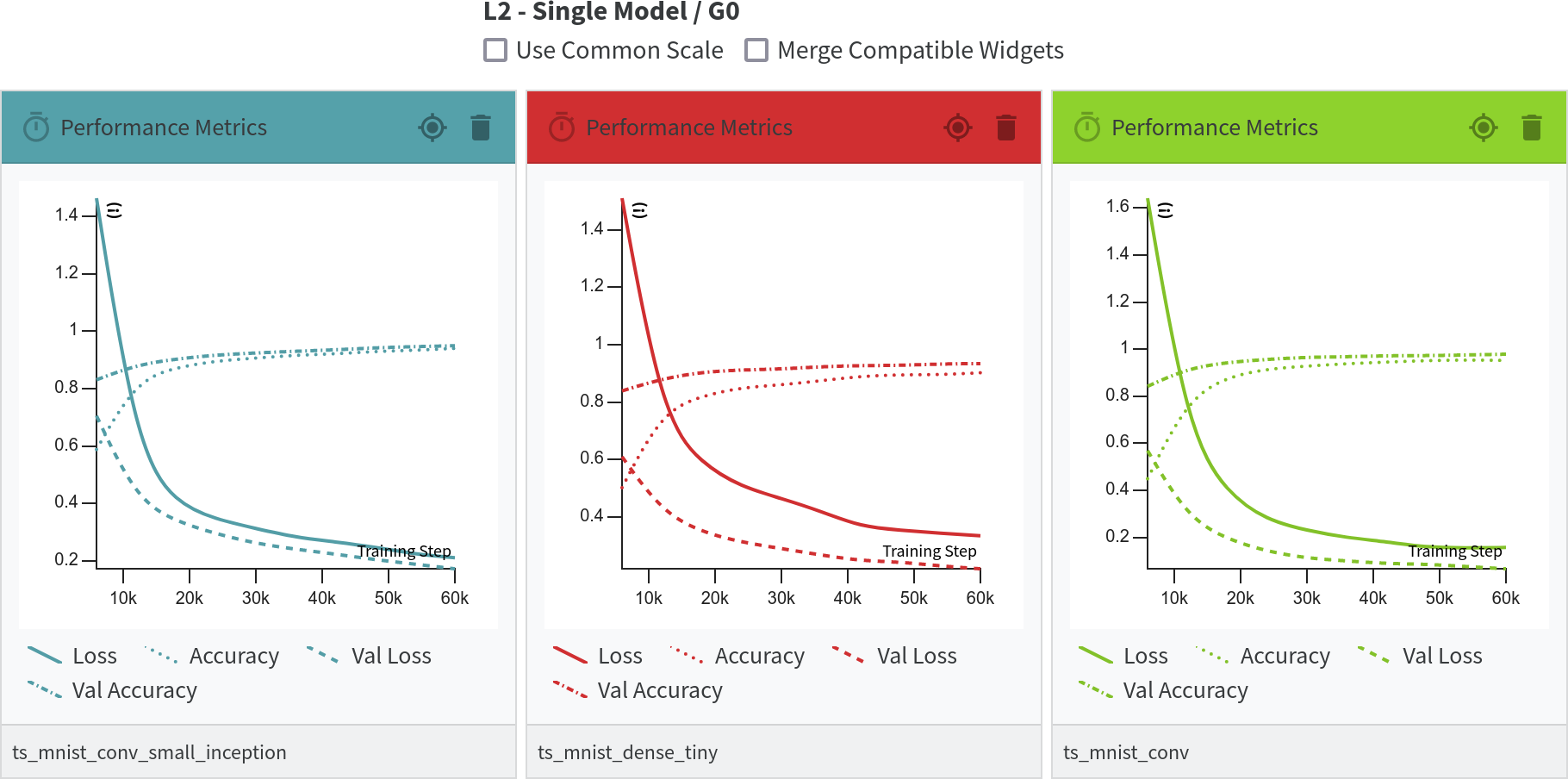}
        \caption{Group of three widgets.}
        \label{inn:fig:widget-group-view}
    \end{subfigure}
    \hfill
    \begin{subfigure}[t]{0.26\linewidth}
        \includegraphics[width=\linewidth]{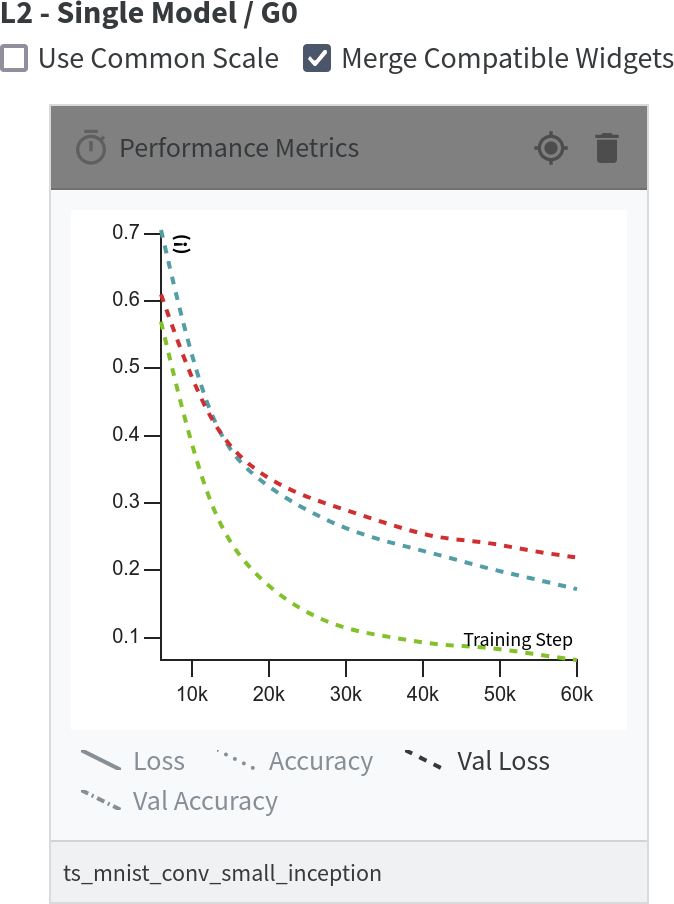}
        \caption{Merged.}
        \label{inn:fig:widget-group-view-merged}
    \end{subfigure}
    \caption[iNNspector's widget group view]{
        iNNspector's widget group view.
        (\subref{inn:fig:widget-group-view}) Grouped widgets can be opened in a detail view, showing them side-by-side in an enlarged page overlay.
        (\subref{inn:fig:widget-group-view-merged}) Widgets featuring compatible data can be chosen to either use a common scale or be merged into a single meta-widget.
    }
\end{figure}

\dbvisparagraph{Widget Panel}
Widgets are organized in the \textbf{widget panel}, depicted in~\cref{inn:fig:innspector-widget-panel}.
The panel is implemented as a vertical accordion UI element, with each accordion flap representing one level of abstraction.
Upon their creation, widgets are inserted into the flap of their corresponding unit of analysis lies on, reflecting the structural hierarchy also in the widget panel and, hence, helping the user to locate searched-for widgets.
This effect could be proven in our evaluation study (see~\cref{inn:subsec:evaluation-study-results}).
To prevent the cumulative cluttering of the widget panel during the advance of the analysis process, widgets can be \textbf{grouped} by dragging and dropping them onto other widgets or groups.
Besides visually sorting semantically related widgets, groups also can be opened in a detail view, showing them side-by-side in an enlarged page overlay.
In the group view, depicted in~\cref{inn:fig:widget-group-view}, widgets featuring compatible data can be chosen to either use a \textbf{common scale} or be \textbf{merged} into a single meta-widget.
Scaling and merging makes the data better comparable across units of analysis.
\Cref{inn:fig:widget-group-view-merged} shows the same group as~\cref{inn:fig:widget-group-view} but with the widgets merged into a single meta-widget.

\dbvisparagraph{Linking and Brushing\label{inn:system:linking-and-brushing}}
To emphasize connections between related entities, we consistently implement linking and brushing over all parts of the system.
On the component level, this affects links between UoAs, widgets, and filters.
E.g., hovering a layer in a model's architecture results in a visual highlight of all widgets relating to the layer, and vice versa.
On the data level, different representations of the same data-point or dataset sample are linked across the system.
E.g., when hovering the projected activation for a certain dataset sample in a scatterplot widget, the datapoints relating to the same dataset sample are highlighted in all other widgets, enabling an interactive comparison across layers or models.

\begin{figure}
    \centering
    \captionsetup[subfigure]{justification=centering}
    \begin{minipage}{0.228\linewidth}
        \begin{subfigure}{\linewidth}
            \includegraphics[width=\linewidth]{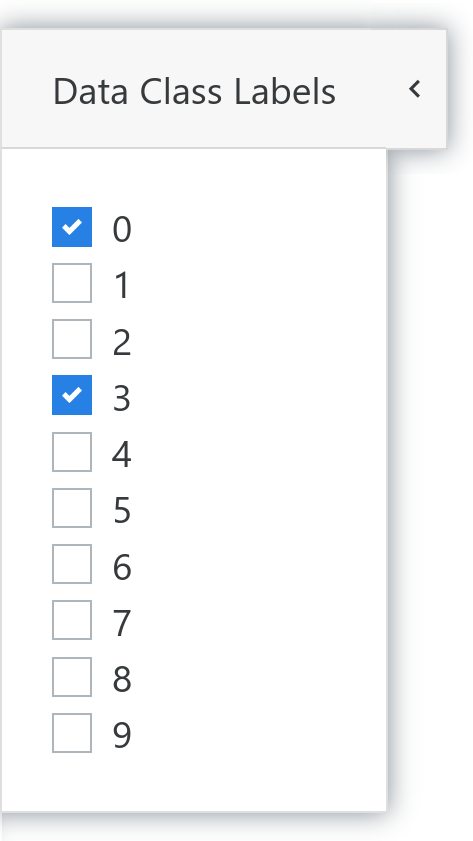}
            \caption{Class selector.}
            \label{inn:fig:innspector-class-selector-panel}
        \end{subfigure}
    \end{minipage}
    \hspace{2em}
    \begin{minipage}{0.5\linewidth}
        \begin{subfigure}{\linewidth}
            \includegraphics[width=\linewidth]{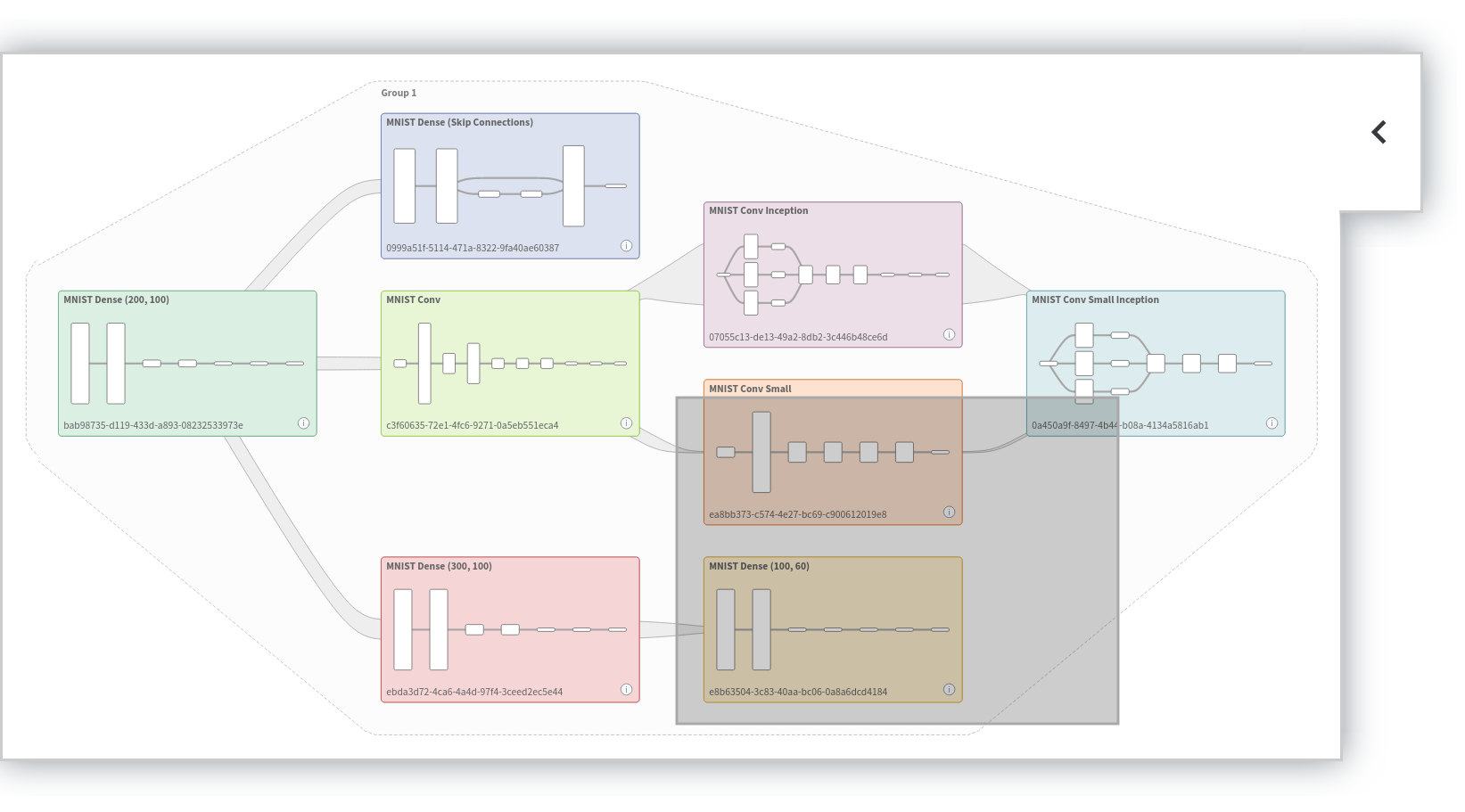}
            \caption{Minimap.}
            \label{inn:fig:innspector-minimap}
        \end{subfigure}
        \vskip 2em
        \begin{subfigure}{\linewidth}
            \includegraphics[width=\linewidth]{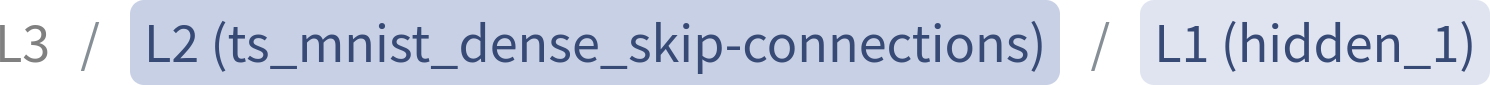}
            \caption{Breadcrumbs.}
            \label{inn:fig:innspector-breadcrumbs}
        \end{subfigure}
    \end{minipage}
    \caption[Configuration and navigation panels in iNNspector]{
        Configuration and navigation panels in iNNspector.
        (\subref{inn:fig:innspector-class-selector-panel}) The class selector panel lists all classes occuring in the dataset, allowing to toggle them.
        The class selection affects the data of all widgets in the widget panel that are class-dependent.
        (\subref{inn:fig:innspector-minimap}) The minimap indicates the viewport of the iNNspection panel, guiding the user in large experiments or models.
        (\subref{inn:fig:innspector-breadcrumbs}) Breadcrumbs show the path to the UoA currently focussed in the iNNspection panel over the different levels of abstraction.
    }
    \label{fig:class-selector-minimap-breadcrumbs}
\end{figure}

\dbvisparagraph{Class Selector Panel\label{inn:paragraph:class-selector-panel}}
The \emph{class selector panel} lists all classes occuring in the dataset, allowing to toggle arbitrary classes.
After modifying the class selection, all components of the system are updated to adhere to the selected classes.
Particularly, this affects (1) the appearance of the iNNspection panel on L1, constraining the structural and high-dimensional data displayed to the selected classes, and (2) the data of the widgets in the widget panel.
The only exceptions are widgets showing data where the datapoints are not directly associated to a class (like they are, e.g., for activations), but which would still change under different classes (compared to, e.g., weights, which are class independent).
For such widgets and $n$ dataset classes, we would have to pre-compute and store $2^{n} - 1$ possible combinations.
An example for such a widget is created by the ``Performance-Metrics''-tool: metrics like loss or accuracy change under different dataset classes, however, since those values are computed at training time, we would either have to pre-compute the data for arbitrary combinations, resulting in huge runtime and storage overheads, or re-evaluate the model just-in-time, affording access to the full training and testing dataset, as well as information on the computed metrics.
We mark those widgets with a \inlinegraphics{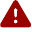}-symbol in the widget header.

\dbvisparagraph{Navigation Functionalities}
For efficient exploration of the structural data space, we include additional functionalities that help the user to keep an overview and locate themselves in the levels and graphs.
A minimap, depicted in~\cref{inn:fig:innspector-minimap}, indicates the viewport of the iNNspection panel, i.e., the section of the panel that is currently visible in the browser window.
By clicking a point on the minimap, the vieport automatically centers at that position, allowing the user to quickly jump to an area of interest, which is particullarly useful for large model architectures on L2.
Besides the minimap providing orientation in the x-y-plane, breadcrumbs show the path to the UoA currently focussed in the iNNspection panel over the different levels of abstraction.
\Cref{inn:fig:innspector-breadcrumbs} shows the breadcrumbs while having the layer ``hidden\_1'' of model ``ts\_mnist\_dense\_skip-connections'' focused on L1.
Clicking an element in the breadcrumb path directly teleports the user to the respective level of abstraction.
A similar functionality is offered by the \inlinegraphics{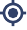}-button in the header of all widgets.
By clicking it, the system automatically jumps to the level of abstraction and the UoA the widget belongs to, regardless of the current position.

\begin{figure}
    \centering
    \captionsetup[subfigure]{justification=centering}
    \begin{subfigure}[t]{0.45\linewidth}
        \includegraphics[width=\linewidth]{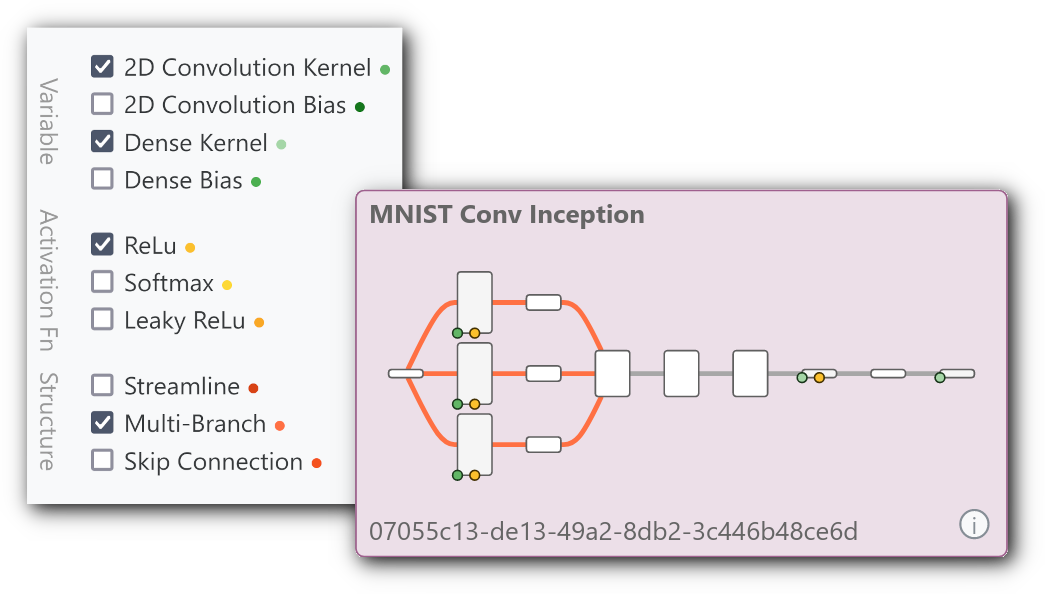}
        \caption{UoA types and --badges.}
        \label{inn:fig:filters-badges}
    \end{subfigure}
    \hspace{1em}
    \begin{subfigure}[t]{0.405\linewidth}
        \includegraphics[width=\linewidth]{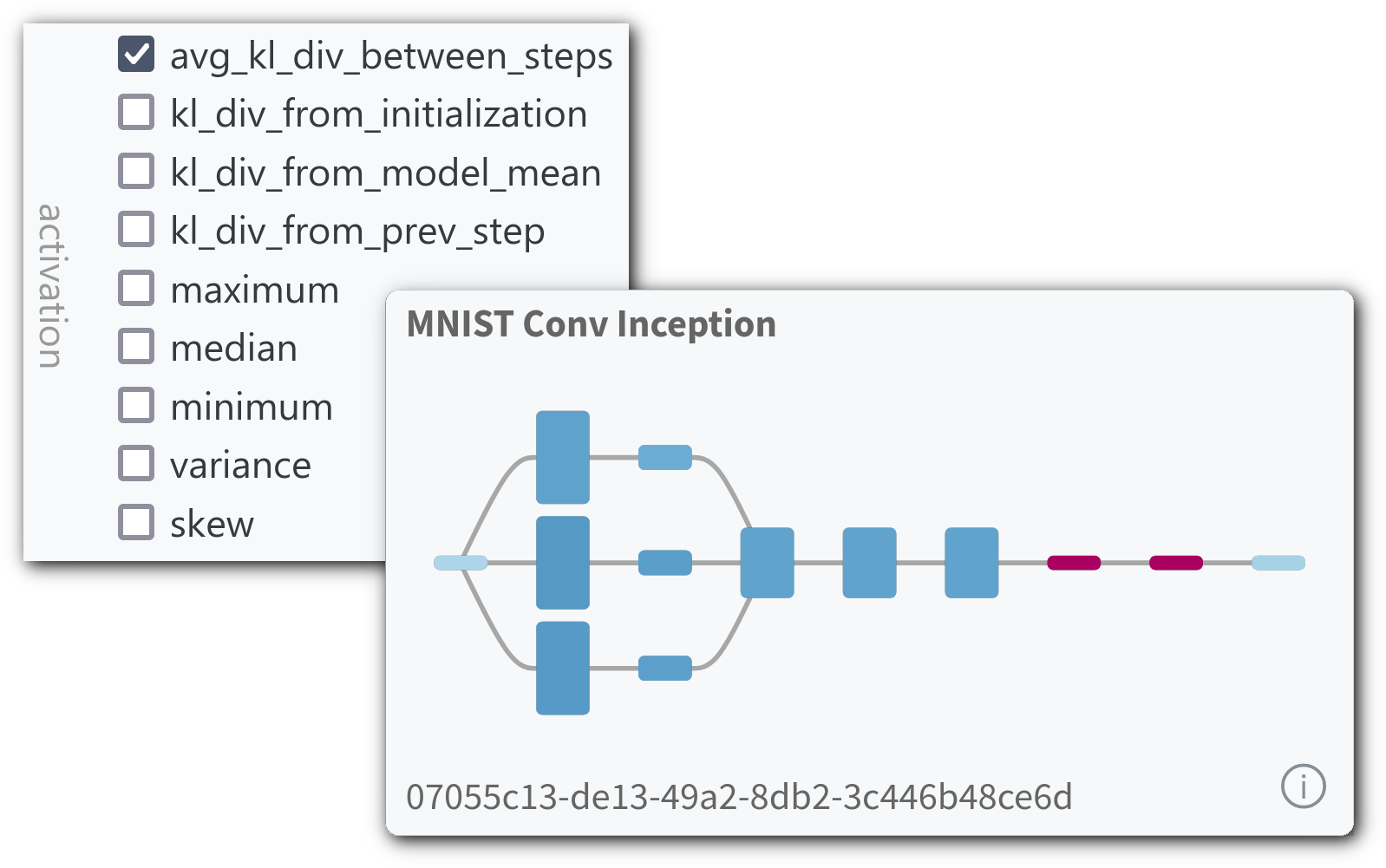}
        \caption{Interestingness types and --annotations.}
        \label{inn:fig:interestingness}
    \end{subfigure}
    \caption[Localization and interestingness functionalities in iNNspector]{
        Localization and interestingness functionalities in iNNspector.
        (\subref{inn:fig:filters-badges}) The localization panel lists different UoA types occuring in the currently visible models.
        By toggling them, small badges in the inspection panel indicate the location of respective UoAs in the structural data.
        (\subref{inn:fig:interestingness}) The interestingness panel colors UoAs according to their interestingness scores.
    }
    \label{fig:filters-badges-interestingness}
\end{figure}

\dbvisparagraph{Localization}
The structural backbone~\textbf{(\ref{inn:mechanism:structural-backbone})} might grow to significant size and depth for large experiments and complex model architectures.
Therefore, iNNspector includes a panel allowing to search~\textbf{(\ref{inn:mechanism:search})} for and locate certain UoA types and architectural structures.
\Cref{inn:fig:filters-badges} shows a segment of the localization panel, listing different UoA types occuring in the currently visible models.
By toggling them, small badges in the inspection panel indicate the location of respective UoAs in the structural data.
For UoAs hidden in lower levels of abstraction, badges are propagated upwards to the current level, helping the user to track down UoAs of interest.
Besides UoA types, the user can also search for common structures in the model architecture~\citep{Wang2019VisualGenealogyDeep}, such as multi-branches (c.f., Inception, \citet{Szegedy2017InceptionV4Inception}) or skip-connections (c.f., ResNet, \citet{He2016DeepResidualLearning}).

\dbvisparagraph{Interestingness}
Complementing the filters, which work exclusively on a structural level, we implement automated \textbf{interestingness measures}~\textbf{(\ref{inn:mechanism:interestingness})} on a data distribution level.
Particularly, we compute a variety of statistical descriptors over the high-dimensional data~\textbf{(\ref{inn:data:high-dim})}, including \emph{skew}, \emph{variance}, \emph{minimum}, and \emph{maximum}, as well as the divergence of the distribution shape from different baselines.
The interestingness panel in \cref{inn:fig:interestingness} allows to toggle the different descriptors, which are then visualized in the inspection panel.
Interestingness is computed per UoA on the lowest-level where data occurs, and then propagated upwards over the structural backbone.
Since the statistical descriptors are only comparable within one variable type (i.e., activation, dense kernel, conv2d bias), we compute the values for each variable type separately.
The values are normalized and aggregated over layers and models.
Visually, we annotate the interestingness by re-coloring the UoAs on a linear color scale, ranging from blue for low to red for high interestingness values.
Interestingness can directly point the analyst to annomalies in the data without the tedious, manual inspection of large parts of the data space, providing an entry-point to the distribution analyis.

\subsection{Design Rationales}
Throughout our two-plus-year development process, we went through various design iterations until the system reached its current form.
In the following, we will elaborate on the more prominent design rationales that shaped the iNNspector frontend.

\dbvisparagraph{Widgets}
As proposed by our framework and, consequently, implemented by iNNspector, we use the structural backbone to localize UoAs and visualize their underlying data.
Following this principle of connecting abstract data to semantically related, tangible visual representatives, we attached the widgets directly to their respective UoAs in the first version of the system.
While this emphasized the affiliation between widget and UoA, we found several issues with the design.
For example, the interface got increasingly crowded during the inspection, and there was no natural way to create groups of widgets.
Furthermore, the close link between widget and UoA was a disadvantage for comparative tasks over multiple levels of abstraction: where would we put a widget if its corresponding UoA is not currently visible in the inspection panel?
Therefore, we decided to organize the widgets in a global panel and visually link them through labels and colors.
In our evaluation study (see~\cref{inn:sec:evaluation}), our users reported liking how widgets are organized in the widget panel.

\dbvisparagraph{Global Class Selector}
The class selector is implemented as a global panel instead of a per-widget class selection.
In our different design iterations, we experimented with different mechanisms to constrain visualizations to specific classes, advancing from having no class selection over per-widget, single-class selection to the current, global multi-class selector.
It best supported the debugging of classification use cases and prevented confusion with a per-widget class selection when comparing across different widgets.

\dbvisparagraph{Toolbox}
The toolbox is a typical pattern in existing systems (e.g., image editing-- or CAD software).
It provides a straightforward and tidy interface to complex functionalities, is easy to learn and use, and can adapt to changing contexts by only showing the currently applicable tools.

\dbvisparagraph{Levels of Abstraction}
In the first iteration of the system, multiple inspection panels could be spawned simultaneously, each of which could be on a different level of abstraction.
Through this, we tried to resolve the previously discussed problem of simultaneously inspecting widgets in different models and levels of abstraction, back when widgets were still directly attached to their respective UoAs.
However, this led to a loss of global context and crowded the screen with different representations of the same structural data.
Therefore, to keep focused and preserve the global context, we moved to strictly hierarchical navigation over levels of abstraction.
In the current version of iNNspector, navigation over levels of abstraction feels like browsing the z-axis while zooming and panning cover the x-y-plane.
In our user study (see~\cref{inn:sec:evaluation}) we observed the navigation over levels of abstraction to be intuitive to the users.

\subsection{Developer Workflow}
\label{inn:subsec:workflow}

The iNNspector system is designed to be easily integrated into the model-building workflow.
In the following, we will describe the steps needed to bring an experiment into the iNNspector system.
iNNspector provides a Python package, consisting of three main components:
(1) the custom model checkpoint, providing the logging mechanisms,
(2) several pre-defined data generators, implementing convenience functions to retrieve the number of classes and a subset of the evaluation dataset,
and (3) a database connector, providing methods to store and load the model's metadata to and from the model database.

\dbvisparagraph{Creating a Model}
The model creation process is similar to the one in ordinary TensorFlow / Keras.
The only difference is that the model has to be annotated with an iNNspector header function (see~\cref{inn:paragraph:innspector-header},~``\nameref{inn:paragraph:innspector-header}'').
The function has to be called before training the model and returns a dictionary containing the model's meta information.

\dbvisparagraph{Training the Model}
For training, an instance of the customized iNNspector checkpoint (see~\cref{inn:paragraph:innspector-checkpoint},~``\nameref{inn:paragraph:innspector-checkpoint}'') has to be passed as a callback to the \emph{fit} method of the model.
The checkpoint receives a reference to the dictionary created by the iNNspector header function and extends it with information on the location of the checkpoint files and the graph file.

\dbvisparagraph{Storing the Metadata to the Model Database}
After training, the model's metadata has to be appended to the model database using the database connector.

\dbvisparagraph{Creating New Model Iterations}
When refining the initial architecture to generate a new model iteration, the user has to create a new iNNspector header, denoting the new model as a child of the previous one.
This can be done manually or by using the \emph{Branch Model} tool, which automatically generates a new iNNspector header with the selected model as parent.

\dbvisparagraph{Inspecting Models in the User-Interface}
After starting the iNNspector back- and frontend, the stored models are automatically loaded into the system.

\subsection{System Architecture \& Implementation}
\label{inn:subsec:implementation}

The iNNspector system is driven by custom logging mechanisms and a backend providing access to the data accumulated during training.
In the following, we provide a short overview over these components to convey an impression on how the system works and which steps it takes to integrate it into the model building workflow.

\subsubsection{iNNspector Keras Logs}
\label{inn:subsubsec:keras-logs}
We provide two custom components to make ordinary TensorFlow / Keras available in the iNNspector system.
The first one is the \emph{iNNspector} header, which provides metadata about a model, and the second one is the \emph{iNNspector Keras Log}, which modifies the default checkpoints to also include a subset of the testing dataset together with the corresponding activations for each layer.
In the following, we will describe how those two components work together to bring models into the iNNspector system.

\dbvisparagraph{iNNspector Header\label{inn:paragraph:innspector-header}}
The iNNspector header is a small code block in the model definition file encoding the model's meta-information in a key-value structure, such as its name and parent-child relationships.
\Cref{inn:listing:innspector-header} shows an exemplary function to generate the header.

\begin{lstlisting}[
    label={inn:listing:innspector-header},
    caption={[Function to generate the iNNspector header]Function to generate the iNNspector header encoding a model's meta information.},
    style=PythonCustomLst,
    float=b,
    captionpos=b
]
def make_innspector_header():
    return {
        "id": "0999a51f-5114-471a-8322-9fa40ae60387",
        "name": "ts_mnist_dense_skip-connections",
        "label": "MNIST Dense (Skip Connections)",
        "parents": ["bab98735-d119-433d-a893-08232533973e"]
    }
\end{lstlisting}

During model training, the iNNspector header is used by the iNNspector logging mechanism to create a model database.
Besides meta information, the database encodes information on the location of graph files and checkpoints, values of performance metrics, and other high-level statistics, such as the number of trainable parameters, memory consumption, and creation time of the model.
To reduce manual effort, the \emph{Branch Model} tool automatically generates a new iNNspector header with the selected model as parent.

\begin{figure}[!t]
    \centering
    \includegraphics[width=0.8\linewidth]{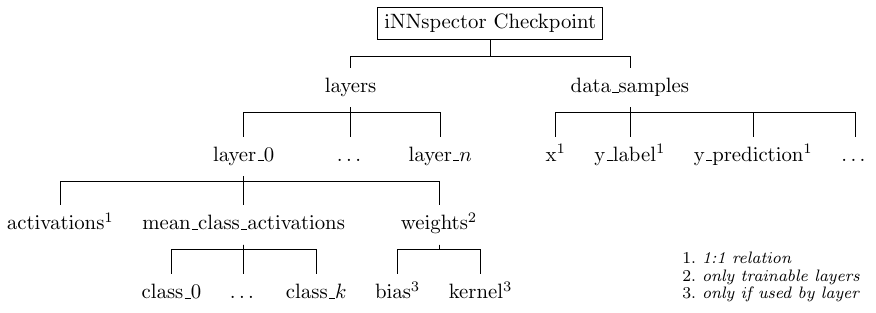}
    \caption[HDF5-structure of iNNspector checkpoint files]{
        HDF5-structure of iNNspector checkpoint files.
        We store weights and activations layerwise, while dataset samples, their labels (if available), and their predictions are stored globally per checkpoint.
    }
    \label{inn:fig:innspetor-checkpoint-hierarchy}
\end{figure}

\dbvisparagraph{iNNspector Checkpoint\label{inn:paragraph:innspector-checkpoint}}
The iNNspector checkpoint extends the default Keras checkpoint by dataset samples and sample activations.
For that, we re-structure the HDF5 file according to the hierarchy diagrammed in \cref{inn:fig:innspetor-checkpoint-hierarchy}.
Dataset samples ($\sim$ \emph{x}), dataset labels ($\sim$ \emph{y\_label}), and model predictions ($\sim$ \emph{y\_prediction}) are stored globally per checkpoint in the \emph{data\_samples} group, since they do not change over layers.
In contrast, weights and activations have to be stored layerwise.
Therefore, the \emph{layers} group contains an entry for every layer of the model, referenced by the Keras layer name.
Since activations and predictions are computed per dataset sample, inputs, labels, predictions, and activations share a one-to-one relationship, reflecting in equal size of their first matrix dimension.
Furthermore, each layer stores pre-computed averaged activations for the individual classes of the dataset ($\sim$ \emph{mean\_class\_activations}).
This avoids that the backend has to repeatedly load and aggregate all activations, since this data is frequently requested by the iNNspector frontend.
Layers with trainable variables also have a \emph{weights} group, containing the values of the layer's \emph{kernel} or \emph{bias}, respectively.
It should be noted that storing pre-computed activations might not be viable for huge models or models featuring convolutions with large output sizes due to storage size issues.
In this case, we would have to fall back to loading and executing the saved model instance on the fly with the drawback of higher response times.
To give an estimate: for our three use cases, we trained eight image classification models and ten variational auto-encoders with different architectures on the MNIST dataset.
Each model had eleven iNNspector checkpoints saved over the training, resulting in an overall size of $\approx\SI{70}{\giga\byte}$.

\subsubsection{iNNspector Backend}
\label{inn:subsec:innspector-backend}
The iNNspector backend exposes the data stored in the iNNspector system over different REST API routes for access from the iNNspector frontend.
It constantly monitors the contents of the logging directory and reloads the model database upon changes.
The API endpoints can be grouped into the following categories:
\begin{description}
    \item[Model IDs] A single \get-endpoint returning the IDs of all models in the database.
    \item[Model Info] Different \get-endpoints taking the model ID as a parameter and returning information about the model, such as architecture graph, performance metrics, or a catalog of checkpoints. All this information is stored in the model database.
    \item[Checkpoints] Different \post-endpoints taking the model ID and a JSON body as parameter and returning (transformed) values stored in-- and extracted from checkpoints, such as weights, activations, or interestingness values.
\end{description}
The API is built using FastAPI\footnote{\url{https://fastapi.tiangolo.com/}} and provides detailed OpenAPI\footnote{\url{https://swagger.io/specification/}} documentation, accessible under the backend URL.
To speed up repeated queries with similar parameters, we cache results in Redis for later re-use.

\dbvisparagraph{Data Transformation}
High-dimensional data~\textbf{(\ref{inn:data:high-dim})} necessarily has to undergo task-specific processing~\textbf{(\ref{inn:dim:processing})} to create human-interpretable representations~\textbf{(\ref{inn:mechanism:data-representation})}.
Therefore, all backend routes returning high-dimensional data take an additional \textbf{transform specification}, allowing to express arbitrary transformations that should be applied to the data.
The transform specification is defined in a grammar which allows specifying a list of operations that are subsequently applied.
Those operations map functionalities of Pandas and Numpy into our grammar.
Furthermore, our grammar provides \emph{production rules}, allowing, e.g., to branch and merge data columns.

By using transformations to filter and aggregate the requested data aggressively, we can mitigate the issues with the huge size of some high-dimensional data entities, facilitating the smooth exchange of data between backend and frontend over HTTP requests.
The transformations allow us to trade memory size for computational effort; combining them with caching allows us to utilize both advantages, leading to quicker overall response times.

\subsubsection{iNNspector Frontend}

Interface and functionalities of the frontend are described exhaustively in \cref{inn:subsec:application-walkthrough}.
The frontend is built using React\footnote{\url{https://reactjs.org/}} and TypeScript\footnote{\url{https://www.typescriptlang.org/}}.
The modularity enabled by the React framework and the strong type system enforced by TypeScript allow new developers to quickly find their way around the code, supporting our goal of system extensibility.

\section{Use-Cases}
\label{inn:sec:use-cases}
We show the applicability of our approach based on three use cases, which we derived from our model developer interviews and described in~\cref{inn:apx:subsec:implications}.
We selected these three use cases, since, during the interviews, the developers emphasized their importance.
All mechanisms proposed in~\cref{inn:subsec:mechanisms} and their corresponding implementations described in~\cref{inn:sec:system} will be utilized to solve the use cases, providing an impression on how their integration in the everyday model building, model diagnosis, and model refinement workflow enables systematic analysis and debugging of the connection between model behavior and its architectural parts.
Use case 1 is relatively universal, demonstrating how the system facilitates the everyday model building and --refinement workflow.
Subsequently, use cases 2 and 3 go into increasing detail on a specific debugging task.
We utilize these use cases in our user study, as described in~\cref{inn:sec:evaluation}.

\subsection{Use Case 1: Model Assessment, --Comparison, and --Refinement}

A model developer wants to build a model for image classification, which should eventually run in real-time on an embedded device.
The hardware has low computational power and limited storage size.
\begin{figure}[t]
    \centering
    \includegraphics[width=0.8\linewidth]{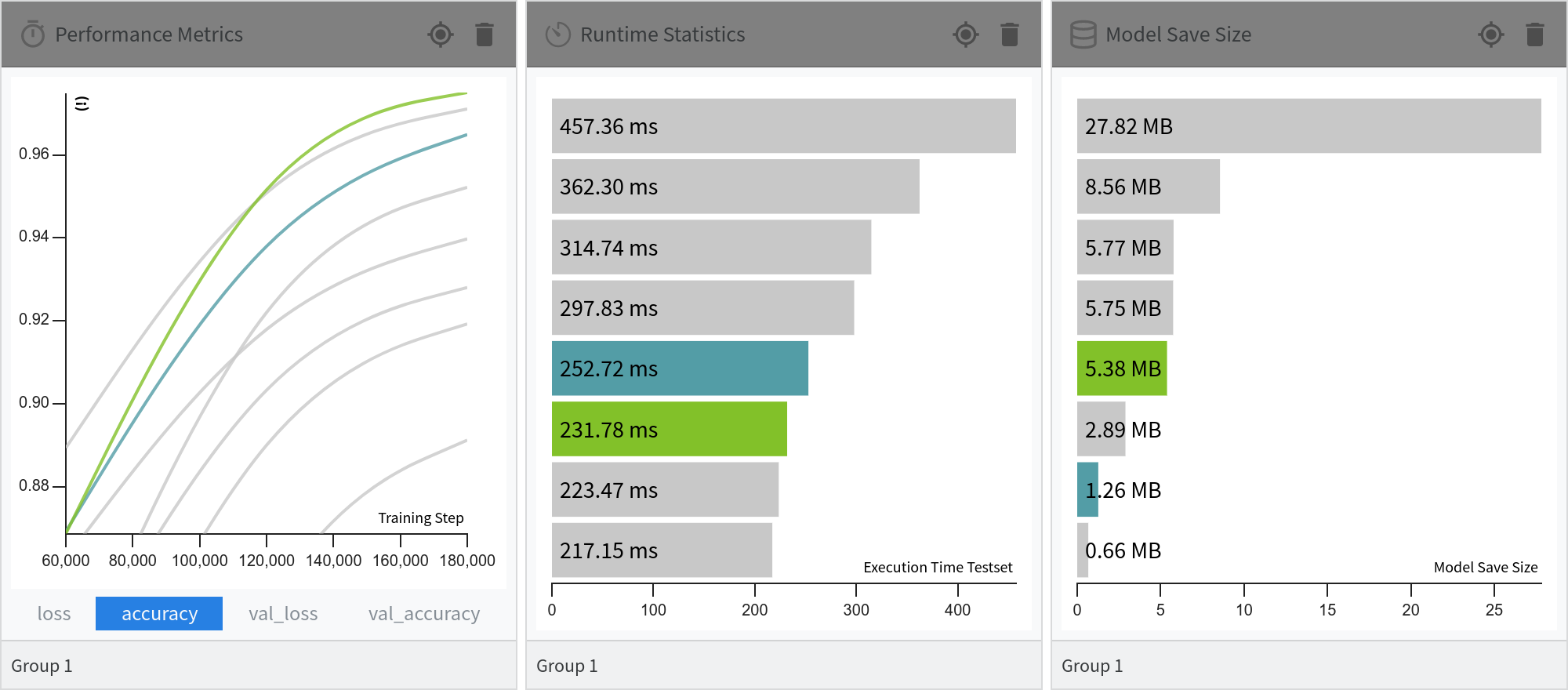}
    \caption[UC1: iNNspector widgets to assess model performance]{
        iNNspector widgets to assess model performance over an experiment.
        The \textit{Performance Metrics} widget shows the accuracy of each model,
        the \textit{Runtime Statistics} widget shows their expected execution time,
        and the \textit{Model Save Size} their size on disk.
    }
    \label{inn:fig:uc1-widgets}
\end{figure}
Therefore, execution time and model save size must be minimized while maximizing the classification performance.
Based on his prior experiences in deep learning, the developer creates an initial model architecture.
During training, the model already shows promising results, reaching an accuracy of 88\% after 20 epochs.
However, he wants to refine architecture and hyperparameters further to boost the model's performance while keeping the hardware requirements constant.
In an iterative approach, the developer creates multiple model variations, experimenting with 2D convolutions, network structures, and layer sizes.
Thereby, he uses the iNNspector logging mechanism to track and debug his experiments in the iNNspector system.
Using L3, the \emph{tree of models} indicates the variations between the models in the abstract model architecture representation while also confirming the change in trainable parameters between model variants.
By applying the \emph{performance metrics} tool to the group of trained models, he can evaluate and compare their classification accuracy.
Since computational complexity matters in the real-time scenario, the \emph{runtime statistics} tool provides a good estimate of the expected model execution time.
Respectively, the \emph{model save size} tool gives an overview of the model size.
Comparing accuracy, save size, and execution times, two models seem promising, as highlighted in~\cref{inn:fig:uc1-widgets}.
The developer adds single-model \emph{performance metrics} widgets for those two models individually.
To estimate their tendency to overfit, he merges the widgets in the group view to compare training-- and validation loss, showing that one model seems to better generalize than the other.
Outgoing from this superior model variant, the developer wants to further tweak the performance by assessing the configuration of its layers and inspecting each layer's value distribution.
He navigates to L2, exposing more architectural details and providing access to the model's variables.
He creates widgets visualizing convolutional and dense kernels, enabling him to compare the individual data distributions.
In the second layer, the variance increases significantly over the training epochs, while the third layer shows many values staying close to zero, indicating a bottleneck in layer two, which limits the information flow.
With this insight, the developer has a clear leverage point for further model tweaking.
He creates a new generation of models with improved performance by re-balancing the capacity between layers two and three.

\subsection{Use Case 2: Balancing Losses}
A developer wants to build a variational auto-encoder (VAE)~\citep{Kingma2014AutoEncodingVariational} for image generation.
Constraining the latent distribution to resemble a normal distribution ensures that when sampling the latent vector from a normal distribution, valid decoder output is generated.
\begin{figure}[t]
    \centering
    \captionsetup[subfigure]{justification=centering}
    \begin{subfigure}{0.277\linewidth}
        \includegraphics[width=\linewidth]{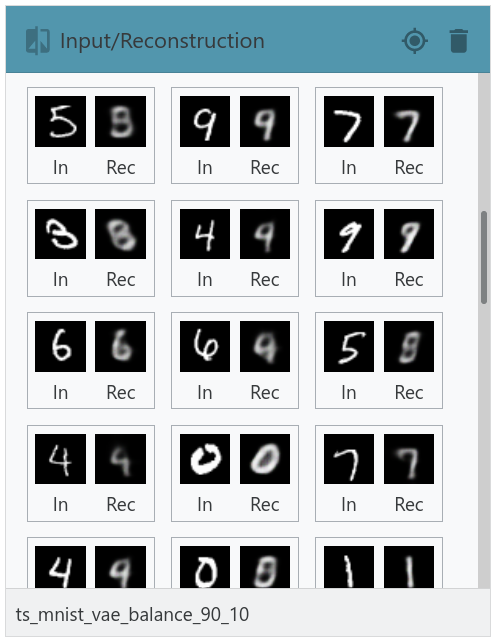}
        \caption{Image reconstructions for different dataset samples.}
        \label{inn:fig:uc2-widgets_1}
    \end{subfigure}
    \hspace{1em}
    \begin{subfigure}{0.567\linewidth}
        \includegraphics[width=0.48\linewidth]{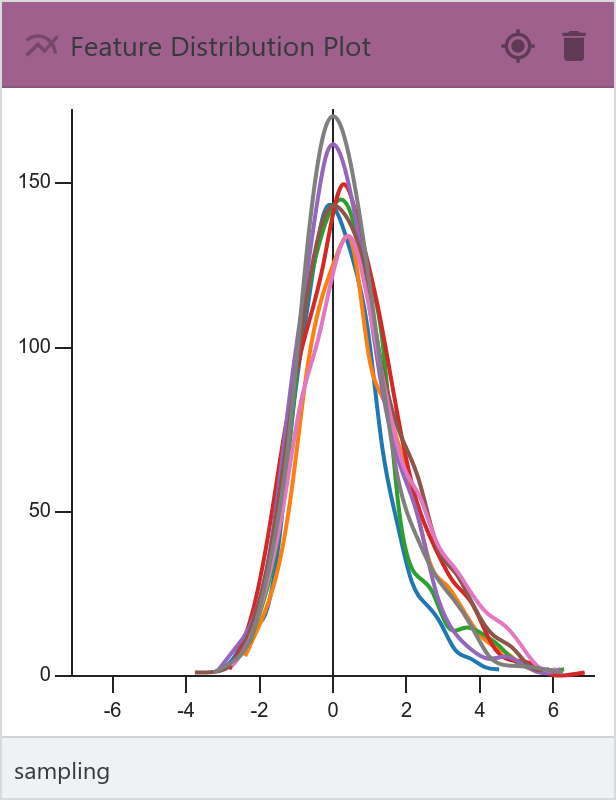}
        \hfill
        \includegraphics[width=0.48\linewidth]{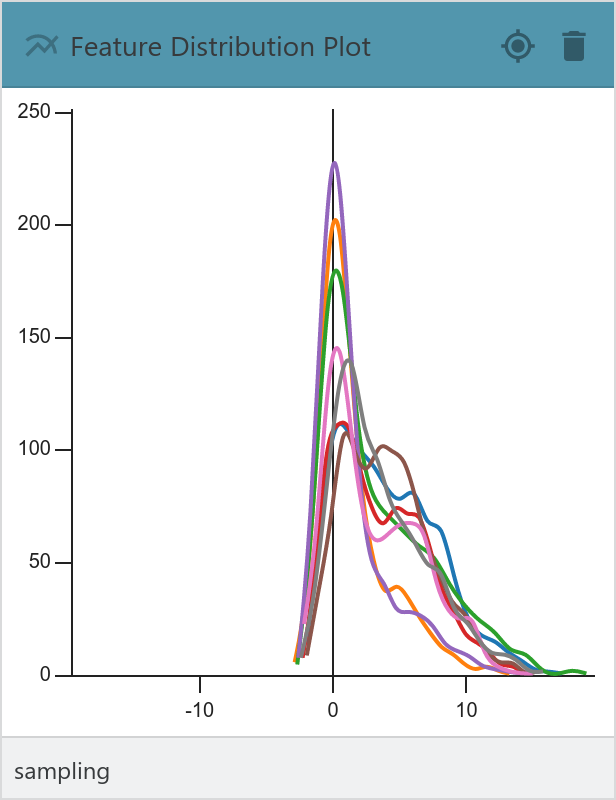}
        \caption{Feature distribution ($\sim$ probability density function) for each dimension in the latent variable of two models.}
        \label{inn:fig:uc2-widgets_2}
    \end{subfigure}
    \caption[UC2: iNNspector widgets to balance losses in a VAE]{
        iNNspector widgets to balance losses in a VAE.
        (\subref{inn:fig:uc2-widgets_1}) The \textit{input/reconstruction} widget shows the input images and their reconstructions for a subset of the test dataset.
        (\subref{inn:fig:uc2-widgets_2}) The \textit{feature distribution} widget shows the shape of the latent distribution, suggesting the left model to be better suited for sampling.
    }
    \label{fig:uc2-widgets}
\end{figure}
The developer uses the Kullback–Leibler (KL) divergence to calculate the difference between the actual distribution in the latent space and the desired normal distribution.
To optimize both reconstruction results and shape of the latent distribution, he must precisely balance reconstruction loss and KL divergence against each other.
However, due to this parameter being dependent on dataset, model, and human perception, there is no analytical solution to this problem.
Therefore, the developer creates multiple variations of the network, altering the loss balancing factor between models.
The customized Keras model overrides the \texttt{train\_step} and \texttt{test\_step} functions to also return KL-- and reconstruction loss, which are, thereby, automatically tracked by the iNNspector logging mechansism.

\noindent While the reconstruction loss can be visually evaluated by simply displaying the reconstructed images using the \emph{input/reconstruction} tool shown in~\cref{inn:fig:uc2-widgets_1}, the KL loss is a rather abstract factor.
To assess it and examine the value distribution in the latent space, the developer descends into L2, enabling him to visualize how the \emph{overall} distribution of activations $l$ in the latent layer developed over the training epochs.
Using automated interestingness measures, he finds that the model with a $25:75$ balance between reconstruction-- and KL loss has a much lower variance of $\Var{\left(l_{25:75}\right)} = 2.02$ than the model with a $90:10$ balance, where the variance is $\Var{\left(l_{90:10}\right)} = 12.38$.
Since, in VAEs, uniformity of each dimension in the latent variable is important, the developer uses the \emph{feature distribution plot} to plot the probability density function for each dimension individually, depicted in~\ref{inn:fig:uc2-widgets_2}.
It shows that the model with more weight on the KL loss has a more uniform distribution in all latent dimensions.
For the following iterations of the model development and refinement process, the inspection results enable the model developer to make well-founded decisions over the loss balance.

\subsection{Use Case 3: Debugging Distributions}

Extending use case 2, while weighting reconstruction and KL loss, the developer seems not to be able to find a satisfying balance.
Instead, either the reconstructed image or the latent distribution deviates significantly from the desired results.
\begin{figure}[t]
    \centering
    \captionsetup[subfigure]{justification=centering}
    \begin{subfigure}{0.2\linewidth}
        \includegraphics[width=\linewidth]{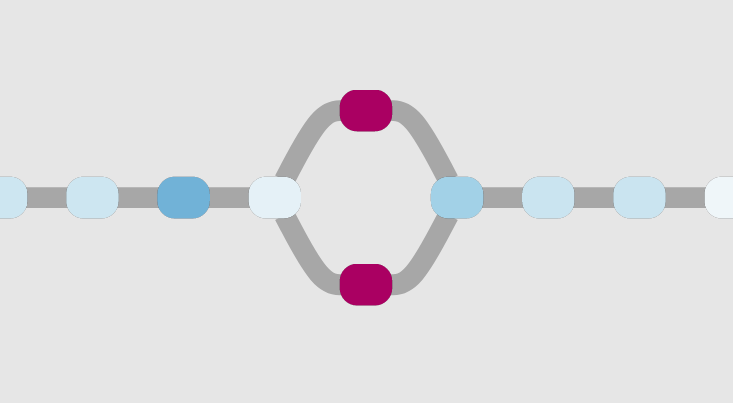}
        \caption{Skewness in the latent block.}
        \label{inn:fig:uc3-interestingness}
    \end{subfigure}
    \hspace{1em}
    \begin{subfigure}{0.3\linewidth}
        \includegraphics[width=\linewidth]{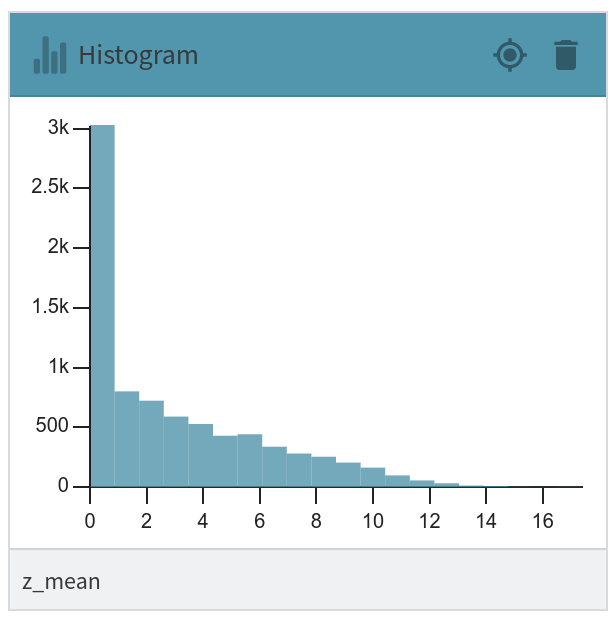}
        \caption{Histogram.}
        \label{inn:fig:uc3-histogram}
    \end{subfigure}
    \hspace{1em}
    \begin{subfigure}{0.4\linewidth}
        \includegraphics[width=\linewidth]{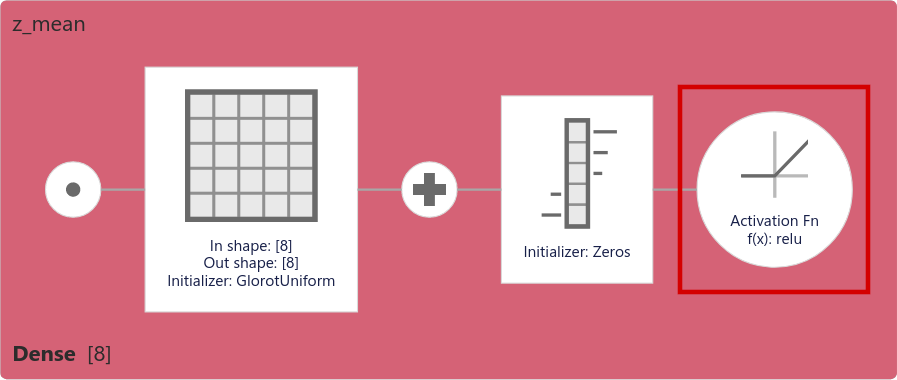}
        \caption{Layer Graph.}
        \label{inn:fig:uc3-layer}
    \end{subfigure}
    \caption[UC3: iNNspector widgets to investigate the distribution of activations]{
        Widgets to investigate the distribution of activations in the latent block.
        (\subref{inn:fig:uc3-interestingness}) Interestingngess annotations points the developer to the skewness of the activations in the latent block.
        (\subref{inn:fig:uc3-histogram}) The \textit{Histogram} widget provides a detailed view on the distribution of activations in the latent layer.
        (\subref{inn:fig:uc3-layer}) The \textit{Layer Graph} reveals the erroneous activation function, leading to the skewed distribution.
    }
    \label{fig:uc3-widgets}
\end{figure}
The strong skew of the the activations in the latent space, visualized by the widgets in~\cref{inn:fig:uc2-widgets_2}, prompts further investigation.
Using the \emph{Activation Skew} interestingness measure on L3, he notices even stronger skew in two layers of the latent block, depicted in~\cref{inn:fig:uc3-interestingness}.
To trace down the detected abnormality to its root cause, the developer descends to L2 and closely investigates the respective layers, namely the \texttt{z\_mean} and \texttt{z\_log\_var} layers, which determine the parameters of the normal distribution.
Applying the \emph{Histogram} tool confirms this observation, as shown in~\cref{inn:fig:uc3-histogram}; there are only positive activations on that layer.
Upon closer inspection of the layer's inner elements, the developer identifies the reasons for that behavior:
due to a careless mistake when defining the \emph{Dense} layer in Keras, the layer's output is passed through a \emph{ReLu} activation function depicted in~\cref{inn:fig:uc3-layer}, resulting in all negative values being cut off.
The developer creates a new model generation using the \emph{Branch Model} tool and removes the suspicious activation function.
Re-training the new model proves the refinement's success.

\section{Expert User Study}
\label{inn:sec:evaluation}
To evaluate the iNNspector system, we conduct a user study with four participants.
The prior objectives of the evaluation study are to assess (1) whether the system does fulfill its claim of being a universal tool for model debugging, (2) the usability and effectiveness of the system, and (3) identify open challenges and future work.

\subsection{Study Setup}

In this section, we describe our study setup, guided by the methodology proposed by~\citet{Sperrle2021SurveyHumanCentered}.
First, we explain our study procedure for each session; second, we describe the participants of our study; and lastly, we discuss the data used in our pair analytics sessions.

\dbvisparagraph{Study Procedure}
We divide each study session into four blocks.
Each session takes approximately two hours and is recorded for later transcription and qualitative evaluation.

\begin{description}
    \item[Current Workflow ($\approx\SI{5}{\minute}$):]
    We start by capturing personal information, such as the amount of experience our participants have with deep learning.
    We continue with questions on the participant's usual development workflow, including questions on experiment tracking, model assessment, and potential debugging routines.
    This block is closed by open-ended questions on deficiencies of current debugging tools and wishes towards a system for universal deep model inspection.
    \item[Formal Framework ($\approx\SI{10}{\minute}$):]
    By reference to \cref{inn:tab:data-types} and \cref{inn:fig:mechanisms}, we briefly describe the parts of our conceptual framework being most relevant for the implementation of a debugging system, namely the data categories (\ref{inn:subsec:data}) and the mechanisms to make this data accessible (\ref{inn:subsec:mechanisms}).
    Following, we capture feedback on the framework in a semi-structured interview, covering its completeness and anticipated applicability.
    \item[iNNspector System \& Use Cases ($\approx\SI{45}{\minute}$):]
    This is the central part of the interview, taking up about one hour of the overall two-hour interview slot.
    In a guided exploration phase, we explain the iNNspector user interface to the participant, similar to the application walkthrough in~\cref{inn:subsec:application-walkthrough}.
    When the participants confirm they feel confident with the interface, they are successively asked to solve use cases 1 to 3 described in~\cref{inn:sec:use-cases} in a pair-analytics~\citep{AriasHernandez2011PairAnalyticsCapturing} session.
    The participants receive a handout explaining the setting of each use case and defining its corresponding analysis tasks, which we also summarize verbally.
    The participants are supposed to solve the use cases mainly on their own; however, when they articulate that they are looking for a particular tool or have questions on either task or system, we jump in to help.
    To capture impressions and feedback on the system, we encourage the participants to speak out loud about their reasoning during this process.
    \item[Questionnaire and Feedback ($\approx\SI{15}{\minute}$):]
    Finally, we collect quantitative and qualitative user feedback on the system.
    Based on a customized questionnaire, inspired by the NASA Task Load Index~\citep{Hart1988DevelopmentNasaTlx} and the System Usability Scale~\citep{Brooke1996SusQuickDirty}, we evaluate iNNspector's usability, usefulness, task load, and effects on the debugging behavior.
    Furthermore, the questionnaire captures the usefulness of specific system components, such as the different levels of abstraction or the toolbox.
    The interview finishes with open-ended questions and broader feedback on the system, including missing components, ideas for future extensions, and general thoughts on the system.
\end{description}

\dbvisparagraph{Participants}
All four participants (P1, P2, P3, and P4) have an educational background in information science on masters level.
P1, P2, and P4 have specialized in deep learning and show multiple years of experience with deep learning development in both research and industry.
P3 has had prior experiences in deep learning as part of his bachelors and master’s degrees.
P1 and P4 were already part of the requirements study, while P2 and P3 were only part of the evaluation study.
Neither the study leader nor the participants are native English speakers; nevertheless, the study sessions are conducted in English to achieve the most accurate transcription and citation.

\dbvisparagraph{Data}
The models available in the iNNspector system are pre-trained on the MNIST dataset.
For each use case, a selection of models supporting the use case is presented to the user.
The models simulate a representative real-world model building and refinement workflow, with model iterations implementing architectural changes based on observations from past generations.
For use case 1, the models are classifiers, while for use cases 2 to 4 the models are variational auto-encoders.

\subsection{Study Results}
\label{inn:subsec:evaluation-study-results}
While the participants showed different exploration strategies, they all reached the analysis goals of use cases 1 and 2 in similar times (P1: \SI{40}{\minute}, P2: \SI{36}{\minute}, P3: \SI{22}{\minute}, and P4: \SI{30}{\minute}).
P3 was the fastest participant; however, also being the least experienced, he did not reach an analysis depth comparable to the other participants, despite in the end drawing the same conclusions.
Use case 3 was introduced as an optional, open-ended task, where we only added a new model generation with the suspicious activation function removed and asked the participant to determine (1) possible issues with the previous generation and (2) how the new generation changed and why this resolves the issue.
Here, the participants focussed on comparing the architectures and, subsequently, found the difference in the latent block.
However, they did not recognize a major improvement, which might be due to relatively blurry reconstruction results of both model variants, caused by the limited capacity of the latent variable ($\dim = 10$).

In the following, we summarize the feedback of our participants.
Due to the limited number of participants, we present quantitative and qualitative results together.

\subsubsection{Feedback on the Framework}
Before facing our participants with the iNNspector system, we collect feedback on our conceptual framework.
Particularly, we introduce the data categories (\cref{inn:subsec:data}) and global mechanisms (\cref{inn:subsec:mechanisms}) and collect feedback on the framework's correctness, completeness, and usefulness.

\dbvisparagraph{Data}
In general, our participants confirmed the data categories to make sense and match their mental model about the data arising in machine learning experiments (P1, P2, P3, P4).
P1 was missing a data category for meta-parameters about the experiment, such as batch size or optimization parameters.
P2 had similar concerns, but readily proposed an integration of those parameters into our categorization as \emph{constant scalar} data, which perfectly complies with our model.
We updated the categorization accordingly.
P2 also mentioned that he sees functions (\ref{inn:data:function}) as part of the structural architecture graph (\ref{inn:data:structural-ag}) since they ``define the geometry of the graph.''
We added a respective pointer to~\cref{inn:tab:data-types}.

\dbvisparagraph{Mechanisms}
When introducing the mechanisms, our participants instinctively started to compare the framework to state-of-the-art tools for model debugging.
E.g., P4 identified a discrepancy between usefulness and convenience in existing tools: ``either you need to self-make your tools, which are very low level [...], or you can use TensorBoard but they have only very general statistical evaluations.''
Therefore, he valued the framework as an ``improvement over existing tools [...], if applied [implemented] correctly.''
With regards to the proposed data representation (\ref{inn:mechanism:data-representation}), P2 stated that ``scalar values and images can be visualized in existing tools, but that's mostly it.''.
Also, he mentioned to not use the graph view in TensorBoard very often, since it yields ``not much new information.''
In contrast, our framework proposes the entities of the structural backbone (\ref{inn:mechanism:structural-backbone}) as a navigational component and a hook to access underlying data.
Overall, our participants agreed that the mechanisms were reasonable (P1, P2, P3, P4) and expressed their anticipation for the system implementation.

\subsubsection{Feedback on the System}

Gathering feedback on the system is divided into three phases: first, we introduce our participants to the iNNspector system in a joint exploration phase.
We describe the interface similar to~\cref{inn:subsec:application-walkthrough} while the participant already has full control over the system to discover and explore its functionalities.
In the second phase, the user solves the use cases 1 to 3 described in~\cref{inn:sec:use-cases}.
Finally, we gather feedback via a questionnaire containing both quantitative and qualitative questions.
In the following, we summarize the feedback structured by our main evaluation goals \emph{usability}, \emph{usefulnes}, and the \emph{claim of the system to be a generalizing tool for systematic model debugging.}

\begin{figure}[!t]
    \centering
    \includegraphics[width=\linewidth]{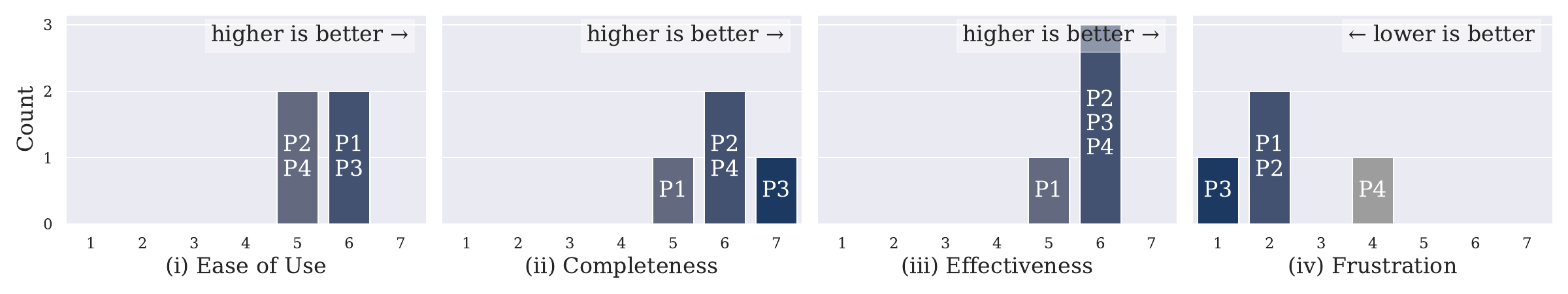}
    \caption[Quantitative study results on the usability of the iNNspector system]{
        Quantitative study results on the \textbf{usability} of the iNNspector system.
        The bars show the count of participants rating the system with the respective score.
    }
    \label{inn:fig:study-usability}
\end{figure}

\dbvisparagraph{Usability}
Usability is captured in terms of \emph{ease of use}, \emph{completeness}, \emph{effectiveness}, and \emph{frustration}.
\Cref{inn:fig:study-usability} shows the quantitative results of the evaluation, which are substantiated by qualitative feedback in the following.

Overall, the system was described as ``intuitive'' (P1, P3, P4) and ``very usable'' (P2).
Particularly, functionalities like the navigation over levels of abstraction, filters, and widget groups were intuitively used by all our participants.
In contrast, some participants needed a quick reminder that the class selector panel could help solve use case 2.
The two major points of critique were that the system is quite complex and would need training to fully exploit all functionalities (P1, P2, P4) and that the tool icons and --descriptions could be improved (P1, P3, P4).
In general, the participants were pleased with the completeness of the system.
The primary point of critique were missing tools, with a confusion matrix being the most frequently requested (P1, P2, P4).
As a response, we added the tool after the study.
The effectiveness of the tool was rated between okay and good without further comments.
The frustration factor was low for most participants (P1, P2, P3), despite a memory leakage bug occurring for P2 causing significant lag.
However, P2 stated that ``[the bug] was not a big deal.''
In the meantime, the error has been fixed.
P1 told the frustration to be ``actually very low''; however, he also experienced a bug causing the class selector to malfunction:
``[if there wasn't the bug], I would put it at 1, as [the system] works quite nice.''
P4 stated medium frustration, caused by a lack of tools to solve use case 2.
Particularly, he was missing a confusion matrix tool and linking and brushing between misclassified samples across different models.

\begin{figure}[!t]
    \centering
    \begin{subfigure}[t]{\linewidth}
        \centering
        \includegraphics[width=\linewidth]{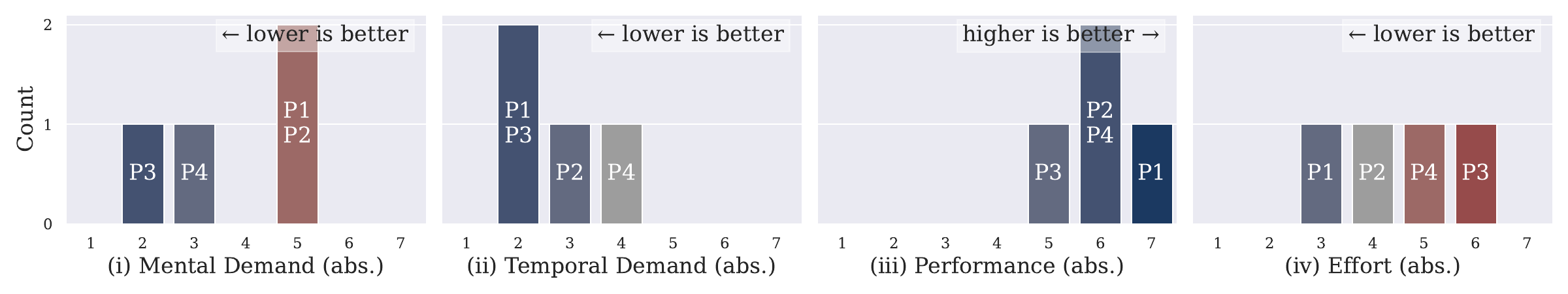}
        \caption{%
            \textbf{Absolute} usefulness rating, i.e., without considering the complexity of the tasks or how it would be solved using traditional tools.
        }
        \label{inn:fig:study-usefulness-a}
    \end{subfigure}\\
    \vspace{1em}
    \begin{subfigure}[t]{\linewidth}
        \centering
        \includegraphics[width=\linewidth]{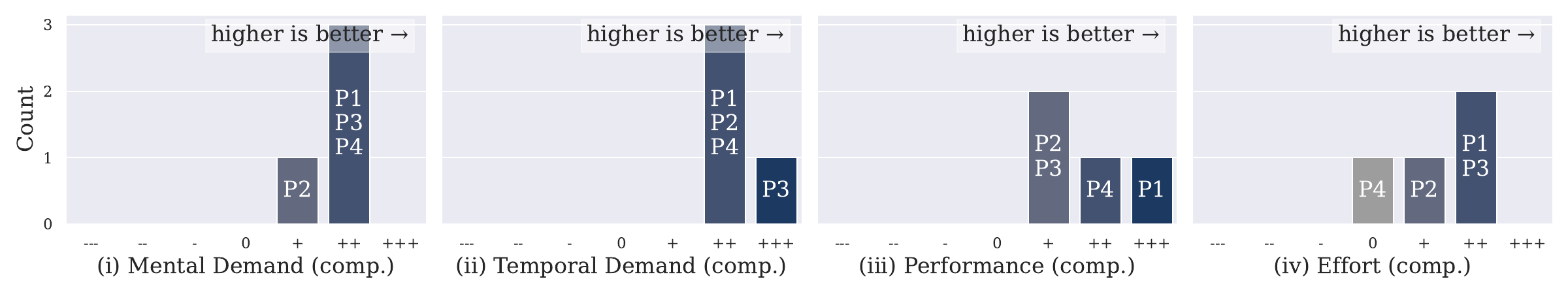}
        \caption{%
            Relative usefulness rating, comparing to the \textbf{usual workflow as a baseline}.
            ``--\,(--\,--)'' means a (strong) negative and ``+(++)'' a (strong) positive influence of the iNNspector system.
        }
        \label{inn:fig:study-usefulness-b}
    \end{subfigure}
    \caption[Quantitative study results on the usefulness of the iNNspector system]{
        Quantitative results on the \textbf{usefulness} of the system.
        While the mental demand and effort are rated high on the absolute scala, the system is rated as a great improvement in comparison to the usual workflow.
    }
\end{figure}

\dbvisparagraph{Usefulness}
Usefulness is captured in terms of \emph{mental demand}, \emph{temporal demand}, \emph{performance}, and \emph{effort}.
Each question has two parts, with the first one asking absolute values (\cref{inn:fig:study-usefulness-a}) and the second one comparing to the usual workflow of our participants as a baseline (\cref{inn:fig:study-usefulness-b}).
In the following, we will report both the quantitative and qualitative feedback on the system's usefullness.

Despite the mental demand of use cases 1 to 3 being rated relatively high on the absolute scale (P1, P2), all participants aggreed that the system significantly simplified the tasks and ``helped a lot so that the mental demand was not very high.''
However, P2 mentioned concerns about mental overload when using the system: ``it's more stuff at once; for example, in a Jupyter notebook or in my normal workflow it's more sequential.''
Our participants agreed on the system strongly decreasing the temporal demand, rating the time gain high (P1, P2, P4) to very high (P3).
P2 noted that ``doing all of [the debugging in the use cases] would definitely take some time, configurations, and stuff.''
P1 particularly mentioned that the system makes it ``much easier to compare different architectures.''
All participants were satisfied with the performance they achieved when solving the use cases.
In comparison, P2 at first expected similar performance with traditional tools but on second thought revised his opinion: ``with the latent distributions it would have taken way longer and would not be that easy.''
Generally, the absolute effort to solve the tasks was rated mid (P1, P2) to high (P3, P4).
P4 ascribed this mainly to being ``new to the system.''
However, in comparison, the system still was rated as an improvement (P2, P3, P4).
Only P4 was sceptical: ``[the system did not reduce the effort] in comparison with statistical evaluation.''

\dbvisparagraph{Adequacy for Systematic Model Debugging}
Concluding our questionnaire, we capture general impressions of the system based on the following questions.
\begin{enumerate}[label=(\roman*),topsep=0.5em]
    \item \emph{How much has the system increased or simplified your access to the data arising in machine learning experiments?}\\[0.25em]
    All our participants agreed that iNNspector simplifies access to data greatly (P1, P2, P3) to very greatly (P4).
    P4 confirmed the coverage: ``you can look at everything you want.''
    Overall, during the exploration-- and use case phases of our sessions, none of the participants requested data that was not available in the system; critique was only mentioned about the representation of data, i.e., our participants wanted to customize tools and widgets according to their needs.

    \item \emph{How much time would iNNspector save you in your everyday model development and debugging workflow?}\\[0.25em]
    The time saving was rated between neutral (P2) over good (P1, P4) to very high (P3).
    P2 expressed his uncertainty: ``I can not really say yet. I would have to use it a bit more.''
    Since the models were pre-trained preliminary to the study due to time constraints, our participants gave to consideration that ``this depends on how much time you need to set up the model[s] like this'' (P3).
    After explaining the logging mechanism described in~\cref{inn:subsubsec:keras-logs}, P3 added that ``this sounds like it is really easy; so I think it would save really much time.''

    \item \emph{In your opinion, to what extent does iNNspector fulfill the claim of being a universal tool for systematic deep model debugging?}\\[0.25em]
    The universality of the system was rated from okay (P1) over good (P2, P3) to very high (P4).
    Here, customizability and expandability were the decisive points for all participants.
    However, P1 added that the system ``already has a lot to offer.''
    P4 justified his distinct vote: ``I think I couldn't come up with a use case that I couldn't solve with the system.''

    \item \emph{How likely are you to deploy iNNspector in your everyday model development and debugging process?}\\[0.25em]
    P1, P2, and P4 stated they would probably integrate the system into their everyday workflow.
    P3 not being confronted with deep leraning on a regular basis reasoned ``[he] would like to use this if [he] would develop machine learning models.''
    Again, the simplicity of the logging mechanism was identified as crucial: ``if I could run it easily with my models, I most likely would use it.''
\end{enumerate}

\begin{figure}
    \centering
    \includegraphics[width=\linewidth]{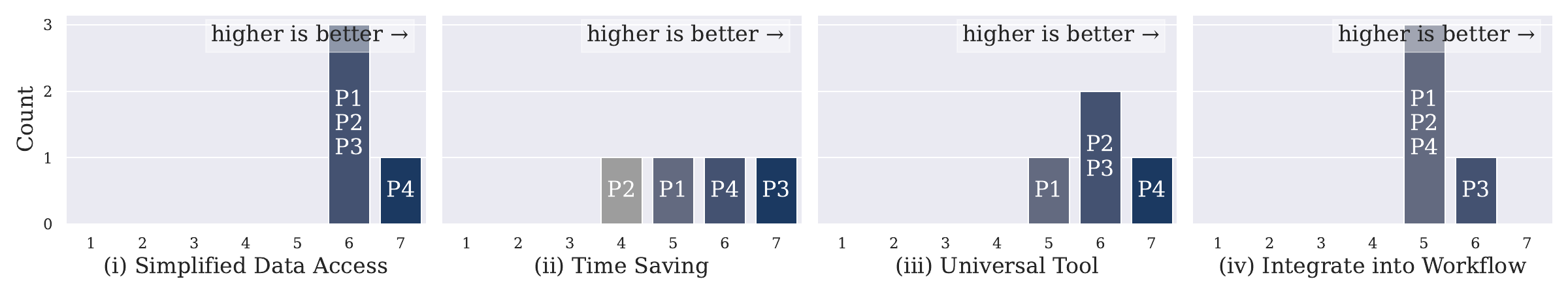}
    \caption[Quantitative study results on iNNspector's goal to enable systematic model debugging]{
        Quantitative results on the \textbf{adequacy} of the iNNspector system for \textbf{systematic model debugging}.
        Particularly the system's ability to simplify access to the data arising in deep learning experiments was rated high.
    }
    \label{inn:fig:study-adequacy}
\end{figure}

\section{Discussion}
\label{inn:sec:discussion}
Summarizing our work, we identify concrete recommendations guiding the design of real-world systems for the systematic debugging of machine learning experiments.
Furthermore, we discuss current limitations of our approach and how they will be addressed in future work.

\subsection{Take-Home Messages}

This section summarizes the major take-home messages of our approach and distills general recommendations for the design of systems for systematic network debugging.

\dbvisparagraph{Consistent Logging of Appropriate Data\label{inn:th1}}
All relevant data generated in machine learning experiments should be logged for later use in the debugging stage.
Model evolution should be actively tracked by encoding parent-child relationships in the model metadata, allowing later reconstructions of the experiment's progression.
Logging sections of the training and testing dataset is crucial in two ways: firstly, single samples and their model output can be inspected in the debugging stage, e.g., to investigate miss-classifications.
Secondly, the activations of a model only exist under input data.
Therefore, the activations should be computed and stored for each model checkpoint and data sample, allowing the inspection of activation distributions and single neurons.

Notably, the so-logged data can grow to a significant size.
However, we argue that the vast majority of physical storage requirements trace back to checkpoints, which are in most cases logged anyway to restore a particular model state.
If the amount of logged data exceeds a feasible storage size, pre-aggregations before storing the data can significantly reduce its size.
Since this comes at the cost of reduced flexibility in the debugging stage, the advantages and disadvantages of pre-aggregations should be carefully weighed.
For architectures producing particularly large activation data (e.g., convolutional networks on large image sizes), the activations can be computed on the fly with the downside of slower query times.

\dbvisparagraph{Modular Approach for Systematic Debugging\label{inn:th2}}
Often, there is no ideal default representation of the data to inspect.
Our proposed framework tackles this challenge by structuring the dimensions that influence and determine the data representation, maximizing flexibility while simultaneously resolving mutual dependencies.
The presented system shows how different visualization and interaction techniques can work together to combine these dimensions in a reasonable way.

\dbvisparagraph{Global Mechanisms\label{inn:th3}}
The global mechanisms to navigate the data space elaborated in \cref{inn:subsec:mechanisms} provide entry points for implementing a system for the systematic debugging of deep learning models.
While not all proposed mechanisms have to be strictly followed to make such a system work, they highlight the challenges that must be considered and suggest potential solutions.
The iNNspector system demonstrates, how the mechanisms can be instantiated into a real-world application, giving the developer access to the vast data space of machine learning experiments.

\subsection{Limitations and Future Work}
\label{inn:subsec:limitations-and-future-work}

In more complex experiments, both the \textbf{tree of models} and the  \textbf{architecture graph} might rapidly grow to an unmanagable size, leading to information overload, as also noted by several of our study participants.
To mitigate this problem we will include mechanisms for semantic block detection in future versions of the system, leveraging common properties of the iterative model refinement process and architecture graphs.
Particularly, model variations often share common architecture or hyper-parameter configurations, e.g., we could visually group models that only contain changes in one parameter, such as the batch size.
Likewise, architecture graphs frequently contain repetitions of similar structures, e.g., stacked residual blocks \citep{He2016DeepResidualLearning}, or common semantic blocks, e.g., encoder/decoder structures \citep{Vaswani2017AttentionIsAll}.
Again, such blocks could be visually abstracted, significantly reducing the complexity of advanced model architectures.

\dbvisparagraph{Extensibility}
The iNNspector system is designed to be easily extendable by own tools and widgets.
To fully determine tool and corresponding widget, the following aspects have to be defined:
\begin{description}
    \item[Tool Defintion ---] The appearance of the tool, including icon, name, description, and category.
    Furthermore, the UoA type to which the tool is applicable has to be specified.
    \item[Data Source ---] The kind of data to query from the backend, i.e., the API endpoint together with its corresponding paramters.
    This includes model id and the kind of data, but also the transformation specification.
    \item[Visual Appearance ---] Finally, the appearence of the widget, including visualizations and interactions, has to be configured in a standardized way.
\end{description}
While at the moment, this still affords programming skills in React and TypeScript, future versions of the system will support specifying tools as external plugins.
Particularly, tool definition and data source can easily be defined in a standardized data-serialization format, such as YAML\footnote{\url{https://yaml.org/}} or JSON\footnote{\url{https://www.json.org/json-en.html}}.
To also support such definition for the visual appearance of the widget, we rely on the Vega-Lite grammar of interactive graphics~\citep{Satyanarayan2017VegaLiteGrammar}, allowing to define complex interactive visualizations in plain JSON\@.

\dbvisparagraph{Specialized Default Tools}
\label{inn:paragraph:limitations-and-future-work-specialized-default-tools}
Complementary to the extensibility of the system, we will provide additional, specialized tools for common debugging tasks.
Particularly, as identified in our conceptual framework (see~\cref{inn:subsubsec:design-dimensions}, ``\nameref{inn:subsubsec:design-dimensions}''), we consider formal verification and correctness checking as important debugging tasks.
Therefore, as a first step towards formal verification, e.g., through SMT~\citep{Katz2017ReluplexEfficientSmt}, we will provide tools to investigate the decision boundaries of neural networks, e.g., through counterfactuals~\citep{Looveren2021InterpretableCounterfactualExplanations} and to check its robustness, e.g., through adversarial examples~\citep{Goodfellow2015ExplainingHarnessingAdversarial,Carlini2017TowardsEvaluatingRobustness}.

\dbvisparagraph{Transferability}
Its flexible design, the described extensibility, and the formal guidelines provided by the framework make iNNspector generalize well to various types of data, models, and tasks.
However, particular application domains might afford extensions going beyond defining new widgets and tools.
E.g., to debug models for deep reinforcement learning, closely-integrated tools to visualize state space and agent policies might be beneficial~\citep{Saldanha2019RelvisVisualAnalytics,Metz2022ComprehensiveWorkflowEffective}.
Another example are language transformers~\citep{Vaswani2017AttentionIsAll}, where a tight integration of attention visualizations into the structural backbone seems useful.

\dbvisparagraph{Explainability}
A direct integration of systems for model explainability~\citep{Spinner2020ExplainerVisualAnalytics, Kokhlikyan2020CaptumUnifiedGeneric} into the iNNspector system facilitates local, model-agnostic debugging.
This enriches the iNNspector toolbox with various state-of-the-art XAI techniques, such as LIME~\citep{Ribeiro2016WhyShouldI}, LRP~\citep{Bach2015PixelWiseExplanations}, or Integrated Gradients~\citep{Sundararajan2017AxiomaticAttributionDeep}.

\dbvisparagraph{Comparative Analytics}
Model comparison is an essential task for deep learning developers, which was confirmed in our requirements-- and evaluation interviews.
While iNNspector already supports comparison of scalar-- and high-dimensional data through multi-model-- and side-by-side widgets, a dedicated tool for architecture comparison can enhance the analysis of structural data.
For this, in future versions, we generate node embeddings for the architectural blocks of a model, allowing to compute a bi-directional mapping between two model graphs.
This mapping is used to align the graphs and visualize them side-by-side, highlighting commonalities and differences in the two architectures.

\section{Conclusion}
\label{inn:sec:conclustion}
With this work, we facilitate the systematic debugging of deep learning experiments.

As a formal foundation for implementing systems enabling this, we structure the design space of such systems and compile it into a \textbf{conceptual framework}.
Particularly, the framework
(1) categorizes the data arising in machine learning experiments,
(2) captures the design dimensions that have to be considered when building such a system, and
(3) defines concrete mechanisms to incorporate data and design dimensions into a comprehensive system for systematic debugging.
The considerations that went into the system are derived from (1) open challenges in the field and (2) requirement interviews with real-world deep learning developers.

Our framework proposes modular UI elements, called \emph{debugging components}, which can be instantiated in a toolbox-like approach.
The user determines the appearance of a debugging component through prioritizing characteristics of the available design dimensions.
E.g., the user can choose between an \emph{assessment} or \emph{comparison task}, or a \emph{visual} or \emph{verbal data representation}.
Debugging components can be attached to different \emph{units of analysis} in the model architecture.
We propose to make the model architectures explorable on different \emph{levels of abstraction}, ranging from a multi-model view down to single weights and neurons.

We transfer our conceptual work into a ready-to-use \textbf{system implementation} called \textbf{\emph{iNNspector}}.
Analogously to the framework, iNNspector lets the user explore models and their architectures over multiple levels of abstraction.
Different tools are available in a toolbox, which can be applied to units of analysis in the architecture representation.
The application of a tool results in creating a widget displaying the data.
The widget's appearance is determined by the tools specification and the UoA type it is applied on, automatically resolving dependencies between design dimensions.
Various additional functionalities, such as interestingness measures, localization of UoA, or class selection, enrich the system on a global scale.
Notably, iNNspector goes beyond demonstrating the feasibility of our framework; upon publication of this work, it will be released as open-source software and, hopefully, find its way into the everyday model building and --debugging workflow of deep learning developers.

The evaluation of the iNNspector is twofold.
First, we present three use cases demonstrating how it can be used to debug a variety of problems occurring in real-world model-building scenarios.
Second, we conduct a user study with three deep learning developers and a data scientist to evaluate the system's usability, usefulness, and versatility.

We argue that a global understanding of models and their evolution can only be achieved by systematically debugging, building the foundation for a well-informed diagnosis, verification, and refinement process.
All data arising in machine learning experiments are relevant for the debugging stage and should be logged and made explorable.
iNNspector enables this kind of data tracking and systematic debugging.

\clearpage
\printbibliography

\clearpage
\appendix

\section{Capturing Developer Workflows and –Challenges}
\label{inn:apx:sec:requirements-study}
In the following, we report additional insights we gained from our expert user study, structured into observations of the current workflows and their implications for model debugging.
Overall, we perceive the workflows of our developers as disjointed, with the individual tasks unconnected and a variety of separate tools involved.

\subsection{Current Workflows and Observations}
\label{inn:apx:subsec:current-workflows}

\sumbox{\ref{inn:apx:subsec:current-workflows}}{
    Developers in our study report disjointed workflows in deep learning model development and debugging.
    Their development is usually based on existing architectures and iterative trial-and-error refinement, hindered by a lack of guidelines and difficulties in detecting mistakes.
    They use various tools for tracking iterations and assessing model performance.
    Systematic debugging workflows are rarely employed.
}

As a baseline, we capture the interviewed developers' conventional model development-- and debugging practices, which we summarize in the following.

\dbvisparagraph{Current Model Development Workflow}
All developers report beginning their workflow by researching promising architectures from related scientific resources (e.g., papers, blogs) and existing approaches (e.g., GitHub repositories).
Then, based on best practices and personal experience, they adapt these architectures to create and train an initial architecture for the intended task.
Following an iterative trial-and-error approach, this initial architecture is refined and re-trained until the results are satisfying.
As a reason for this fuzzy procedure, the developers mention a lack of guidelines, best practices, and suitable tools for model inspection.
Furthermore, hard-to-detect careless mistakes are identified as the most hindering obstacle in the model building and refinement process.
To track the different model variants and --iterations arising in this workflow, the developers rely on folder structures and file names, Jupyter notebooks, custom shell scripts, or the output of automated hyperparameter optimizers, c.f., Hyperopt \citep{Bergstra2013MakingScienceModel}.
For model assessment, they reportedly rely on performance metrics (e.g., loss or accuracy), which are logged and inspected using graphical tools (e.g., TensorBoard), Jupyter notebooks, or as raw (textual) command-line outputs.

\dbvisparagraph{Current Model Debugging Workflow}
To deduce actual needs and requirements for our system's design, we capture our developers' current debugging practice, including possible frustration factors and discontents with their current tool chain.
Only one of our participants reports that he ``occasionally'' systematically debugs his models, usually when there are fundamental problems with the model's performance.
When doing so, he would debug in a top-down approach, inspecting questionable model outputs, checking the activations of layers and neurons, and investigating other experiment factors, such as domain-specific loss functions.
The other participants would still rely on their trial-and-error-based process, change entirely to a new architecture, or, sometimes, use local XAI methods on individual dataset samples.
The quality assessment of a model in the trial-and-error process is mainly based on the local inspection of dataset samples combined with attribution methods.
As problems in this process, one of our participants names missing best practices and a lack of leverage points to refine models.

\subsection{Implications for Model Debugging}
\label{inn:apx:subsec:implications}

\sumbox{\ref{inn:apx:subsec:implications}}{
    In our study, we let our participants identify crucial data types for model debugging, common pitfalls, and use cases where systematic debugging tools would be beneficial.
    Despite the apparent benefits, some developers resist approaches for systematic debugging, favoring personal experience.
    Our findings advocate integrating systematic debugging into the deep learning workflow to improve model performance and decision-making.
}

As the core of our requirements study, we collect wishes and open challenges from our model developers, which they experience in their daily workflows.
From these insights, we derive implications and actionable suggestions on how a system enabling the systematic debugging of deep learning models should look like.

\dbvisparagraph{Important Data for Model Debugging}
\label{inn:paragraph:data-for-model-debugging}
The debugging of deep learning models involves a variety of data arising in the machine learning experiment (c.f., \cref{inn:subsec:data}).
Therefore, our interviews capture the data that our model developers consider important for model assessment and --comparison.
Textual and graphical representations of performance metrics were rated by far the most relevant.
For time-critical use cases, this also includes the computational complexity of the network.
For high-level model comparison, abstract hyperparameters like the number of layers and number of trainable parameters were named important.
For more detailed architecture inspection, layer types, activation functions, loss functions, optimizers, and learning rates were of interest for our developers.
Particularly important for debugging was the possibility to log and visualize activations for data samples.

\dbvisparagraph{Use Cases Suggested by Study Participants\label{inn:paragraph:requirements-use-cases}}
We collect a variety of use cases, where either in-depth debugging or the ability to inspect particular data in the model post-training would have been useful.

Careless mistakes were reported to be the most prevalent frustration factor, often affording time-consuming narrowing down of the problem.
One developer mentioned the use of a sigmoid activation in the last layer as cause of such error, naturally preventing the model to adapt to a regression task.
However, often the root cause of such seemingly simple problems is hard to identify due to the abundance of possible error sources.
With easy, simultaneous access to model architecture and the shape of activation distributions, this issue could likely be resolved much quicker.
Another developer listed diverse similar situations, including vanishing gradients through the wrong choice of activation functions, erroneous data preprocessing (wrong features boosted, timestamps not converted to UTC), or confused order of Numpy matrix indices.
Notably, the latter situations emphasize a fundamental challenge in the debugging of deep learning models: their tremendous ability to adapt to arbitrary data makes the model learn \emph{something}, even if the data is wrongly (pre)processed.
In conjunction with the model's opaqueness, issues are hard to detect and locate since no strict exception occurs; instead, the results often just get (slightly) worse.

Aside from these everyday-mistakes, several other use cases for systematic debugging were identified, most common of which was to balance losses (e.g., the generator and discriminator in GANs) or to assess the influence of frequently used techniques (e.g., dropout, activation functions, or batch normalization).

Complementing the expert requirement interviews, we captured additional use cases voiced by data science-- and machine learning researchers outside our expert study group.
One researcher questioned the usefulness of $1 \times 1$ convolutions typically part of inception blocks.
Observing the layer in-- and output and statistics on the layer activations could be helpful to evaluate their use as a dimensionality reduction technique~\citep{Szegedy2015GoingDeeperConvolutions}.
Another researcher mentioned beginners mistakes as a significant frustration factor when entering the field of deep learning.
Particularly, he got confused when experimenting with auto-encoder tutorials and mistakenly used class labels as optimization target.
One machine learning engineer had a particular interest in observing the values of different kernel initializations post-training to evaluate their influence on the formation of capable sub-networks~\citep{Frankle2018LotteryTicketHypothesis}.
Finally, one deep learning developer experienced stagnant training and poor reconstruction quality when building a variational auto-encoder.
After hours of manual debugging, he realized that a leaky relu activation function in the latent space interfered with the gaussian shape of the latent distribution.
This could have been easily recognized with suitable tools to inspect the distribution of latent variables in combination with a detailed architecture representation.

\dbvisparagraph{The Importance of Systematic Debugging}
Surprisingly, our participants occasionally show a very conservative opinion about using systematic debugging to substantiate their decision-making in the model building and verification workflow.
For example, when asked whether they would like to inspect and attribute the responses of single neurons, e.g., to determine the ideal size of a GAN's latent space, they instead wanted to stick to best practices and personal experiences: ``in a GAN, always use 100.''

We argue that this example highlights how model inspection and debugging should be a fundamental part of every informed DL workflow;
the factors playing a crucial role in the performance of a model should not be chosen based on guesses, leading to serendipitous results and leaving the potential of an architecture unexplored.

\end{document}